\pdfoutput=1
\documentclass[aps,prd,amsmath,floats,floatfix,onecolumn,superscriptaddress,nofootinbib,showpacs]{revtex4-2}

\usepackage{comment}
\usepackage{ragged2e}
\usepackage[T1]{fontenc}
\usepackage[utf8]{inputenc}
\usepackage{lmodern}
\usepackage{lipsum}

\usepackage{mathabx}
\usepackage{physics}
\usepackage{mathabx}
\usepackage{array}
\usepackage[dvipsnames]{xcolor}
\definecolor{linkcolor}{rgb}{0.0,0.3,0.5}
\usepackage[hypertexnames=false, unicode, colorlinks=true, linkcolor=linkcolor,
citecolor=linkcolor, filecolor=linkcolor,urlcolor=linkcolor,
pdfusetitle]{hyperref}

\usepackage[all]{hypcap}
\usepackage{graphicx}
\usepackage{xspace}
\usepackage[normalem]{ulem} 
\usepackage{bm} 
\usepackage{appendix}
\usepackage[small]{caption}
\usepackage{subcaption}

\usepackage{microtype}

\usepackage[english]{babel}
\usepackage{blindtext}

\graphicspath{%
  {figs/}%
}

\usepackage{tikz}
\usetikzlibrary{shapes,snakes}
\usetikzlibrary{arrows,intersections,patterns,decorations.markings,shapes.geometric}
\usetikzlibrary{calc,fadings,decorations.pathreplacing,positioning}
\usetikzlibrary{decorations.pathmorphing}
\tikzset{snake it/.style={decorate, decoration=snake}}

\usepackage{pgfplots}
\usepgfplotslibrary{colormaps}
\usetikzlibrary{intersections,patterns,pgfplots.fillbetween}

\tikzset{->-/.style={decoration={
  markings,
  mark=at position .5 with {\arrow{>}}},postaction={decorate}}
}
\tikzset{-<-/.style={decoration={
  markings,
  mark=at position .5 with {\arrow{<}}},postaction={decorate}}
}

\tikzset{%
  >=latex, 
  inner sep=0pt,%
  outer sep=2pt,%
  mark coordinate/.style={inner sep=0pt,outer sep=0pt,minimum size=3pt,
    fill=black,circle}%
}

\DeclareMathAlphabet{\mathpzc}{OT1}{pzc}{m}{it}

\definecolor{darkred}{RGB}{175,0,0}
\definecolor{darkblue}{RGB}{14,0,185}
\definecolor{salmon}{RGB}{255,160,105}

\definecolor{darkblue}{RGB}{14,0,185}

\newcommand*{\mH}{\mathcal{H}}
\newcommand*{\bk}{\mathbf{k}}
\newcommand*{\bq}{\mathbf{q}}
\newcommand*{\bp}{\mathbf{p}}

\newcommand*{\bx}{\mathbf{x}}

\newcommand*{\fnl}{f_{\rm NL}}

\newcommand*{\deltam}{\delta_{\rm M}}

\newcommand{\beq}{\begin{equation}}
\newcommand{\eeq}{\end{equation}}
\newcommand{\cell}[3]{\parbox[c][#2][c]{#1}{\centering #3}}

\begin{document}

\title{Optimal marked statistics from the Effective Field Theory of Large Scale Structure}

\author{Marco Marinucci}
\affiliation{Institute for Theoretical Physics, ETH Zurich, 8093 Zurich, Switzerland}
\affiliation{Dipartimento di Fisica Galileo Galilei, Universit\` a di Padova, I-35131 Padova, Italy}
\affiliation{INFN Sezione di Padova, I-35131 Padova, Italy}
\author{Michele Liguori}
\affiliation{Dipartimento di Fisica Galileo Galilei, Universit\` a di Padova, I-35131 Padova, Italy}
\affiliation{INFN Sezione di Padova, I-35131 Padova, Italy}
\author{Federico Semenzato}
\affiliation{Dipartimento di Fisica Galileo Galilei, Universit\` a di Padova, I-35131 Padova, Italy}
\affiliation{INFN Sezione di Padova, I-35131 Padova, Italy}
\author{Benjamin D. Wandelt}
\affiliation{Department of Physics and Astronomy, Johns Hopkins University, 3400 North Charles Street, Baltimore, MD, 21218, USA}
\affiliation{Department of Applied Mathematics and Statistics, Johns Hopkins University, 3400 North Charles Street, Baltimore, MD, 21218, USA}

\begin{abstract}
We present a simple and robust framework to assess the information content of the power spectrum of a marked density field within the Effective Field Theory of Large Scale Structure, providing a physically motivated regime of validity of perturbation theory for such fields.  We optimize the choice of mark to maximize the constraining power on cosmological parameters, focusing on the resulting reduction of parameter uncertainties. After marginalizing over the small-scale counterterms, we find improvements in the parameter error bars by a factor of approximately $1.30$--$2.3$, $1.3$--$8.3$, and $1.2$--$2.5$ for $\Omega_m$, $\sigma_8$, and $h$, respectively, depending on redshift and smoothing scale. The optimized mark exhibits a simple functional form, for which we provide a physical interpretation in terms of its response to the underlying density field. These results demonstrate the potential of marked statistics as a tool to extract additional cosmological information beyond standard two-point analyses and open a pathway for further exploration of nonlinear transformations of the density field.
\end{abstract}

\maketitle

\tableofcontents

\section{Introduction}

The next generation of large scale structure (LSS) surveys, including Euclid~\cite{EUCLID:2011zbd, Euclid:2024yrr}, DESI~\cite{DESI:2016fyo}, SPHEREx~\cite{SPHEREx:2014bgr}, the Vera Rubin Observatory~\cite{LSSTDarkEnergyScience:2012kar} and the Nancy Grace Roman Space Telescope~\cite{Wang:2021oec}, will map the three-dimensional distribution of matter across an unprecedented volume of the observable Universe. A central challenge is to extract as much cosmological information as possible from the resulting datasets, particularly on mildly nonlinear scales where the signal is rich but the theoretical modeling is demanding. Doing so will substantially sharpen constraints on the full set of cosmological parameters, including the matter density $\Omega_m$, the amplitude of fluctuations $\sigma_8$, the Hubble constant $H_0$, neutrino masses, and primordial non-Gaussianity.

The workhorse observable of LSS analyses is the matter power spectrum $P(k)$, the two-point function of the density contrast $\delta(\mathbf{x})$ in Fourier space. At linear order, this is a near-complete statistic for Gaussian initial conditions, but gravitational evolution generates non-Gaussianities that lock significant information into higher-order correlators. Recovering this information directly, via the bispectrum and beyond~\cite{Scoccimarro:1999ed, Scoccimarro:2000sn, Bernardeau:2001qr, Sefusatti:2006pa, Sefusatti:2010ee, Eggemeier:2021cam, Beyond-2pt:2024mqz}, is computationally expensive and theoretically challenging in the nonlinear regime.

A growing body of work has therefore explored \emph{alternative summaries} of the density field that are cheap to compute, analytically tractable, and yet capable of compressing information from multiple higher-order correlators simultaneously. Among the most promising are \emph{marked statistics}~\cite{Sheth:2005aj, White:2016yhs, Valogiannis:2017yxm, Philcox:2020fqx, Philcox:2020srd, Massara:2020pli, Massara:2022zrf, Karcher:2024twr, Cowell:2024wyl, Cowell:2025mov, Ebina:2024zkv, Ebina:2026qzf, Marinucci:2024bdq, Massara:2024cvu, Bianchi:2026liy}: one constructs a nonlinearly weighted version of the density field and computes its power spectrum. The resulting \emph{marked power spectrum} (MPS) is sensitive to the clustering environment and can encode higher-order information while retaining the simplicity of a two-point function.

The most widely studied mark was introduced in~\cite{White:2016yhs}, which proposed weighting the local density field by a function of the overdensity $\delta_R$ smoothed on a scale $R$:
\begin{equation}
    m(\delta_R(\mathbf{x})) = \left(1 + \frac{\delta_R(\bx)}{1 + \delta_s}\right)^{-p}.
    \label{eq:white_mark}
\end{equation}
Positive values of $p$ up-weight underdense regions (voids), while negative $p$ enhances overdense ones. Ref.~\cite{Massara:2020pli} showed that this mark provides significant improvements on $\sigma_8$, neutrino masses, and modifications of gravity, with the common interpretation that underdense environments are particularly sensitive to variations of these parameters.

A natural question then arises: is the mark in Eq.~\eqref{eq:white_mark} optimal, or could alternative choices capture more of the available information, for instance on the standard parameters or on primordial non-Gaussianity? This question was initially addressed in~\cite{Cowell:2024wyl}, where the functional form $m(\delta_R)$ was treated as a free function and optimized to maximize the Fisher information on a given parameter set, using Gaussian processes with empirical covariances and derivatives estimated from $N$-body simulations. Significant improvements were found across multiple parameters. An important observation in that work is that the Fisher matrix is invariant under \emph{affine transformations} of the mark function, $m \to A\,m + B$. As a consequence, a void-enhancing mark and an overdensity-enhancing mark carry identical information when the full data vector $\{P, C, M\}$ is used. The true driver of information gain is therefore the \emph{environmental contrast} modulated by the mark, rather than the preferential weighting of voids specifically. This result relies on the assumption of a Gaussian likelihood, which is a good approximation for most standard observables in the absence of strong degeneracies in the parameter space. Significant deviations from Gaussianity could in principle modify the picture.
Recent machine learning frameworks have generalized the mark optimization from a fiducial cosmology to a broad parameter prior, extending the functional form of the mark through rotationally equivariant neural architectures to isolate local morphological features of the cosmic web~\cite{Semenzato:2026stt}.

The MPS can also be studied \emph{analytically}, using the Effective Field Theory of Large Scale Structure (EFTofLSS)~\cite{Carrasco:2012cv, Baumann:2010tm, Senatore:2014via, Senatore:2014vja}. Refs.~\cite{Philcox:2020fqx, Philcox:2020srd, Ebina:2024zkv, Marinucci:2024bdq} showed that expanding the mark as a power series in $\delta_R$ and computing the marked power spectrum at one-loop order yields predictions in good agreement with $N$-body simulations up to $k \lesssim 0.3\,h\,\text{Mpc}^{-1}$.

In this work, we combine the analytical control of EFTofLSS with the flexibility of the mark optimization program. The key idea is to expand the mark as a polynomial in powers of $\delta_R$, as is usually done in perturbation theory, and treat the coefficients of the expansion as free parameters:
\begin{equation}
    m(\delta_R) = c_0 + c_1\,\delta_R + c_2\,\delta_R^2 + c_3\,\delta_R^3 + \mathcal{O}(\delta_R^4)\,,
\end{equation}
and optimize them to maximize the Fisher information. This turns the EFT expansion into a flexible polynomial basis, playing a role analogous to the Gaussian process (GP) approach of~\cite{Cowell:2024wyl} but within a fully analytical theoretical framework. Affine invariance of the Fisher matrix is a necessary ingredient of the optimization: without it, the problem is degenerate, since a constant shift of the mark leaves all observables invariant up to a linear rescaling. Imposing such invariance reduces the free parameter space to $\{c_1, c_2, c_3\}$ (with $c_0 = 1$ fixed by convention). A central aspect of our framework is the restriction of the optimization domain to the perturbative regime, where higher-loop contributions remain suppressed with respect to the one-loop result~\cite{Braganca:2023pcp, Spezzati:2025zsb}. This physically motivated constraint, which we derive and discuss in detail below, substantially restricts the space of viable marks and ensures the theoretical consistency of the optimization. The Fisher information serves as the cost function, and the optimization runs in minutes on a laptop.

We validate our optimal marks against 1000 realizations of the Quijote simulation suite~\cite{Villaescusa-Navarro:2019bje} and find that our analytical predictions reproduce the measured power spectra at the percent level. The resulting Fisher forecasts show improvements in the parameter error bars by a factor of approximately ${\sim}1.30$--$2.3$ on $\Omega_m$, ${\sim}1.3$--$8.3$ on $\sigma_8$, and ${\sim}1.2$--$2.5$ on $h$, depending on redshift and smoothing scale, in good agreement with~\cite{Cowell:2024wyl}. The convergence of an analytical, simulation-free approach with a GP-based one validates both the EFTofLSS description of marked statistics and the broader picture that a polynomial expansion is a sufficient functional basis for mark optimization in the perturbative regime. Because the pipeline is fast and fully analytic, it is a natural ingredient for future MCMC analyses of actual survey data, as well as a stepping stone towards optimized marks for primordial non-Gaussianity, where simulation-based approaches are considerably more expensive.

The paper is organized as follows. In Sec.~\ref{sec:mark_theo} we introduce the marked field and review its perturbative description within EFTofLSS, defining the general polynomial mark and the one-loop expressions for $P(k)$, $C(k)$, and $M(k)$. In Sec.~\ref{sec:PTprior} we derive the perturbativity constraints on the $c_i$ coefficients. In Sec.~\ref{sec:forecast_details} we describe the Fisher matrix formalism, the covariance model, and the optimization procedure. In Sec.~\ref{sec:results} we present our main results: the optimal marks, the comparison with Quijote simulations, and the forecasted parameter improvements. We conclude and outline future directions in Sec.~\ref{sec:conclusions}.

\section{Theoretical model}
\label{sec:mark_theo}
\subsection{General mark function}
In this work we will focus on the dark matter density contrast field $\delta(\bx)$, modeled in real space. Extensions to redshift space and biased tracers would bring the framework into direct contact with galaxy survey data and are left for future work.
The dark matter density field $\delta(\bx)$ is defined as
\begin{equation}
    \delta(\bx) = \frac{\rho (\bx)}{\bar{\rho}} -1\,
    \label{eq:deltadef}
\end{equation}
where $\rho(\bx)$ is the matter density and $\bar{\rho} \equiv \langle\rho(\bx)\rangle$ is its mean value at a fixed time. Given a density perturbation field, one can always define its smoothed version on a comoving scale $R$ as 
\begin{equation}
    \delta_R (\bx) = \int \, d^3 \mathbf{z} W_R(|\mathbf{z}|) \delta(\bx - \mathbf{z})\,, \quad \delta_R(\bk) = W_R(k) \delta(\bk)\,,
\end{equation}
where $W_R$ is a generic smoothing function that we will assume to be gaussian in Fourier space\footnote{In Fourier and in configuration space it reads\begin{equation}
    W_R(k) = e^{- \frac{k^2 R^2}{2}}\,,\quad W_R(x) \equiv \int\frac{d^3\bk}{(2\pi)^3} W_R(k) =  \frac{1}{(2\pi)^{3/2}}\frac{e^{-\frac{x^2}{2R^2}}}{R^3}\,.
\end{equation}}. Previous works~\cite{Sheth:2005aj, White:2016yhs, Valogiannis:2017yxm, Philcox:2020fqx, Philcox:2020srd, Massara:2020pli, Massara:2022zrf, Karcher:2024twr, Cowell:2024wyl, Cowell:2025mov, Ebina:2024zkv, Ebina:2026qzf, Marinucci:2024bdq, Massara:2024cvu} have shown that this quantity can be exploited to construct a new, nonlinear map of the density field itself, in order to extract more information from the full dark matter distribution and environment. The technical details of the construction of the marked field are presented in Sec.~\ref{sec:forecast_details}. The mark function usually adopted in literature is~\cite{White:2016yhs, Massara:2020pli, Massara:2024cvu}
\begin{equation}
    m(\delta_R(\bx)) = \left(1 + \frac{\delta_R(\bx)}{1 + \delta_s}\right)^{-p}\,.
    \label{eq:massara_mark}
\end{equation}
This choice of the mark (not being the only one) is particularly suited to amplify, in a very intuitive way, the specific features of the density field environment: positive values of $p$ enhance under-dense regions (\textit{voids}), while over-dense regions (\textit{density peaks}) are magnified when $p<0$. 
This functional form of the mark was introduced based on the hypothesis that low-density regions are more sensitive to physics beyond the $\Lambda$CDM model, particularly in the context of massive neutrinos~\cite{Philcox:2020fqx, Philcox:2020srd, Massara:2020pli, Massara:2022zrf}, modified gravity~\cite{White:2016yhs, Valogiannis:2017yxm, Karcher:2024twr}, and primordial non-Gaussianities~\cite{Marinucci:2024bdq}. The marked field is then defined as
\begin{equation}
    \rho_M(\bx) = m(\delta_R(\bx)) \rho(\bx)\,,
\end{equation}
and, by defining the mean mark as 
\begin{equation}
    \overline{\rho}_M = \langle\rho_M(\bx)\rangle = \langle m\big(\delta_R(\bx)\big) \rho(\bx)\rangle \equiv \overline{m} \,\overline{\rho}\,,
\end{equation}
one can obtain the expression for the marked overdensity field
\begin{equation}
    \delta_M(\bx) = \frac{\rho_M(\bx) - \overline{\rho}_M}{\overline{\rho}_M} = \frac{m(\delta_R(\bx))}{\overline{m}} \left[1 + \delta(\bx)\right] - 1\,.
    \label{eq:deltaM}
\end{equation}
This mapping effectively provides a nonlinear transformation of the density field: this operation creates a new (weighted) field with additional nonlinear and non-Gaussian features in its two-point correlation function. This is the reason why one should expect the marked density field to encapsulate more information about the surrounding environment when the same scales are analyzed with the power spectrum of the unmarked density field. This is similar to the findings of~\cite{Marinucci:2024bdq}, where it was highlighted that the combination of the standard power spectrum and bispectrum has, in general, more information on cosmological parameters compared to an analysis with the marked power spectrum. We will investigate this point further in the following Sections. 

A related concern is that the nonlinear mapping may produce a field with large nonlinearities, potentially compromising the applicability of perturbation theory. To preserve perturbativity, we therefore impose a set of conditions on the weighting scheme. This constitutes one of the main novelties of this work and will be discussed in more detail in Sec.~\ref{sec:PTprior} and App.~\ref{app:PTloop}.

\subsection{EFTofLSS for the marked power spectra}

Perturbative methods based on the Effective Field Theory of Large Scale Structure~\cite{Baumann:2010tm, Carrasco:2012cv} have been shown to be flexible and accurate enough to reproduce the galaxy power spectrum and bispectrum at one-loop in redshift space with sufficient accuracy up to mildly nonlinear scales. Much effort is being put in to push the EFTofLSS to higher perturbative orders~\cite{DAmico:2022ukl, Bakx:2025pop, Bakx:2025cvu, Bakx:2025jwa, Anastasiou:2025jsy} and for higher n-point functions~\cite{Steele:2021lnz, Gualdi:2020eag, Gualdi:2021yvq, Spezzati:2025zsb}. Refs.~\cite{Philcox:2020fqx, Philcox:2020srd, Ebina:2024zkv, Marinucci:2024bdq} showed that this analytical procedure can be safely applied to marked correlators, producing results that are in good agreement with numerical simulations, both for dark matter and halos. Here, we briefly review the methodology and the main results obtained in this framework. Throughout the analysis, we work in real space and consider the dark matter density field. Extensions to redshift space and biased tracers are natural next steps, left for future work.

With the same spirit as previous works, which adopted the mark function in Eq.~\eqref{eq:massara_mark}, we start by expanding $m(\delta_R)$ as a power series of $\delta_R$ 
\begin{equation}
    m(\delta_R(\bx)) = \sum_{n=0}^{\infty} (-1)^n \frac{p(p+1)\dots (p+n-1)}{n!} \frac{\delta_R^n(\bx)}{(1 + \delta_s)}\equiv \sum_{n=0}^{\infty}c_n(p, \delta_s) \delta_R^{n}(\bx)\,,
    \label{eq:m_exp}
\end{equation}
with 
\begin{equation}
    c_n(p, \delta_s) \equiv (-1)^n\frac{p(p+1)\dots (p+n-1)}{n!\,(1 + \delta_s)^n}\,.
\end{equation}
For this series to be convergent, the first requirement is that $|\delta_R(\bx)/(1 + \delta_s)|\lesssim 1$, which can be translated to the condition on the fluctuation of the smoothed field
\begin{equation}
    \langle\delta_R^2(\bx)\rangle \equiv \sigma_{RR}^2(z) \lesssim \frac{1 + \delta_s}{D(z)}\,,
\end{equation}
where $D(z)$ is the growth function for the linear matter density, $\delta_R(z)\simeq (D(z)/D(z_{in}))\delta (z_{in})$. When considering the mark function employed in~\cite{Massara:2020pli}, the $c_n$ parameters that appear in the perturbative expansion of Eq.~\eqref{eq:m_exp} are functions of the parameters $p$ and $\delta_s$. In this work, we will consider a more general form of the mark function by leaving these parameters free to vary. In practice, since we are interested in the one-loop power spectrum, we will consider the mark function expanded up to third order in $\delta_R$
\begin{equation}
    m(\delta_R) = c_0 + c_1 \delta_R + c_2 \delta_R^2 + c_3 \delta_R^3 + \mathcal{O}(\delta_R^4)\,,
    \label{eq:markCi}
\end{equation}
with $c_n$'s now free to vary. We still need this expansion to be convergent and meaningful in a perturbative sense. We will investigate more about this in the following Section. 
We can now proceed with the perturbative expansion up to the third order of the marked field in Eq.~\eqref{eq:deltaM}
\begin{equation}
    \delta_M(\bx) = \sum_{n=0}^{3} \delta_M^{(3)}(\bx)\,,
\end{equation}
where $(n)$ indicates the $n$-th perturbative order, and we have defined
\begin{equation}
    \begin{aligned}
        \delta_M^{(1)} (\bx) =&\, c_0 \delta^{(1)}(\bx) + c_1 \delta_R^{(1)}(\bx)\,,\\
        \delta_M^{(2)} (\bx) =& \,c_0 \delta^{(2)}(\bx) + c_1 \delta^{(1)}(\bx)\delta_R^{(1)}(\bx) + c_1 \delta_R^{(2)}(\bx) + c_2\delta_R^{(1)}(\bx)\delta_R^{(1)}(\bx)\,\\
        \delta_M^{(3)}(\bx) =&\, c_0 \delta^{(3)}(\bx) + c_1\delta_R^{(1)}(\bx)\delta^{(2)}(\bx) + c_1 \delta_R^{(2)}(\bx)\delta^{(1)}(\bx) + c_1 \delta_R^{(3)}(\bx)\\
        &+c_2 \delta_R^{(1)}(\bx)\delta_R^{(1)}(\bx)\delta^{(1)}(\bx) + 2 c_2 \delta_R^{(1)}(\bx)\delta_R^{(2)}(\bx) + c_3 \delta_R^{(1)}(\bx)\delta_R^{(1)}(\bx)\delta_R^{(1)}(\bx)\,.
    \end{aligned}
\end{equation}

In order to find the perturbative solution for the dark matter field and its smoothed version, it is useful to start from the dynamics of the dark matter fluid, which can be described by the standard continuity and Euler equations in combination with the Poisson equation
\begin{align}
\label{eq:continuity}
    &\dot{\delta} + a^{-1}\partial_i\big((1 + \delta)v^i\big) = 0\,,\\
\label{eq:Euler}
    &\dot{v}^{i} + H v^i + a^{-1} v^{j} \partial_j v^i + a^{-1}\partial_i \Phi = -\frac{\partial_j \tau^{ij}}{a \rho}\,,\\
\label{eq:poisson}
    &\frac{\nabla^2 \Phi}{a^2 H^2} = \frac{3}{2}\Omega_m(a) \delta\,,
\end{align}
where $\rho$ is the density of the energy of dark matter, $\delta$ is the density contrast already introduced in Eq.~\eqref{eq:deltadef}, $v^i$ is the perturbation of the velocity of dark matter, and $\tau_{ij}$ is the effective stress tensor. A dot here represents a derivative with respect to the cosmic time $t$, while $H = d\log{a}/dt$ is the Hubble rate and $\Omega_m = \bar{\rho}_m/\rho_{\rm crit}$ is the (non-relativistic) matter density parameter. These functions of time are analytically defined in a $\Lambda$CDM universe with a cosmological constant and no curvature as
\begin{equation}
    H(t) = H_0\sqrt{\Omega_{m,0} \, a(t)^{-3} + \Omega_{\Lambda,0}}\,,\qquad \Omega_m(t) = \frac{\Omega_{m,0}\,a(t)^{-3}}{\big(H(t)/H_0\big)^2}\,,\qquad \Omega_\Lambda(t) = 1 - \Omega_m(t)\,\,
    \end{equation}
with $H_0$, $\Omega_{m,0}$, and $\Omega_{\Lambda,0}$ respectively being the values of the Hubble parameter, the matter density parameter, and the cosmological constant density parameter today~\cite{Planck2018}.  The right-hand side of the Euler equation in Eq.~\eqref{eq:Euler} represents the effective stress-energy tensor, which describes the effect of short nonlinear modes on the dynamics of the long modes. The EFTofLSS provides a theoretical framework that allows us to expand it as a perturbative series of the long wavelength gravitational potential, resulting, eventually, in the addition of the so-called \textit{counterterms}, which parameterize the cut-off dependence of the theory, effectively marginalizing and renormalizing the uncontrolled short scales. Here we will momentarily neglect the stress-energy tensor as we will briefly discuss its effect later. It is convenient to rewrite these equations in Fourier space and consider the scale factor $a(t)$ as the new time variable, such that $dt = da/\mH$, where we have defined the conformal Hubble rate $\mH  = aH$. Moreover, we also introduce the rescaled velocity divergence, defined as
\begin{equation}
    \theta \equiv -\frac{\partial_i v^i}{f \mH}\,,
\end{equation}
with $f \equiv d\log{D}/d\ln{a}$ being the growth factor. Neglecting vorticity modes, the continuity and Euler equations now become 
\begin{align}
\label{eq:continuity_Four}
&a\frac{d\delta(\bk, a)}{d a} - f(a) \theta(\bk, a) =  \,f(a)\int_{\bq_1}\int_{\bq_2}(2\pi)^3 \delta_D(\bk - \bq_{12})\alpha(\bq_1, \bq_2) \theta (\bq_1, a)\delta (\bq_2, a)\,,\\
&a \frac{d \theta(\bk,a)}{da} - f(a) \theta (\bk, a) + \frac{3}{2}\frac{\Omega_m(a)}{f(a)}(\theta(\bk, a) - \delta(\bk, a)) =  \,f(a)\int_{\bq_1}\int_{\bq_2}(2\pi)^3 \delta_D(\bk - \bq_{12})\beta(\bq_1, \bq_2) \theta (\bq_1, a)\theta (\bq_2, a)\,,
\label{eq:Euler_Four}
\end{align}
where we have already included the Poisson equation in the Euler equation and made use of the well known evolution equation for $f$
\begin{equation}
    a \frac{d f(a)}{da} + f^2 + \left( 2 + \frac{d\ln{H}}{d\ln{a}}\right)f - \frac{3}{2} \Omega_m(a) = 0\,.
\end{equation}
The functions of momenta $\alpha(\bq_1, \bq_2)$ and $\beta(\bq_1, \bq_2)$ are the standard dark matter interaction vertices~\cite{Bernardeau:2001qr}
\begin{equation}
    \alpha(\bq_1, \bq_2) = 1 + \frac{\bq_1,\cdot \bq_2}{q_1^2}\,,\qquad \text{and} \qquad \beta(\bq_1, \bq_2) = \frac{|\bq_1 + \bq_2|^2 \bq_1\cdot \bq_2}{2 q_1^2 q_2^2}\,
\end{equation}
and we have used the notation $\int_\bq \equiv \int \frac{d^3\bq}{(2\pi)^3}$ and $\bq_{1\dots n} \equiv \bq_1 + \dots + \bq_n$. As anticipated above, we can write the dark matter density and velocity contrasts as a perturbative expansion of the form
\begin{equation}
    \delta (\bk, a) = \sum_{n} \delta^{(n)}(\bk, a)\,, \quad \text{and} \qquad \theta (\bk, a) = \sum_{n} \theta^{(n)}(\bk, a)\,,
\end{equation}
which allows us to solve the equations~\eqref{eq:continuity_Four} and \eqref{eq:Euler_Four} order by order. The perturbative solution of these equations can then be expressed as a convolution in Fourier space of $n$ linear fields over momentum kernels
\begin{align}
\label{eq:Fn}
    \delta^{(n)}(\bk, a) =& \int\frac{d^3\bq_1}{(2\pi)^3}\dots\int\frac{d^3\bq_n}{(2\pi)^3} F_n(\bq_1, \dots, \bq_n) \delta^{(1)}(\bq_1, a) \dots \delta^{(1)}(\bq_n, a)\,,\\
    \theta^{(n)}(\bk, a) =& \int\frac{d^3\bq_1}{(2\pi)^3}\dots\int\frac{d^3\bq_n}{(2\pi)^3} G_n(\bq_1, \dots, \bq_n) \delta^{(1)}(\bq_1, a) \dots \delta^{(1)}(\bq_n, a)\,,
\label{eq:Gn}
\end{align}
where the $F_n$ and $G_n$ are the well known kernels for matter density contrast that can be obtained by imposing the so-called Einstein-de Sitter approximation for time evolution~\cite{Bernardeau:2001qr}, which translates to $\Omega_m = f^2$. In the same fashion, we can express the marked density field with the new convolution kernels as
\begin{equation}
    \delta_M^{(n)}(\bk,a) = \int\frac{d^3 \bq_1}{(2\pi)^3} \dots \int\frac{d^3 \bq_n}{(2\pi)^3} H_n(\bq_1, \dots, \bq_n) \delta^{(1)}(\bq_1, a)\dots \delta^{(1)}(\bq_n, a)\,,
\label{eq:Hn}
\end{equation}
where the expressions for the $H_n$ kernels up to third order are given by
\begin{align}
    H_1(\bq_1) &= C_{\deltam}(q_1) F_1(\bq_1) = C_{\deltam}(q_1)\,,\nonumber\\
    H_2(\bq_1, \bq_2) &= C_{\deltam}(q_{12}) F_2(\bq_1, \bq_2) + C_{\deltam^2}(q_1, q_2)F_1(\bq_1)F_1(\bq_2)\,, \label{eq:kern}\\
    H_3(\bq_1, \bq_2, \bq_3) &=  C_{\deltam}(q_{123}) F_3(\bq_1, \bq_2, \bq_3)  + 2 C_{\deltam^2}(q_1, q_{23})F_1(\bq_1)F_2(\bq_2, \bq_3) \nonumber\\
    &\quad+C_{\deltam^3}(k_1, k_2, k_3) F_1(\bk_1)F_1(\bk_2)F_1(\bk_3)\,, \nonumber
\end{align}
and the third order kernel has to be symmetrized over the momenta. Moreover, we have defined the functions
\begin{align}
    C_{\deltam}(q_1) = &\, c_0 + c_1 W_R(q_1)\,, \label{eq:Cd1}\\
    C_{\deltam^2}(q_1, q_2) = &\, c_2 W_R(q_1)W_R(q_2) + \frac{c_1}{2}[W_R(q_1) + W_R(q_2)]\,,\label{eq:Cd2}\\
    C_{\deltam^3}(q_1, q_2, q_3) = &\, c_3 W_R(q_1)W_R(q_2)W_R(q_3) \nonumber\\
    &\, +\frac{c_2}{3}[W_R(q_2)W_R(q_3) + W_R(q_1)W_R(q_3) + W_R(q_1)W_R(q_2)]\,,
    \label{eq:Cd3}
\end{align}
which encode the perturbative structure of the mark function through the coefficients $c_n$.

With all these ingredients, we are now ready to compute the theoretical prediction for the observables we will use in this work, the matter power spectrum, the marked power spectrum and the cross power spectrum among the two fields
\begin{equation}
    P(k) = \langle\delta(\bk)\delta(\bk')\rangle'\,,\quad {C}(k) = \langle\delta(\bk)\delta_M(\bk')\rangle'\,,\quad M(k) = \langle\delta_M(\bk)\delta_M(\bk')\rangle'\,,
\end{equation}
where $\langle\dots\rangle'$ represents the correlator normalized by the usual $(2\pi)^3 \delta_D(\bk + \bk')$ factor, and we have left the redshift dependence implicit to avoid clutter. Inserting the perturbative expansions into these expressions, we obtain the following expressions up to one-loop order
\begin{equation}
    \begin{aligned}
        P(k) &= P_{11}(k) + P_{22}(k) + P_{13}(k)\,,\\
        \overline{m}\,C(k) &= C_{11}(k) + C_{22}(k) + C_{13}(k)\,,\\
        \overline{m}^2\,M(k) &= M_{11}(k) + M_{22}(k) + M_{13}(k)\,.
    \end{aligned}
\end{equation}
For the standard power spectrum, the one-loop terms are the usual ones
\begin{equation}
\begin{aligned}
    P_{11}(k) &= P_{\rm lin}(k)\equiv \langle\delta^{(1)}(\bk)\delta^{(1)}(\bk')\rangle'\,,\\
    P_{22}(k) &= 2 \int_\bq \Big[F_2(\bk - \bq, \bq)\Big]^2 P_{\rm lin}(|\bk - \bq|)P_{\rm lin}(q)\,,\\
    P_{13}(k) &= 6 P_{\rm lin}(k) \int_\bq F_3(\bk, -\bq, \bq)P_{\rm lin}(q)\,,
\end{aligned}
\end{equation}
for the cross power spectrum, we obtain
\begin{equation}
    \begin{aligned}
        C_{11}(k) &= H_1(\bk) P_{\rm lin}(k)\,,\\
        C_{22}(k) &= 2 \int_\bq F_2(\bk - \bq, \bq)H_2(\bk - \bq, \bq) P_{\rm lin}(|\bk - \bq|)P_{\rm lin}(q)\,,\\
        C_{13}(k) &= 3 P_{\rm lin}(k)\int_\bq \Big[H_1(\bk) F_3(\bk, \bq, -\bq) + H_3(\bk, \bq, -\bq)\Big]P_{\rm lin}(q)\,,
    \end{aligned}
\end{equation}
and, finally, for the marked power spectrum, we have
\begin{equation}
    \begin{aligned}
        M_{11}(k) &= H_1^2(\bk) P_{\rm lin}(k)\,,\\
        M_{22}(k) & = 2 \int_\bq \Big[H_2(\bk - \bq, \bq)\Big]^2 P_{\rm lin}(|\bk - \bq|)P_{\rm lin}(q)\,,\\
        M_{13}(k) &= 6 H_1(\bk) P_{\rm lin}(k) \int_\bq H_3(\bk, -\bq, \bq) P_{\rm lin}(q)\,.
    \end{aligned}
\end{equation}
As already highlighted in~\cite{Marinucci:2024bdq}, the one-loop terms of the marked correlators (both the cross and the auto power spectra) present additional contributions due to the presence of the smoothed field in the mark function. The additional pieces are convolutions of bispectrum- and (disconnected) trispectrum-like terms with some kernels given by the marking procedure. These terms arise because the marked field is, by definition, more non-Gaussian, and already at the level of its two-point function, it contains information from higher order correlation functions of the un-marked field. This is why it is expected to gain more constraining power using marked correlators compared to the usual power spectrum analysis. However, the information gain is not expected to surpass that from a combination of higher order functions, such as the bispectrum and the trispectrum\cite{Spezzati:2025zsb}. This is particularly true if one requires the marked field to still preserve perturbativity. We leave details about this in Section~\ref{sec:PTprior} and Appendix~\ref{app:PTloop}. 

We now take into account the back-reaction of small-scale physics on the long wavelength modes. For this, we need to consistently reintroduce the stress-energy tensor in Eq.~\eqref{eq:Euler_Four}. This has been achieved with the EFTofLSS~\cite{Baumann:2010tm, Carrasco:2012cv}, which represents a systematic way to account for the UV cut-off dependence of the loop integrals. The stress-energy tensor is written as a Taylor expansion of the long-wavelength quantities, imposing the Equivalence Principle and solving the equations perturbatively. In practice, for the standard one loop power spectrum in real space, EFTofLSS introduces an additional term at third perturbative order, which, in Fourier space, reads $\delta^{\rm ct}(\bk) = - c_s^2 k^2 \delta^{(1)}(\bk)$. The $\sim k^2$ dependence can be easily understood as a consequence of the Equivalence Principle, while the \textit{speed of sound} $c_s^2$ is a parameter that renormalizes the UV dependence of the theory. In data analysis, $c_s^2$ is treated as a free parameter of the theory that has to be fitted with simulations or observational data. Ref.~\cite{Philcox:2020fqx, Philcox:2020srd, Marinucci:2024bdq} showed that when marked correlators are considered, no additional counterterms are required, and the relevant terms for the cross and marked power spectra are
\begin{align}
    C_{\rm ct}(\bk) &= - 2 c_s^2 k^2 H_1(k) P_{\rm lin}(k)\,\\
    M_{\rm ct}(\bk) &= - 2 c_s^2 k^2 H_1^2(k) P_{\rm lin}(k)\,.
\end{align}
For the shot noise and the IR-resummation of the power spectra, we follow~\cite{Marinucci:2024bdq}.

Summarizing, the one loop EFT model for the standard, cross, and marked power spectrum of matter in real space has the following forms\footnote{Notice that in this work we will limit our analysis to the matter field in real space in order to compare our optimization procedure to previous results. The analysis can be easily extended to biased tracers (galaxies) in redshift space and will be investigated in future works.}
\begin{align}
    P(k) &= P_{11}(k) + P_{22}(k) + P_{13}(k) + P_{\rm ct}(k)\,,\label{eq:P1loop_EFT}\\
    \widebar{m}\, C(k) &= C_{11}(k) + C_{22}(k) + C_{13}(k) + C_{\rm ct}(k)\,,\label{eq:C1loop_EFT}\\
    \widebar{m}^2 M(k) &= M_{11}(k) + M_{22}(k) + M_{13}(k) + M_{\rm ct}(k)\,, \label{eq:M1loop_EFT}
\end{align}
which eventually contain only the free parameter $c_s^2$. 

\subsection{Perturbative prior}
\label{sec:PTprior}
One of the key assumptions of our work is that one can safely apply perturbation techniques to marked fields as described above. Before moving on with the comparison to simulations and with the optimization of such weighting maps, we need to accurately consider if this is the case.
It is well known that in standard perturbation theory~\cite{Spezzati:2025zsb} the key quantity that controls nonlinearities is the variance of the density field at a given scale $k$, defined as
\begin{equation}
    \Delta^2(k) = \frac{1}{2\pi^2}\int_0^kdq\, q^2 P_{\rm lin}(q)\,,
\end{equation}
where $P_{\rm lin}(k)$ is the linear power spectrum.
Crucially, for perturbation theory to be accurate, this parameter must be small when evaluated at the maximum value of the wavenumber considered, $k_{\rm max}$. As an example, in a $\Lambda$CDM cosmology at redshift $=0$, the variance of the field is $\Delta^2\simeq 0.22$ at $k = 0.12\, h/\text{Mpc}$ and $\Delta \simeq 0.53$ at $k = 0.2\, h/\text{Mpc}$, getting smaller at higher redshift and larger for higher values of $k$, as expected. For perturbation theory to be valid, it is essential that not only the relative size of each individual term in perturbative calculations is of order $\sim\Delta^2(k_{\rm max})$, but also that the \textit{total} loop contribution to the correlation functions has the same size. This assumption may seem obvious, but it has nontrivial consequences: for a biased tracer, such as halos or galaxies, the total $m$-loop contributions to a given $n$-point function can have a large number of terms multiplied by a large number of different nuisance parameters. Assuming that the variance of the density fields controls the perturbative expansion and that it is small corresponds to the requirement that all the bias parameters and counterterms, despite being of order $\sim \mathcal{O}(1)$, cannot contribute to large loop corrections for a realistic tracer. In this work, we are considering the matter field; hence, these assumptions should always be valid, at least on large scales well above the nonlinear scale. However, the marking procedure produces a new field that is, in general, a biased and more nonlinear tracer of the underlying dark matter field. The new parameters introduced in the mark function, the $c_i$'s, distort the original dark matter field in a specific way, and we need to verify if the assumptions we made so far are still valid for the marked field. A way to estimate if the perturbative treatment of the $n$-point functions is under control is to use the following (approximate) signal-to-noise (SNR) ratio formula for a connected $n$-point function~\cite{Spezzati:2025zsb}
\begin{equation}
    (\text{SNR})_n^2 \simeq \int^{k_{\rm max}}\frac{d^{3}\bk_1}{(2\pi)^3}\dots \int^{k_{\rm max}}\frac{d^{3}\bk_n}{(2\pi)^3}\frac{\langle\delta(\bk_1)\dots\delta(\bk_n)\rangle_c^2}{P(k_1)\dots P(k_n)}\sim N_{\rm pix.} \Delta^{2(n-2)}(k_{\rm max})\left[1 + \mathcal{O}(\Delta^2(k_{\rm max})\right]\,,
    \label{eq:SNR}
\end{equation}
where $N_{\rm pix.} \equiv V\int^{k_{\rm max}}d^3\bk/(2\pi)^3$ is the number of pixels of Fourier modes in a given survey. This expression is approximate in two respects: first, the covariance is evaluated at leading (Gaussian) order, retaining only the disconnected contribution $P(k_1)\dots P(k_n)$ in the denominator; second, the final scaling assumes standard perturbation theory power counting, $\langle\delta^n\rangle_c\sim \Delta^{2(n-1)}$, which holds in the weakly nonlinear regime. Moreover, numerical prefactors of order $\sim \mathcal{O}(1)$ arising from the angular integration over the momentum structure of the correlators are not tracked. Note that the same parameter $\Delta^2$ controls both the SNR for the $n$-point functions and the corrections to the SNR due to higher loop contributions. This is expected in perturbation theory where all nonlinearities are indeed  controlled by $\Delta^2$\footnote{Notice that there are some special cases where new parameters can play a non-negligible role, causing a failure of the perturbative expansion, see the discussion in~\cite{Cabass:2023nyo}. For the marked power spectrum, similar effects can emerge due to the nonlinear remapping, as described in the main text.}. This result shows two main things: on one hand, the signal-to-noise in each higher order $n$-point function is $\sim\mathcal{O}(\Delta^2)$ smaller than the previous one, meaning that the information content of the nonlinear field is almost saturated by the leading summary statistics; on the other hand, Eq.~\eqref{eq:SNR} shows that, analogously, the $\ell$-loop contribution to the signal-to-noise ratio of the $n$-point function scales as $\sim\mathcal{O}\left(N_{\rm pix.}\Delta^{2(n+\ell-2)}\right)$. This scaling implies that higher-loop contributions are increasingly suppressed, reflecting the perturbative nature of the nonlinear fields up to a cutoff scale $k_{\rm max}$.

The same arguments must hold for the marked field since we are interested in using a perturbative approach to predict the shape of its summary statistics. Note that these assumptions are strictly needed within the EFTofLSS and are not required in other approaches, such as those employed in simulation based analyses or machine learning. For a marked field $\delta_M$, the parameter controlling the perturbative expansion is
\begin{equation}
    \Delta_M^2(k) = \frac{1}{2\pi^2}\int^{k_{\rm max}}dq \, q^2 M_{11}(q)\,,
    \label{eq:DeltaM}
\end{equation}
with $M_{11}(k) = C_{\delta_M}^2(k) P_{\rm lin}(k)$. We can expand Eq.~\eqref{eq:DeltaM} and obtain
\begin{equation}
\begin{aligned}
    \Delta^2_M(k) &= \frac{1}{2\pi^2}\int^{k_{\rm max}}dq \, q^2 P_{\rm lin}(q) - \frac{ c_1}{\pi^2}\int^{k_{\rm max}}dq \, q^2 W_R(q)P_{\rm lin}(q) + \frac{c_1^2}{2\pi^2}\int^{k_{\rm max}}dq \, q^2 W_R^2(q)P_{\rm lin}(q) \\
    &\equiv \Delta^2(k) - c_1 \Delta_{\rm W}^2(k) + c_1^2\Delta_{{\rm W}^2}^2(k)\,,    
\end{aligned}
\end{equation}
where we have assumed $c_0=1$, as we will do in the rest of the paper, and we have defined the integrals
\begin{equation}
    \Delta_{{\rm W}}^2(k)\equiv  \frac{1}{\pi^2}\int^{k_{\rm max}} dq\, q^2 W_R(q) P_{\rm lin}(q)\,, \qquad
    \Delta_{{\rm W}^2}^2(k)\equiv  \frac{1}{2\pi^2}\int^{k_{\rm max}} dq\, q^2 W_R^2(q) P_{\rm lin}(q)\,,
\end{equation}
which have values $\Delta_{\rm W}^2\simeq 0.07$ and $\Delta_{{\rm W}^2} \simeq 0.015$ for $k_{\rm max} \simeq0.25\,h/\text{Mpc}$ at redshift $z=0$ for $R = 30 \,\text{Mpc}/h$, $\Delta^2_{\text{W}} = 0.26$, and $\Delta_{\text{W}^2}^2 = 0.07$ with $R = 15 \,\text{Mpc}/h$. This means two things: a smaller smoothing radius $R$ makes the theory for the marked field less perturbative; however, the variance of the marked field can still be $\Delta_M^2\lesssim1$ if the value of $c_1$ is adjusted accordingly to respect this condition. As a consequence of this, we expect the assumption of perturbativity to be safely satisfied by the marked field. In addition, one can see that the one loop contribution to the signal-to-noise can produce terms that remain constant even at very large scales, mimicking the effect of a shot-noise-like term~\cite{Karcher:2024twr}. This is somewhat similar to what is discussed in~\cite{Cabass:2023nyo} for the case of biased tracers, where the term presenting the second order bias $b_2$ can give an unwanted non-negligible contribution to the nonlinear trispectrum in the squeezed limit. This kind of term is usually under perturbative control for real life tracers: in the case of marked fields, these terms may arise due to the weighting procedure and could lead to a potential breaking of the perturbation theory assumptions. We leave further details about this in App.~\ref{app:PTloop}

\begin{figure}[h]
    \centering
    \includegraphics[width=0.49\linewidth]{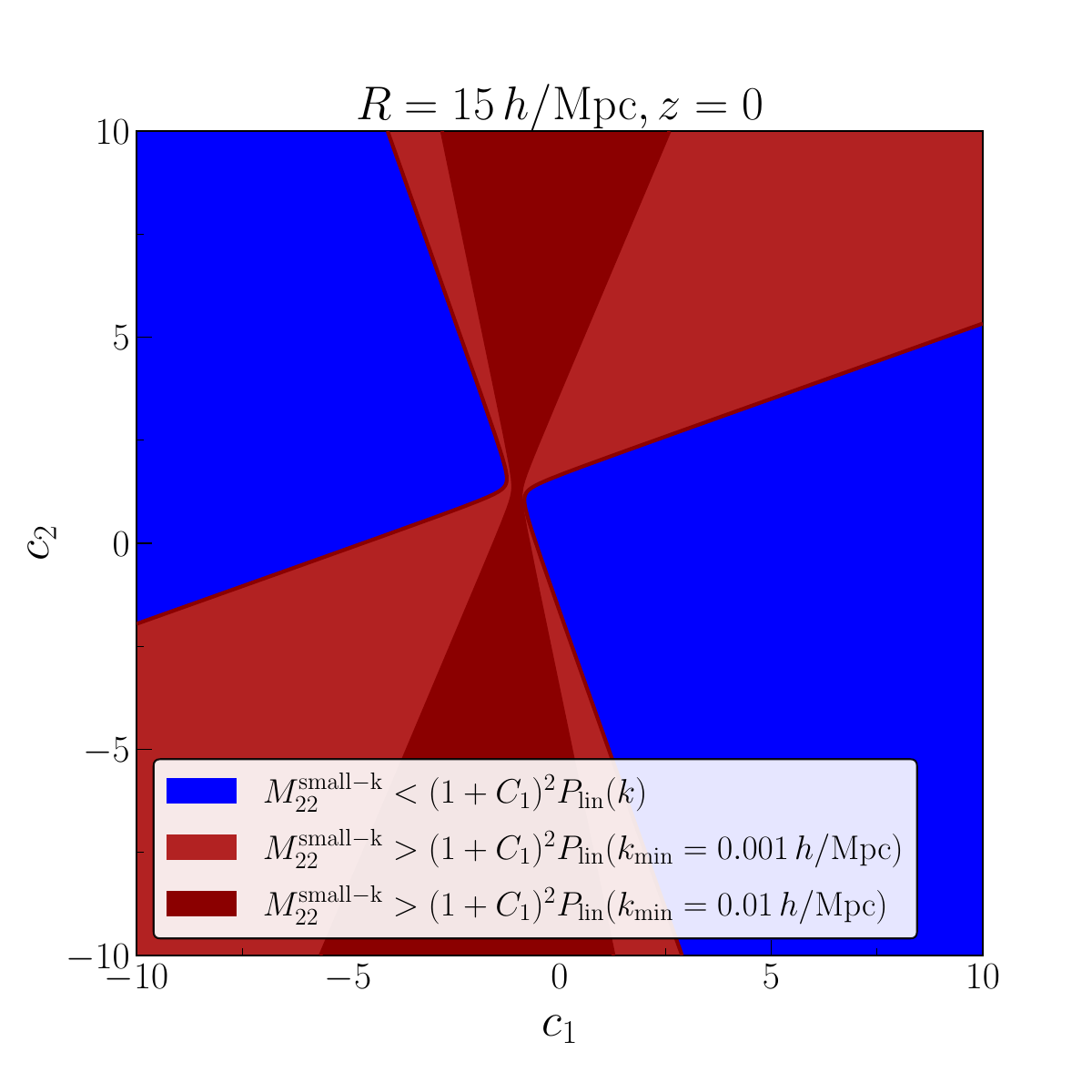}
    \includegraphics[width=0.49\linewidth]{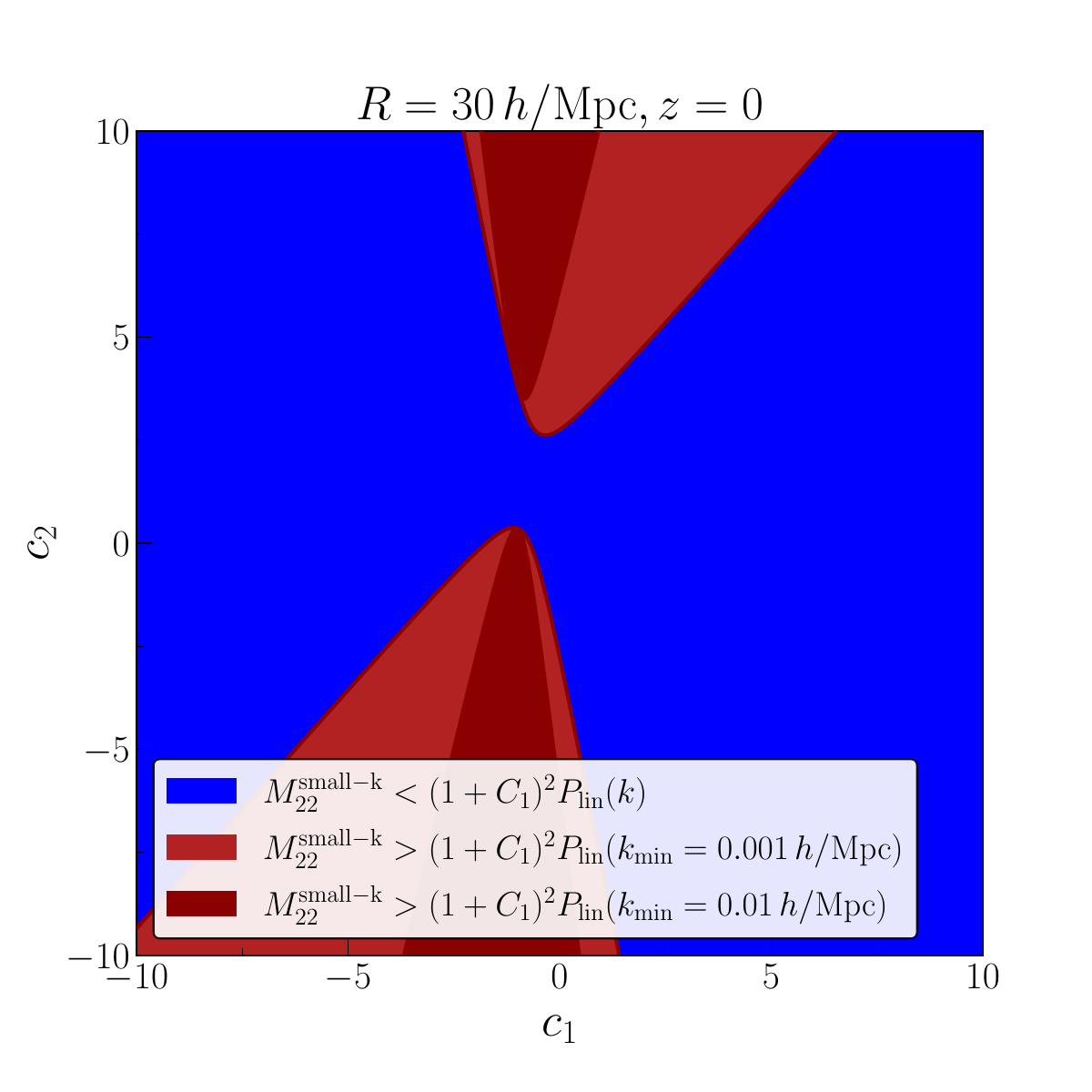}
    \caption{ \justifying Prior on the coefficients $c_1$ and $c_2$ imposed by requiring perturbativity of the marked field, shown for the two smoothing scales $R = 15 \,\mathrm{Mpc}/h$ and $R = 30 \,\mathrm{Mpc}/h$ at $z = 0$. The blue region indicates the allowed parameter space, while the red regions highlight the excluded ones for two different choices of the minimum wavenumber $k_{\rm min}$. The allowed region shrinks as the smoothing radius decreases, reflecting the more nonlinear structure probed at smaller $R$.}
    \label{fig:PTprior}
\end{figure}

Imposing that the numerator of Eq.~\eqref{eq:SNR_M22} is smaller than its denominator translates into a condition for $c_2$ and $c_1$, as shown in Fig.~\ref{fig:PTprior}. What is shown in this figure is somewhat expected: for larger values of the smoothing radius, the perturbative approach is more convergent, which leads to a larger allowed parameter space. 
It would be interesting, though, to explore the contributions coming from even higher loops than the one considered in this work. Perturbation theory requires that, for example, the two loop order is always smaller than the total power spectrum computed at one loop. This condition, usually referred to as \textit{perturbativity prior}~\cite{Braganca:2023pcp}, has been recently exploited in real data analysis~\cite{DAmico:2022osl, DAmico:2025zui} and relies on the arguments presented here. Interestingly, we could impose such perturbativity condition also for the marked power spectrum: in Appendix~\ref{app:PTloop}, we show that terms from the two loop marked power spectrum can give a non-negligible contribution on large scales. This should not surprise us, given the one loop example of $M_{22}$. What is interesting is that pieces coming from the two loop include also the mark parameter $c_3$ and other combinations of $c_2$ and $c_1$. In Appendix~\ref{app:PTloop}, we also show that for scenarios in which the perturbation theory is not valid anymore, nonlinear scales significantly leak into larger scales. For all the reasons stated above, we find it more convenient to impose a prior that is easier to implement, such that each successive term in the expression of the mark function must be suppressed with respect to the previous one when evaluated at the maximum value of the smoothed overdensity field $\delta_{R, {\rm max}}$, measured from the simulations, as reported in table~\ref{tab:delta_range}. This translates into the two conditions 
\begin{equation}
    |c_2| \delta_{R, \rm{max}}< |c_1|\,,\quad |c_3| \delta_{R, \rm{max}}< |c_2|\,.
    \label{eq:prior}
\end{equation}
Points in parameter space that violate either of these conditions are excluded from the scan, as they would correspond to a mark whose higher-order terms dominate the lower-order ones, invalidating the perturbative interpretation of the model. We will additionally require that the mark functions are positive in the range of $\delta_R$ sampled by the simulations. This condition is both physically and operationally motivated: a negative mark would flip the sign of the contribution of the corresponding region to the marked field, making the statistical interpretation of the marked power spectrum as a weighted summary statistic ill-defined. In practice, enforcing positivity ensures that $m(\delta_R)$ can always be interpreted as a well-defined, non-negative weight function assigned to each region of the density field. We further show in 

\begin{table}[t]
\centering
\renewcommand{\arraystretch}{1.3}
\resizebox{0.75\textwidth}{!}{%
\begin{tabular}{ccccccc}
\hline\hline
$R\,[\,h^{-1}\mathrm{Mpc}]$ 
& $\delta_{R,\mathrm{min}}(z=0)$ & $\delta_{R,\mathrm{max}}(z=0)$ 
& $\delta_{R,\mathrm{min}}(z=0.5)$ & $\delta_{R,\mathrm{max}}(z=0.5)$
& $\delta_{R,\mathrm{min}}(z=1)$ & $\delta_{R,\mathrm{max}}(z=1)$ \\
\hline
$30$ & $-0.41$ & $0.68$ & $-0.34$ & $0.49$ & $-0.26$ & $0.42$ \\
$15$ & $-0.66$ & $2.22$ & $-0.57$ & $1.54$ & $-0.45$ & $1.20$ \\
$10$ & $-0.77$ & $5.47$ & $-0.69$ & $3.58$ & $-0.56$ & $2.79$ \\
\hline\hline
\end{tabular}}
\caption{\justifying Minimum and maximum values of the smoothed overdensity field 
$\delta_R$, measured from the Quijote simulations at $z = \{0, 0.5, 1\}$ for different smoothing radii. Density contrasts are progressively suppressed at higher redshift, while smaller smoothing scales probe significantly larger overdensities at all redshifts.}
\label{tab:delta_range}
\end{table}

\section{Minimization procedure}
The aim of this paper is to identify the optimal configuration of the mark function that maximizes the information gain from marked correlators relative to a standard power spectrum-only analysis. In particular, we explore the constraining power on a set of cosmological parameters, $\{\Omega_{m}, \sigma_8, h\}$, while marginalizing over the EFT parameter $c_s^2$.
To this end, we vary the shape of the mark function parametrized by $\{c_0, c_1, c_2, c_3\}$, fixing $c_0 = 1$ as discussed in Sec.~\ref{sec:affine}. We scan a three-dimensional grid for $\{c_1, c_2, c_3\}$ in the range $[-10,10]$ subject to two physically motivated priors. We adopt a grid-based scan as we found gradient-based methods to converge to the same optimal mark at a higher computational cost. The mark function in Eq.~\eqref{eq:markCi} is intended as a controlled expansion around the smoothed density field, and for this reason, we impose a perturbative prior on the $c_i$ coefficients as discussed in~\ref{sec:PTprior}.
For each configuration, we compute the Fisher matrix using the theoretical model outlined above, including both derivatives and covariance as detailed in Secs.~\ref{sec:forecast_details} and \ref{sec:covariances}.
We compare the constraints obtained from the power spectrum alone with those from the combination of power spectrum, cross spectrum, and marked power spectrum, quantifying the improvement through the ratio $r(\theta) = \sigma_P(\theta) / \sigma_{P+C+M}(\theta)$ (or equivalently the FoM, see Sec.~\ref{sec:covariances}) for each of the selected parameters $\theta = \{\Omega_{m}, \sigma_8, h\}$, while marginalizing over the EFT parameter $c_s^2$.
After identifying the optimal mark configurations, we compute the full non-Gaussian covariance using 1000 realizations from the Quijote suite at the fiducial cosmology. This allows us to assess the actual improvement in parameter constraints when a realistic covariance is taken into account. Additional methodological details are provided in the following Sections.

\subsection{Details about the Fisher matrix}
\label{sec:forecast_details}

In this Section, we detail the procedure adopted to perform the Fisher forecast for cosmological parameters, combining the standard power spectrum, the cross power spectrum, and the marked power spectrum for the dark matter field.
We focus on the fiducial $\Lambda$CDM cosmology adopted in Quijote and consider variations in the matter density $\Omega_{m}$, the amplitude of fluctuations $\sigma_8$, and the reduced Hubble parameter $h=H_0/100$. These parameters have been shown to drive the largest variations in large scale structure observables at low redshift and hence are the most promising in terms of constraining power from LSS. In addition, we include and marginalize over the EFT counterterm $c_s^2$, which parametrizes the impact of small-scale physics on the one-loop power spectrum through a $k^2 P_{\rm lin}(k)$ correction; we will discuss possible degeneracies among these parameters and the role played by the marked power spectrum in breaking them in Section~\ref{sec:improvs}.

\noindent
The Fisher matrix provides an extremely useful tool for forecasts in cosmology~\cite{Tegmark:1997rp}. Forecast methods based on the Fisher matrix rely on the Cramér–Rao lower bound, which establishes that the covariance matrix of any unbiased estimator of the parameter set $\theta$ has a lower limit given by the inverse of the Fisher information matrix $F_{\mu\nu}$. The latter is defined as the expectation value of the Hessian of the log-likelihood
\begin{equation}
    F_{\mu\nu} = - \Bigg\langle\frac{\partial^2 \log{\mathcal{L}}}{\partial \theta_\mu \partial \theta_\nu}\Bigg\rangle\,.
\end{equation}
By assuming that, at leading order, the observables considered follow a Gaussian distribution with mean $D$ and covariance $C_D$, the Fisher matrix is
\begin{equation}
    F_{\mu\nu} = \sum_{i,j} \frac{\partial D_i}{\partial \theta_\mu} \big(C_D^{-1}\big)_{ij}\frac{\partial D_i}{\partial \theta_\nu}\,,
    \label{eq:Fisher}
\end{equation}
where the sum runs over the data bins. The inverse Fisher matrix in Eq.~\eqref{eq:Fisher} gives the lower limit for the error bar for a given parameter $\theta_\mu$, which satisfies $\sigma^2(\theta_\mu)\ge (F^{-1})_{\mu\mu}$. These estimates include marginalization over the other parameters considered in the forecast. In order to compute the Fisher matrix, one needs to assume fiducial parameters $\theta_{\rm fid.}$, over which the derivatives are estimated. The observables we consider in this work are the power spectrum alone $D_P = \{P(k)\}$ and the combination of the standard dark matter power spectrum $P(k)$, the cross power spectrum among the unmarked and marked field $C(k)$, and the power spectrum of the marked field $M(k)$, i.e., $D_{PCM} = \{P(k),\,C(k),\,M(k)\}$. We will compute the Cramér–Rao bounds in both cases and compare them to quantify the improvement in the resulting parameter constraints. As already discussed, each of these quantities encodes complementary information about the underlying density field: while $P(k)$ captures the clustering of matter on large scales, $M(k)$ weights the contribution of low- and high-density regions according to the adopted mark function, effectively probing different environmental regimes \cite{Sheth:2005aj, White:2016yhs, Valogiannis:2017yxm, Philcox:2020fqx, Philcox:2020srd, Massara:2020pli, Massara:2022zrf, Karcher:2024twr, Cowell:2024wyl, Cowell:2025mov, Ebina:2024zkv, Ebina:2026qzf, Marinucci:2024bdq, Massara:2024cvu}. When LSS data are considered, the three cosmological parameters examined in this analysis are the ones that are better constrained from the shape of the power spectrum: other $\Lambda$CDM parameters such as the amount of baryonic matter $\Omega_b$ or the primordial tilt $n_s$ are usually fixed using priors from other datasets, such as CMB data or measurements from BBN~\cite{DAmico:2019fhj, Ivanov:2019pdj, DESI:2024hhd}. However, given that our approach relies on a theoretical model, including additional parameters is straightforward; see, for example,~\cite{Marinucci:2024bdq}~\footnote{Notice that in ref.~\cite{Cowell:2024wyl} the authors vary only $\Omega_m$ and $\sigma_8$ due to the high computational power of their pipeline. In Sec.~\ref{sec:results} we show a more direct comparison.}. The choice of the reference values for the cosmological parameters is dictated by the fiducial cosmology of the Quijote simulation suite~\cite{Villaescusa-Navarro:2019bje}
\begin{equation}
\{\Omega_{m},\, \sigma_8,\, h,\, n_s,\, \Omega_b\} = \{0.3175,\, 0.834,\, 0.6711,\, 0.9624,\, 0.049\},
\end{equation}
while the fiducial value for the parameter $c_s^2$ is obtained by fitting the unmarked power spectrum at fixed cosmology. 
The derivatives with respect to a given cosmological parameter $\theta_\mu$ are computed using our analytical model using the symmetric finite-difference scheme:
\begin{equation}
    \frac{d D_i(\theta_\mu)}{d\theta_\mu} = \frac{D_i(\theta_\mu + (\Delta\theta_\mu)/2) - D_i(\theta_\mu - (\Delta \theta_\mu)/2)}{\Delta \theta_\mu}\,.
    \label{eq:derivative}
\end{equation}
 All other parameters are held fixed while varying $\theta$, ensuring that the derivative isolates the physical dependence of the observable on that specific cosmological parameter. This approach has been extensively validated in the context of Fisher forecasts for large scale structure \cite[e.g.][]{White:2016yhs, Yankelevich:2018uaz, Agarwal:2020lov, Braganca:2023pcp}. The finite steps $\Delta \theta$ are chosen to be small enough to accurately estimate the derivatives while avoiding numerical instabilities due to round-off errors: we use $\Delta \Omega_m = 0.02$, $\Delta\sigma_8 = 0.03$, and $\Delta h = 0.01$. We have verified that the derivatives computed using our analytical prescription with these values for the $\Delta$'s agree with those obtained numerically from the Quijote suite, which provides simulations at fiducial and displaced parameter values for derivative estimation, and the agreement is stable across all the parameters, the redshift, and the range of scales considered.

\subsection{Covariance matrices}
\label{sec:covariances}
The evaluation of the Fisher matrix for a given set of parameters requires an estimate of the covariance matrix of the observables involved in the forecasting procedure. Crucially, the covariance matrices of the cross and marked power spectra depend on the choice of the parameters $c_i$'s that enter the mark function as in Eq.~\eqref{eq:markCi}: this means that the covariances have to be recomputed each time during the optimization procedure, where the $c_i$'s are varied, even though the cosmological parameters are kept fixed to their fiducial values. As a first step in our optimization procedure, we use perturbation theory to compute the Gaussian part of the fields involved. As already clarified in Section~\ref{sec:PTprior}, non-Gaussian contributions to the covariance matrix (off-diagonal terms) are expected to be of order $\sim\Delta^2(k)$ and could be safely neglected for the redshift and the scale we are considering in this work.
The expressions we will use for the covariances of the auto-spectra are given by~\cite{Marinucci:2024bdq}
\begin{equation}
\begin{aligned}
    &\text{Cov}[P(k_i)P(k_j)] = 2 \frac{(2\pi)^3}{V V_s}P^2(k_i)\delta_{ij}\,,\\ &\text{Cov}[C(k_i)C(k_j)] = \frac{(2\pi)^3}{V V_s}\Big[P(k_i)M(k_i) + C^2(k_i)\Big]\delta_{ij}\,,\\
    &\text{Cov}[M(k_i)M(k_j)] = 2 \frac{(2\pi)^3}{V V_s}M^2(k_i)\delta_{ij}\,,
\end{aligned}
\end{equation}
while the cross-covariances between the different spectra are 
\begin{equation}
\begin{aligned}
    &\text{Cov}[P(k_i)C(k_j)] = 2 \frac{(2\pi)^3}{VV_s} P(k_i)C(k_i)\delta_{ij}\,,\\ &\text{Cov}[P(k_i)M(k_j)] = 2 \frac{(2\pi)^3}{VV_s} C^2(k_i)\delta_{ij}\,,\\
    & \text{Cov}[C(k_i)M(k_i)] = 2\frac{(2\pi)^3}{VV_s} C(k_i)M(k_i)\delta_{ij}\,.
\end{aligned}
\end{equation}
Specific configurations of the mark parameters could lead to a potentially large correlation between the dark matter density perturbation field $\delta$ and the marked field $\delta_M$: the particular case $\{c_0, c_1, c_2, c_3\} = \{1, 0, 0, 0\}$ would give a singular joint covariance matrix, manifesting itself as an ill-conditioned forecast case, whose inversion is numerically unstable. This is also true, for example, for mark configurations close to the one presented. To mitigate this, we employ the pseudo-inverse of the covariance matrix, defined using the Moore–Penrose formalism, and introduce a threshold parameter $f_{\rm cond} = 10^{-9}$ below which the eigenvalues are truncated, following~\cite{Cowell:2024wyl, 2023ApJ...946..107P}. This regularization ensures that the Fisher matrix defined in Eq.~\eqref{eq:Fisher} remains numerically stable and well defined. We tested the impact of the condition number threshold $f_{\rm cond}$. For larger values, such as $f_{\rm cond}=10^{-6}$, too many eigenvalues are truncated, leading to unphysically large error bars. In the range $f_{\rm cond}=10^{-9}$–$10^{-10}$, the estimated uncertainties are stable, varying at the $\sim10\%$ level, as shown in Fig.~\ref{fig:fcond}. Smaller values of $f_{\rm cond}$ instead lead to unstable and overly optimistic results. We therefore adopt $f_{\rm cond}=10^{-9}$ as a conservative choice, which we expect to yield more reliable results, particularly for smaller values of $R$.
This approach follows standard practice in Fisher analyses involving correlated observables \cite{Carron2013, Tegmark:1998wm}.

Given the relatively low computational power required to estimate the theoretical derivatives and the gaussian covariances, we vary the mark functions in order to identify the optimal configuration of $c_i$'s, which maximizes the cosmological information content. We explore two complementary optimization criteria: 
\begin{itemize}
    \item \textit{Global optimization}: maximization of the Figure of Merit (FoM) defined as
    \begin{equation}
        \text{FoM} = \sqrt{\text{det}(F_{\mu\nu})}\,,
    \end{equation}
    which measures the inverse hypervolume of the confidence ellipsoid  in parameter space. A large FoM indicates a smaller joint uncertainty on the parameters, corresponding to a more constraining dataset;
    \item \textit{Individual optimization}: minimization of the marginalized uncertainty on each parameter $\theta = \{\Omega_m, \sigma_8, h\}$ obtained using the Cramér-Rao bound
    \begin{equation}
        \sigma^2(\theta_\mu) = (F^{-1})_{\mu \mu}\,.
    \end{equation}
    This approach highlights whether the improvement from marked statistics primarily benefits specific cosmological directions (e.g. $\sigma_8$) rather than a global reduction of degeneracies. 
\end{itemize}
In practice, the FoM is sensitive to the degree of degeneracy among parameters, while the marginalized errors are more sensitive to the amplitude of the derivative signals. It is therefore not guaranteed that the mark configuration maximizing the FoM also minimizes the uncertainty on each parameter individually. We therefore explore both optimization strategies in parallel. Once the optimal configurations of the mark functions are obtained using a Gaussian covariance, we proceed to the estimate of the full non-Gaussian covariance using the Quijote simulations: we compute the variance of the different observables among 1000 different realizations at the fixed fiducial cosmology. This refined evaluation is necessary in order to have a more realistic covariance that captures possible mode couplings that are not included in the Gaussian one. 

This two-stage approach follows the methodology of \citet{Cowell:2024wyl} and ensures that our conclusions about the information gain from marked statistics are robust to the inclusion of realistic covariance effects.

The forecast is performed at the redshifts $z = 0$, $z=0.5$, and $z=1$, and for smoothing scales $R = 10\, \text{Mpc}/h$, $R = 15\, \text{Mpc}/h$, and $R = 30\, \text{Mpc}/h$. 
This choice ensures that the derived constraints are also representative of low-$z$ observables, where nonlinearities play an important role. The Fourier-space binning and survey volume are matched to those of the Quijote simulations~\cite{Villaescusa-Navarro:2019bje}, with $V = 1\, (\mathrm{Gpc}/h)^3$ and $ \Delta k = (2\pi)/V^{1/3} \simeq 0.0063\, h/\mathrm{Mpc}$.
Furthermore, we will use a maximum wavenumber of $k_{\rm max} = 0.30\, h/\text{Mpc}$. 

\subsection{Affine invariance and optimal weights}
\label{sec:affine}
One of the key results of~\cite{Cowell:2024wyl} resides in the exploitation of the invariance of the Fisher matrix under \textit{affine transformations} of the mark function, so that
\begin{equation}
    M(\delta_R)\to M'(\delta_R) = A M(\delta_R) + B\,, \qquad (F_{\mu\nu})^{-1}\to (F'_{\mu\nu})^{-1} = (F_{\mu\nu})^{-1}\,.
    \label{eq:affine}
\end{equation}
In our approach, the transformation in Eq.~\eqref{eq:affine} corresponds to a shift in the $\{c_n\}$ parameters
\begin{equation}
    \{c_0, c_1, c_2, c_3\}\to \{c_0', c_1', c_2', c_3'\} = \{A\,c_0 + B, A\,c_1, A\,c_2, A\,c_3\}.
    \label{eq:affineCi}
\end{equation}
One can easily see that the observables considered here, under this affine transformation, are changed as
\begin{align}
    P(k)&\to P'(k) = P(k)\,,\\ C(k)&\to C'(k) =  B\, P(k) + A\, C(k)\,,\\ M(k)&\to M'(k) = B^2 P(k) + 2 AB\, C(k) + A^2 M(k)\,,
\end{align}
and the covariance matrix transforms as $C\to C' = N \cdot C \cdot N^{\dagger}$, with 
\begin{equation}
    N \equiv \begin{pmatrix}
1 & 0 & 0 \\
B & A & 0 \\
B^2 & 2 AB & A^2
\end{pmatrix}\,.
\end{equation}

These transformation laws also hold for our theoretical modeling since perturbation theory only involves linear operations among the different spectra. The Fisher matrix is invariant under this affine transformation. Interestingly, the affine invariance of the Fisher information matrix also extends naturally to any complete set of $n$-point correlators of the density field $\delta$ and marked field $\Delta$ (by complete, we mean including all the available auto- and cross-spectra); this is shown explicitly in Appendix \ref{app:affine}.

Eq.~\eqref{eq:affineCi} indicates that the parameter space of the mark functions can be reduced: the invariance of the Fisher matrix under a constant shift of the mark function implies that the information content is independent of $c_0$, which can therefore be chosen arbitrarily. For convenience\footnote{We fix the value of $c_0$ to 1 and not, for example, to 0, to avoid instabilities in the theoretical modeling during the optimization operation.}, we will fix $c_0 = 1$. The invariance under linear scaling $m(\delta_R) \to A\, m(\delta_R)$ leaves  the ratio of the mark function unchanged for two different values of $\delta_R$, explicitly
\begin{equation}
    \frac{m(\delta_{R, 1})}{m(\delta_{R, 2})} \to \frac{m'(\delta_{R, 1})}{m'(\delta_{R, 2})} = \frac{m(\delta_{R, 1})}{m(\delta_{R, 2})}\,.
\end{equation}
In other words, this implies that if $g(c_1,  c_2, c_3)$ is a linear combination of the three parameters, the relation 
\begin{equation}
    g(c_1',  c_2', c_3') = \kappa(A)g(c_1,  c_2, c_3)\,,
\end{equation}
holds, with $\kappa$ being a generic function of $A$. A clear consequence of this is that what is relevant, in terms of information content, is not the absolute values of the mark parameters ${c_1, c_2, c_3}$, but their relative size, i.e., the ratios $c_2/c_1$ and $c_3/c_1$. Since we impose the perturbativity prior described in Eq.~\eqref{eq:prior} (see also the discussion in Appendix~\ref{app:PTloop}), which directly constrains these ratios (e.g. $|c_2| \lesssim |c_1|/\delta_{R, \rm{max}}$ and similarly for $c_3$), the physically relevant region of parameter space is naturally restricted. In this sense, the prior already reduces the effective redundancy associated with the overall normalization of the mark. This choice simplifies the numerical implementation and allows us to scan the space of mark shapes without explicitly fixing a normalization convention.

A consequence of affine invariance is that the mark function $m(\delta_R)$ itself does not provide physically meaningful information, since it is defined only up to an equivalence class under $m \rightarrow Am + b$. When a specific mark is found to improve constraints on a cosmological parameter, the apparent physical interpretation (for example, that information resides preferentially in underdense or overdense regions) is gauge-dependent and therefore not invariant. As the number of $n$-point correlators included in the data vector grows, the information becomes distributed across an increasing number of cross-correlators, making it progressively harder to attribute the information gain to any particular density environment. The physically meaningful quantity is the equivalence class $[m]$, not any particular representative.

This conclusion holds in the regime in which the likelihood is nearly gaussian and could be approximated with the Fisher matrix: highly non-gaussianities in the likelihood, such as those sourced by large degeneracies in the parameter space~\cite{Piga:2022mge, Carrilho:2022mon, Taule:2024bot, Tsedrik:2025hmj, Reeves:2025bxc}, may alter this picture and lead to different conclusions.

For explicitly rotation-invariant marks of the form introduced in~\cite{White:2016yhs}, or like those considered in this work, which depend only on the scalar smoothed density field $m(\delta_R)$, the full invariant content reduces to two quantities: the smoothing scale $R$, which sets the spatial scale at which the local environment is defined, and the relative weighting between density environments. Concretely, the affine-invariant content is encoded in ratios of the form

\begin{equation}
    \frac{\left|m(\delta_{R,1}) - m(\delta_{R,3})\right|}
         {\left|m(\delta_{R,3}) - m(\delta_{R,2})\right|}
\end{equation}

for fixed density values $\delta_{R,1}$, $\delta_{R,2}$, $\delta_{R,3}$. The sign of this contrast is itself gauge-dependent, since the reflection $m \to -m$ is an affine transformation with $A = -1$, $B = 0$, mapping a void-upweighting mark onto a halo-upweighting one while leaving the Fisher information unchanged.

This analysis also suggests a possible route to interpretability: removing the matter power spectrum $P(k)$ breaks the full affine invariance, since the shift symmetry $m \to m + b$ no longer closes on the reduced data vector, rendering the zero-crossing $\delta_R^\ast$ of the optimal mark a genuine invariant with a physical interpretation as the optimal density threshold separating information-rich from information-poor environments. This of course comes at a price, as becomes clear when examining the Fourier-space content of the marked statistics. At mildly nonlinear scales, the marked power spectrum is equivalent to a particular projection of the matter bispectrum, where the mark function acts as a convolution kernel weighting different triangle configurations. Specifically, $M(k)$ probes squeezed bispectrum configurations $B(k,\,k,\,q)$ with $q \lesssim 1/R$, since the smoothing suppresses small-scale modes in $m(\delta_R)$, while $C(k)$ is sensitive to less-squeezed, more equilateral configurations. Removing $C(k)$ is therefore equivalent to discarding sensitivity to equilateral and folded bispectrum configurations, mostly sourced by gravitational evolution. Nevertheless, this approach is of considerable interest, as the optimal density threshold $\delta_R^\ast$ could provide a direct window into the relationship between the parameter-dependent information content of the field and its morphological properties. We note, however, that breaking affine invariance by removing $C(k)$ requires revisiting several assumptions underlying the present analysis: in particular, $c_0$ is no longer freely fixable to unity by gauge choice and becomes a physically meaningful parameter. The perturbative priors on 
the mark coefficients, which in the current framework constrain the ratios $c_2/c_1$ and $c_3/c_1$ independently of the overall normalization, may likewise need to be reassessed in this new context. We defer a systematic investigation of this approach to future work.

\section{Results and discussion}

\label{sec:results}
We now discuss the application of the analytical optimization procedure described in the previous sections. Exploiting the availability of a fully analytical EFTofLSS prediction, we rapidly evaluate the Fisher information matrix across any mark configuration parametrized by $\{c_0, c_1, c_2, c_3\}$, scanning a three-dimensional grid of $100\times100\times 100$ values for $c_1$, $c_2$, and $c_3$ in the range $[-10,10]$, subject to the perturbativity prior discussed in Sec.~\ref{sec:PTprior}. We focus on three smoothing radii, $R = 10, 15$ and $30\, \text{Mpc}/h$. These are chosen based on the analysis performed in~\cite{Massara:2024cvu}, where the usual mark functional dependence was adopted to analyze the BOSS dataset. 
We also investigate the redshift dependence of the optimal mark considering $z = 0$, to allow comparison with~\cite{Cowell:2024wyl}, $z = 0.5$ and $z = 1$, bracketing the redshift range targeted by realistic surveys such as BOSS~\cite{BOSS:2016wmc}, DESI~\cite{DESI:2016fyo}, and Euclid~\cite{Euclid:2024yrr}. The optimal mark coefficients, obtained using the Gaussian covariance approximation, are reported in Table~\ref{tab:Cs_R30_R15}.

\begin{figure}
    \centering
    \includegraphics[width=\linewidth]{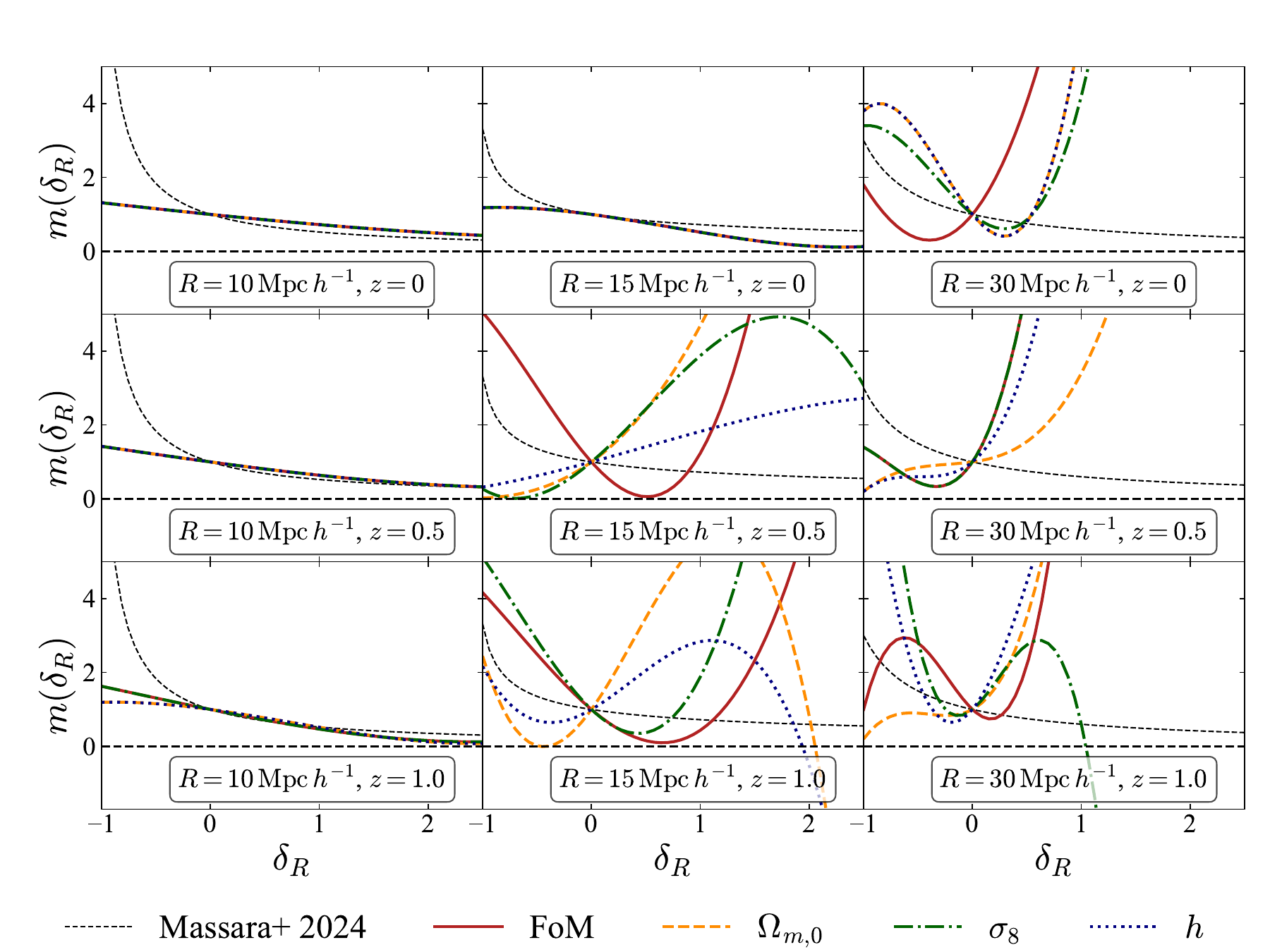}
    \caption{ \justifying Optimal mark functions for different smoothing radii and redshifts. We show the result of the optimization in different scenarios: maximizing the FoM over the three parameters $\{\Omega_{m,0}, \sigma_8, h\}$ jointly, and minimizing the uncertainty on each parameter individually. In dashed black, we also plot the mark functions adopted in the analysis~\cite{Massara:2024cvu}.}
    \label{fig:OptMarkR30}
\end{figure}

The resulting optimal mark functions are shown in Fig.~\ref{fig:OptMarkR30}. The picture that emerges depends significantly on the smoothing scale considered. For $R = 10\,\text{Mpc}/h$, the perturbativity prior restricts the allowed parameter space $\{c_1, c_2, c_3\}$ considerably, and all optimal marks collapse onto a similar, smooth, monotonically decreasing function of $\delta_R$, closely resembling a linearized expansion of the standard mark around $\delta_R \simeq 0$. In this regime, the optimization has little freedom and the resulting mark functions are largely insensitive to the choice of target parameter or redshift. For $R = 15$ and $30\,\text{Mpc}/h$, where the perturbativity prior is less restrictive, the optimal marks display a richer structure: they develop a parabolic shape over the range of $\delta_R$ sampled by the simulations (see Table~\ref{tab:delta_range}), enhancing the differential weighting between high- and low-density environments. The variation with redshift is more pronounced at $R = 15\,\text{Mpc}/h$ than at $R = 30\,\text{Mpc}/h$, which we discuss further below.

This outcome is consistent with the affine invariance of the Fisher matrix: as discussed in Sec.~\ref{sec:mark_theo}, {\em when the full data vector $\{P, C, M\}$ is adopted}, any mark function that up-weights underdense regions carries exactly the same information as one that up-weights overdense regions because the two are related by an affine transformation that leaves the Fisher matrix invariant. The actual information gain in the marked field therefore does not arise from the voids themselves, as sometimes implied in the literature~\cite{Massara:2020pli}, but rather from the environmental modulation of clustering~\cite{Bonnaire:2021sie, Bonnaire:2022ocm}. This result is in agreement with~\cite{Cowell:2024wyl}.

This interpretation is also consistent with the findings of~\cite{Marinucci:2024bdq}, who showed that the bispectrum combined with the standard power spectrum captures a significant fraction of the information recovered by marked statistics. The reason is straightforward: the late-time bispectrum encodes correlations between modes generated by gravitational nonlinearities, even in the absence of primordial non-Gaussianity, and is therefore a probe of environmental structure. The marked power spectrum effectively re-weights two-point statistics to access part of this information in a computationally simpler way; see also the discussion in Appendix~\ref{app:PTloop}.

The strength of the method adopted here is twofold. First, being grounded in perturbation theory, it provides a theoretically motivated and fully analytical framework that can be readily extended to additional cosmological parameters, as demonstrated in~\cite{Philcox:2020fqx, Philcox:2020srd, Ebina:2024zkv, Marinucci:2024bdq}. Second, its computational efficiency and flexibility make it a promising starting point for incorporation into a future MCMC analysis of survey data: the evaluation time for biased tracers in real space is around $t_{\rm eval} \simeq 2.5\,s$ for one given cosmology and mark configuration, and $t_{\rm eval} \simeq 6.5\,s$ when non-local primordial non-Gaussianities are included. These results can be further optimized and extended to redshift space, for which we already have an implementation that needs to be optimized. We leave this to future work.

\subsection{Comparison with simulations}

\begin{figure}
    \centering
    \includegraphics[width = 0.8\linewidth]{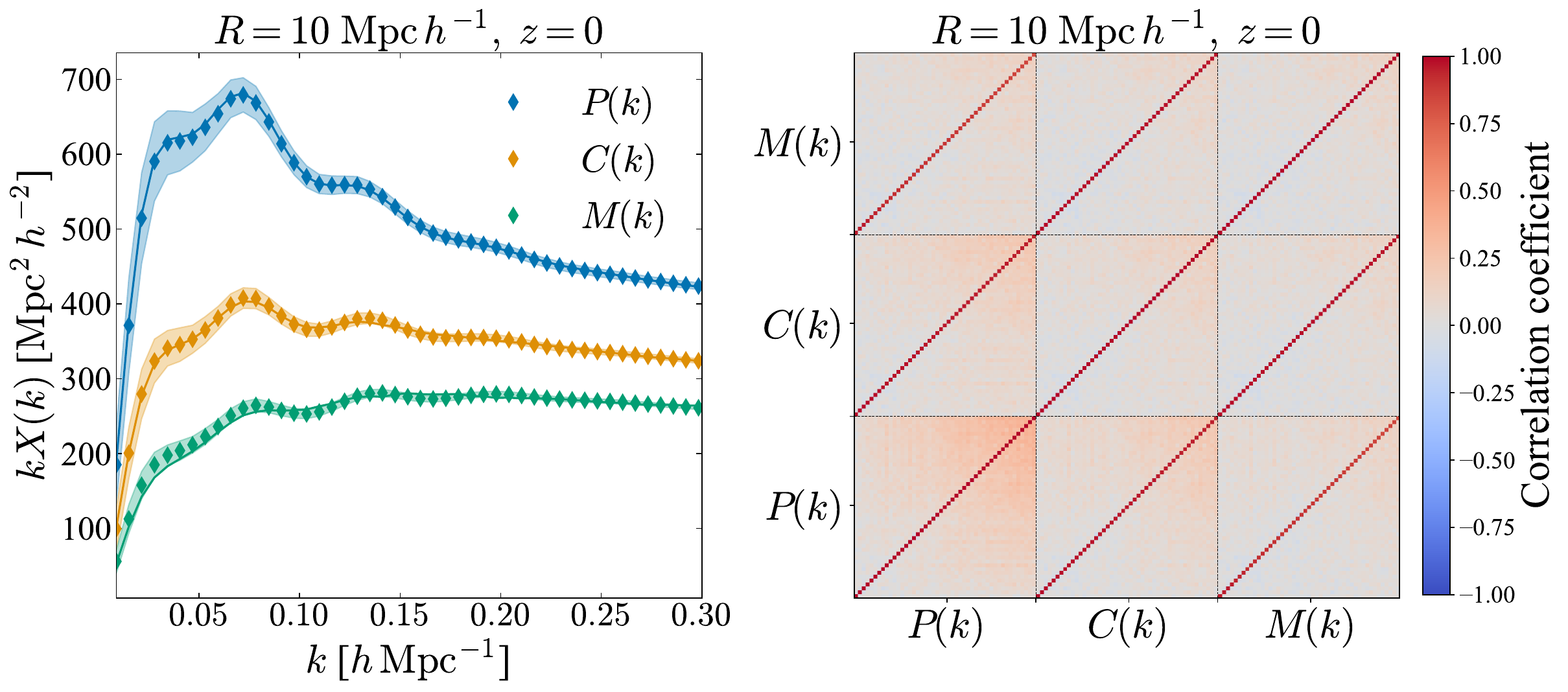}
    \includegraphics[width = 0.8\linewidth]{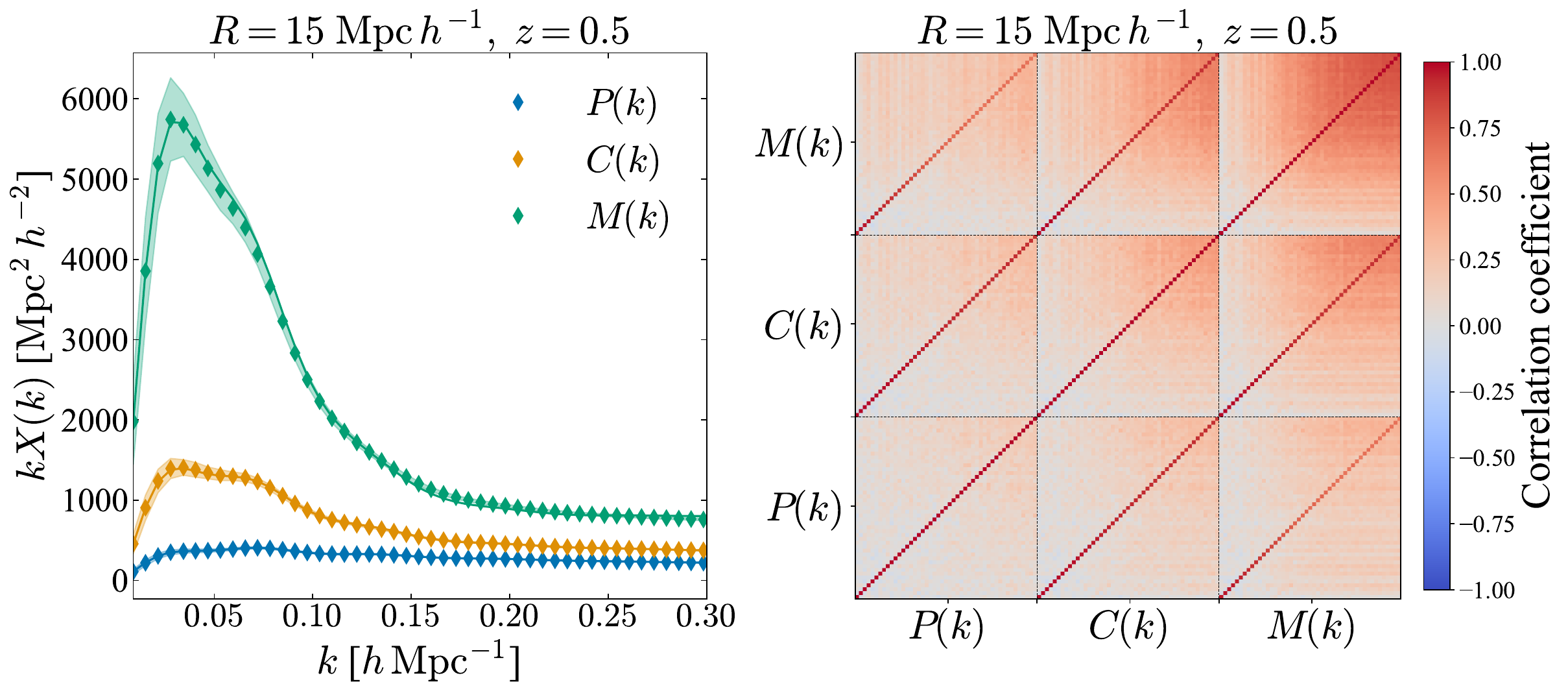}
    \includegraphics[width = 0.8\linewidth]{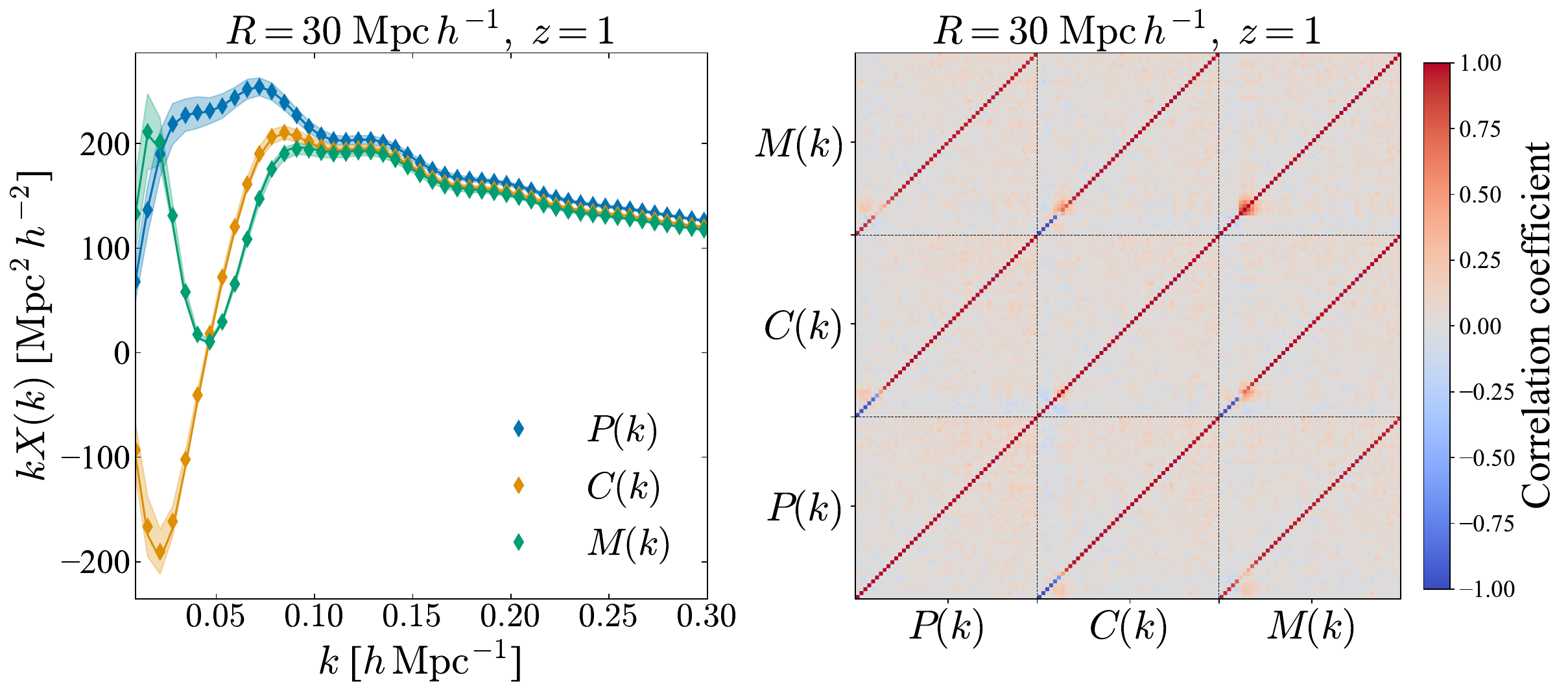}
    \caption{ \justifying Comparison of theoretical predictions with power spectra measured from the Quijote simulations for the mark with $R = 10\, \text{Mpc}/h$ at $z=0$ optimizing over all the parameters (top), $R = 15\, \text{Mpc}/h$ at $z=0.5$ optimizing for $\sigma_8$ (middle) and $R = 30\, \text{Mpc}/h$ at $z=1$ optimizing the FoM (bottom). On the right columns we show the corresponding correlation matrices estimated from 1000 realizations of the Quijote suite at the fiducial cosmology.}
    \label{fig:Pk_comparisons}
\end{figure}

Having identified the optimal marks, we compare the theoretical predictions against power spectra measured from the Quijote simulation suite. Among all the optimal marks obtained at different smoothing scales and redshifts, we select three representative cases that display qualitatively distinct shapes, illustrating the variety of forms that the marked power spectrum can take. Fig.~\ref{fig:Pk_comparisons} shows the matter power spectrum $P(k)$, the cross-spectrum between the marked and unmarked fields $C(k)$, and the marked power spectrum $M(k)$ for these three cases.

The marked power spectrum from each simulation box is estimated at the particle level through a two-step procedure~\cite{Pylians}.  We first assign the N-body particles to a $256^3$ grid via cloud-in-cell (CIC) interpolation to build $\delta(\bx)$~\cite{Hand_2018}, smooth it on the scale $R$ to obtain $\delta_R$, and interpolate $\delta_R$ back to each particle position. Each particle is then re-deposited on the grid carrying the weight $m(\delta_R)$ evaluated at its location, which reconstructs $\rho_M(\bx)$ of Eq.~\eqref{eq:deltaM} directly from the particles.

Across all wavenumbers of interest, the analytical model reproduces the simulation measurements at the percent level, demonstrating that the EFTofLSS framework correctly describes all three spectra up to mildly nonlinear scales $k_{\rm max} \simeq 0.30\, h/\text{Mpc}$.

We next examine the covariance properties of the three spectra. The right panels of Fig.~\ref{fig:Pk_comparisons} show the correlation matrices estimated from 1000 realizations at the fiducial cosmology. For $R = 10\,\text{Mpc}/h$ and $R = 30\,\text{Mpc}/h$ the correlation matrices are close to diagonal and hence well described by the Gaussian approximation, whereas for $R = 15 \, \text{Mpc}/h$ the covariance displays more prominent off-diagonal structure. We show here three very different shapes for the mark function, which translate into different shapes for the power spectra.

The right panels show the correlation matrix of the joint data vector $\{P(k), C(k), M(k)\}$, estimated from the $1000$ Quijote realizations. The structure of the covariance depends sensitively on the smoothing scale. For $R = 10\,\text{Mpc}/h$, the correlation matrix is close to diagonal within each block, indicating that the non-Gaussian contributions to the covariance are small. This is consistent with the fact that at this smoothing scale the optimal mark is a smooth, nearly linear function of $\delta_R$ (see Fig.~\ref{fig:OptMarkR30}), which limits the coupling between different Fourier modes induced by the mark. As $R$ increases to $15\,\text{Mpc}/h$, significant off-diagonal correlations begin to appear, particularly in the $M$--$M$ and $M$--$C$ blocks, reflecting the more complex, non-monotonic shape of the optimal mark at this scale and the larger non-Gaussian contributions to the density field probed. A qualitatively similar pattern is observed for $R = 30\,\text{Mpc}/h$, despite the marked power spectrum having a very different amplitude and shape compared to the $R = 15\,\text{Mpc}/h$ case.

These off-diagonal correlations, while they do not qualitatively alter the Fisher forecasts, must be properly accounted for in any rigorous likelihood analysis. In the following Section we therefore present results based on the full non-Gaussian covariance estimated from simulations, and defer a detailed comparison with the Gaussian approximation to Appendix~\ref{app:further}.

\subsection{Parameter improvements and implications}
\label{sec:improvs}

Finally, we quantify the cosmological information gain from marked statistics. Table~\ref{tab:improv_all} summarizes the forecasted improvement on the marginalized error bars of the cosmological parameters, defined as the ratio $r(\theta_\mu) \equiv \sigma_{P}(\theta_\mu)/\sigma_{P+C+M}(\theta_\mu)$, for the full simulation-based covariances, across three smoothing radii $R = \{10, 15, 30\} \, \text{Mpc}/h$ and three redshifts $z = \{0, 0.5, 1\}$.

The results show two clear and physically motivated trends. First, the improvements are systematically larger at smaller smoothing radii: as $R$ decreases, the mark probes increasingly nonlinear environments, capturing more of the higher-order statistical information encoded in the density field that is inaccessible to the standard power spectrum. Second, the improvement decreases with increasing redshift at fixed $R$, since at higher redshift the dark matter field is more linear and the nonlinear environmental information that the mark is designed to capture becomes progressively less abundant. Both trends are consistent with the expectation that the gain from marked statistics is fundamentally driven by the amount of nonlinear information available in the field.

For $R = 10 \, \text{Mpc}/h$, the improvements are the largest across all configurations. At $z = 0$, the FoM-optimized mark yields $r(\sigma_8) \simeq 8.3$, $r(\Omega_m) \simeq 2.3$, and $r(h) \simeq 2.5$, representing the most significant gains found in our analysis. These large values reflect the fact that at this smoothing scale the mark is sensitive to highly nonlinear overdensities, with $\delta_{R,\mathrm{max}} \simeq 5.5$ at $z = 0$ (see Table~\ref{tab:delta_range}), where the density field carries substantial nonlinear information. The large improvement on $\sigma_8$ can be understood as a direct consequence of marginalizing over the EFT counterterm parameter $c_s^2$. The correlation between these two parameters can be estimated analytically as follows. The power spectrum derivatives with respect to $\sigma_8$ and $c_s^2$ read
\begin{equation}
    \frac{\partial P(k)}{\partial \sigma_8} \simeq 2\,\sigma_8\, P_{\rm lin}(k)\,, \qquad \frac{\partial P(k)}{\partial c_s^2} = -2\,\sigma_8^2\, k^2\, P_{\rm lin }(k)\,,
    \label{eq:derivs}
\end{equation}
where in the first expression we have neglected contributions of order $\mathcal{O}(P_{\rm 1loop}/P_L)$. At leading order, the inverse covariance of the power spectrum is diagonal and given by
\begin{equation}
    \left[{\rm Cov}\big(P(k_i),P(k_j)\big)\right]^{-1} = \frac{V \Delta k}{(2\pi)^2}\,\frac{k_i^2}{P_{\rm lin}^2(k_i)}\,\delta_{ij}\,.
    \label{eq:invCPP}
\end{equation}
The correlation coefficient between $c_s^2$ and $\sigma_8$ is defined as
\begin{equation}
    \rho_F^2(\sigma_8, c_s^2) \equiv \frac{\big(F_{\sigma_8 c_s^2}\big)^2}{F_{\sigma_8\sigma_8}\,F_{c_s^2 c_s^2}}\,,
\end{equation}
where $F_{\alpha \beta}$ denotes the Fisher matrix element associated with parameters $\alpha$ and $\beta$. Substituting eqs.~\eqref{eq:derivs} and~\eqref{eq:invCPP}, we obtain the compact result
\begin{equation}
    \rho_F^2(\sigma_8, c_s^2) = \frac{\big(\sum_i k_i^4\big)^2}{\big(\sum_i k_i^2\big)\big(\sum_i k_i^6\big)} \simeq \frac{21}{25}\left[1 + \mathcal{O}\!\left(\left(\frac{k_{\rm min}}{k_{\rm max}}\right)^{\!3},\; c_s^2 k_{\rm max}^2\right)\right] \simeq 0.85\,,
    \label{eq:corr_s8cs}
\end{equation}
where the higher-order corrections amount to only a few percent. Remarkably, this correlation is nearly independent of $k_{\rm max}$. This behavior follows from the fact that the logarithmic derivative of the power spectrum with respect to $\sigma_8$ is approximately scale-independent, while the one with respect to $c_s^2$ exhibits a simple power-law dependence on $k$. The same argument does not apply to $\Omega_m$ and $h$, whose power spectrum derivatives encode non-trivial $k$-dependence through the transfer function, the BAO features, and the matter-radiation equality scale.

The correlation in Eq.~\eqref{eq:corr_s8cs} drives the degradation of the forecasted error on $\sigma_8$ when only the power spectrum is considered. We find $\sigma(\sigma_8)|_{P\text{-only}} \simeq 0.02$ for fixed $c_s^2$, which degrades to $\sigma(\sigma_8)|_{P\text{-only}} \simeq 0.08$ once $c_s^2$ is marginalized over, with a degradation by a factor of $\sim 4$. Considering a restricted Fisher matrix in which only $\sigma_8$ and $c_s^2$ are varied, we have
\begin{equation}
    \sigma(\sigma_8)\big|_{c_s^2 \text{ fixed}} = \big(F_{\sigma_8\sigma_8}\big)^{-1/2}\,,
\end{equation}
while the marginalized error reads
\begin{equation}
    \sigma(\sigma_8)\big|_{c_s^2 \text{ marg.}} = \Big[F_{\sigma_8\sigma_8}\big(1 - \rho_F^2(\sigma_8, c_s^2)\big)\Big]^{-1/2} \simeq 2.5\,\sigma(\sigma_8)\big|_{c_s^2 \text{ fixed}}\,,
\end{equation}
in agreement with our numerical findings and confirming the robustness of our results. Including the cross-spectrum and the marked power spectrum in the analysis breaks this degeneracy through their distinct $k$-dependence. This is consistent with the findings of~\cite{Marinucci:2024bdq}, where the marked power spectrum was shown to alleviate degeneracies between cosmological parameters and the EFTofLSS bias and counterterm coefficients, acting in a role traditionally played by the bispectrum.

At $z = 0.5$ the improvements reduce to $r \sim 1.5-3$, and at $z = 1$ further to $r \sim 1.4-2$, confirming the expected suppression with redshift. We note that for $R = 10 \, \text{Mpc}/h$ the FoM-based and single-parameter optimizations provide identical results due to the reduced parameter space after imposing perturbativity.

For $R = 15 \, \text{Mpc}/h$, the simulation-based results show a clear and interesting pattern: the improvements obtained from the FoM-optimized mark are usually larger than those from the single-parameter optimization. At $z = 0$, the FoM-optimized mark yields $r(\sigma_8) \simeq 5.6$, $r(\Omega_m) \simeq 1.5$, and $r(h) \simeq 1.6$, while the single-parameter optimization gives more modest values, $r \sim 1.1-1.2$ for $\Omega_m$ and $h$, with the exception of $\sigma_8$ which reaches $r \simeq 5.6$ in both cases. This behavior is somewhat unexpected, since one would naively expect the single-parameter optimization to provide larger improvements on each individual parameter by construction. We attribute this to the fact that the mark functions selected by the single-parameter optimization tend to have larger non-Gaussian contributions to the covariance matrix, which are not captured by the Gaussian approximation used during the optimization but are properly accounted for in the simulation-based covariance. As a result, the gains forecasted during optimization are partially washed out when the more realistic covariance is employed. The FoM-based optimization, by contrast, appears to select mark functions that are less sensitive to these non-Gaussian corrections, making it a more robust figure of merit for the purposes of mark optimization at this smoothing scale. At $z = 0.5$ the improvements are $r \sim 1.8-2.8$ for the FoM optimization, slightly larger than at $z = 1$ where they reduce to $r \sim 1.3-2.0$, consistent with the general redshift trend discussed above.

For $R = 30 \, \text{Mpc}/h$, the improvements are the most modest across all three radii, as expected given that at this smoothing scale the density field is closer to the linear regime and the mark captures less nonlinear information. At $z = 0$, the FoM-optimized mark show improvements $r(\sigma_8) \simeq 1.8$, $r(\Omega_m) \simeq 1.3$, and $r(h) \simeq 1.3$, while the single-parameter optimization gives somewhat larger values, up to $r(\sigma_8) \simeq 2.5$ and $r(\Omega_m) \simeq 1.7$. The fact that $\sigma_8$ and $\Omega_m$ benefit more than $h$ from the addition of marked statistics has a natural explanation in the context of our parameter set $\{\Omega_m, \sigma_8, h\}$ applied to the matter field in real space. Without a linear halo bias parameter $b_1$, the amplitude of the power spectrum is directly sensitive to $\sigma_8$, but the marked power spectrum effectively provides a weighted integral over bispectrum triangles, and this richer environmental information is particularly efficient in breaking the remaining parameter degeneracies. At $z = 0.5$ the improvements are more modest, $r \sim 1.2-1.3$, while at $z = 1$ they recover slightly to $r \sim 1.3-1.7$. 

Taken together, these results confirm that marked statistics provide a robust and significant gain in cosmological constraining power across all smoothing scales and redshifts considered, with the largest improvements achieved at small $R$ and low $z$, where the nonlinear signal is richest. These findings are broadly consistent with recent data analyses combining the one-loop power spectrum and bispectrum, such as~\cite{DAmico:2022osl}, where improvements of $30\%$, $18\%$, and $13\%$ are reported on $\sigma_8$, $h$, and $\Omega_m$, respectively. A direct comparison is not straightforward, since that analysis is performed on galaxy data in redshift space and accounts for a range of observational effects not considered here; nevertheless, the overall trends are recovered. Furthermore, recent power-spectrum-only analyses extended to two-loop order~\cite{Bakx:2025jwa} report improvements of $35\%$, $20\%$, and $15\%$ on $\Omega_m$, $h$, and $A_s$, respectively, which translate to an improvement of $\sim 7.5\%$ on $\sigma_8$. This suggests that extending the power spectrum analysis to higher $k_{\rm max}$ primarily tightens constraints on shape parameters such as $\Omega_m$, whereas the inclusion of higher-order statistics has a proportionally larger impact on amplitude parameters such as $\sigma_8$ and $A_s$. The results presented here are consistent with this picture: the complementary information captured by the marked power spectrum is primarily of higher-order origin, as already argued in~\cite{Marinucci:2024bdq}.

\begin{figure}
    \centering
    \includegraphics[width=\linewidth]{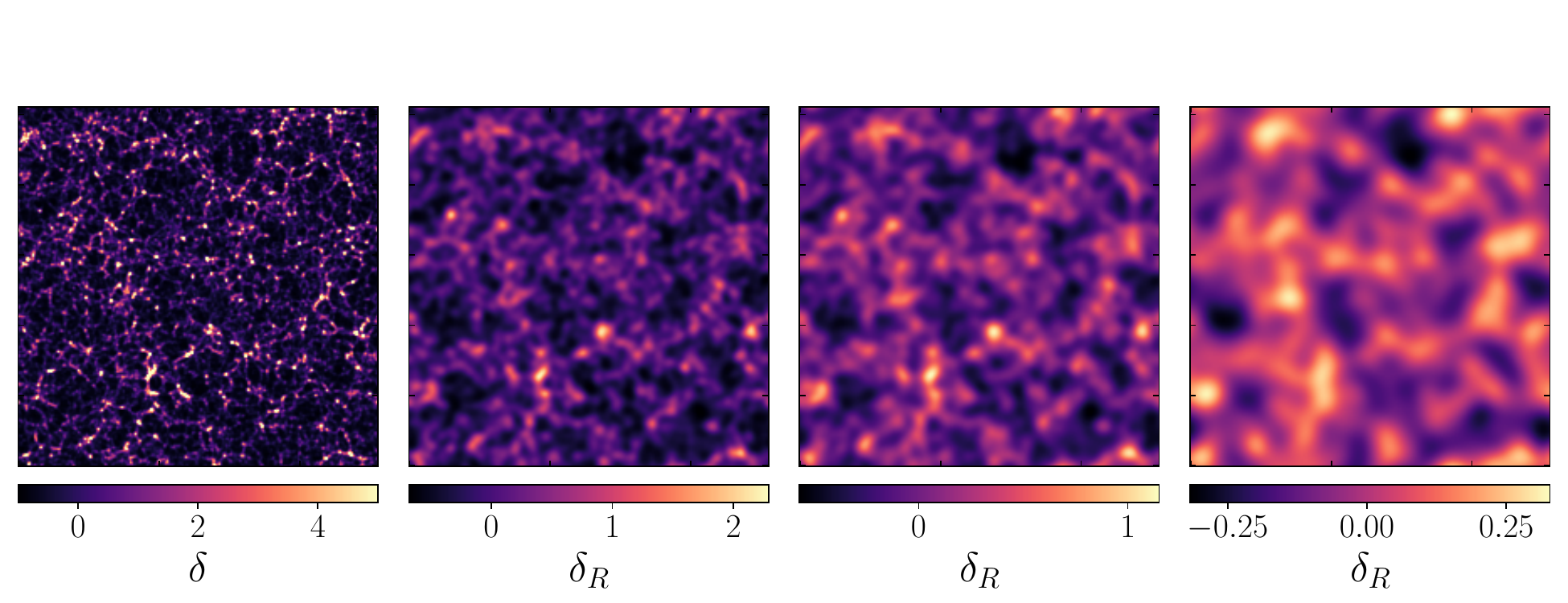}
    \caption{ \justifying  Density field for a slice of one realization of the Quijote simulations at $z = 0$ (left plot). We also show the smoothed density field at the same redshift for the radii used in this work $R = 10\, {\rm Mpc}/h$,$15 \, {\rm Mpc}/h$ and $30 \, {\rm Mpc}/h$.}
    \label{fig:deltas_z0}
\end{figure}

\begin{figure}
    \centering
    \includegraphics[width=0.7\linewidth]{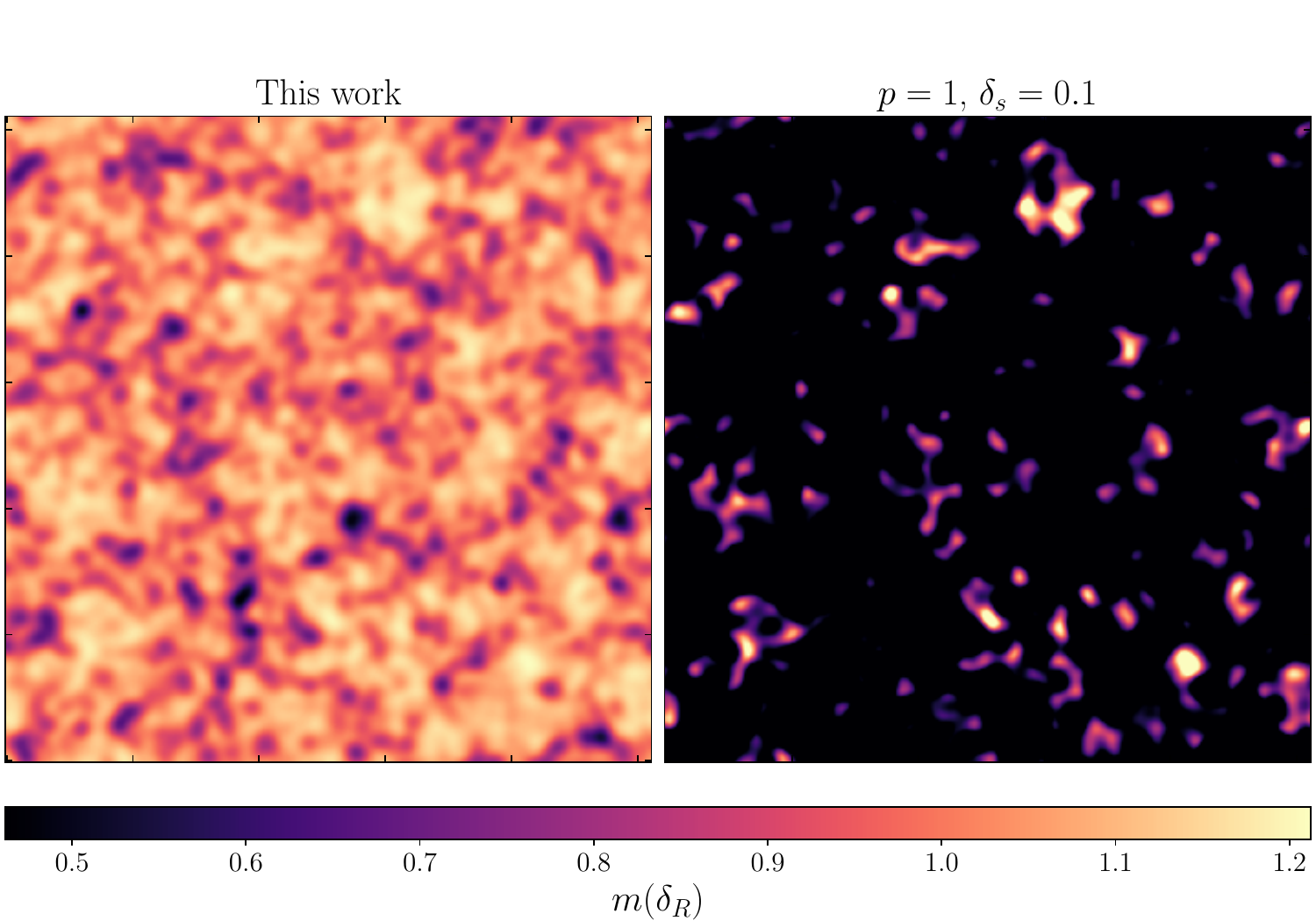}
    \includegraphics[width=0.7\linewidth]{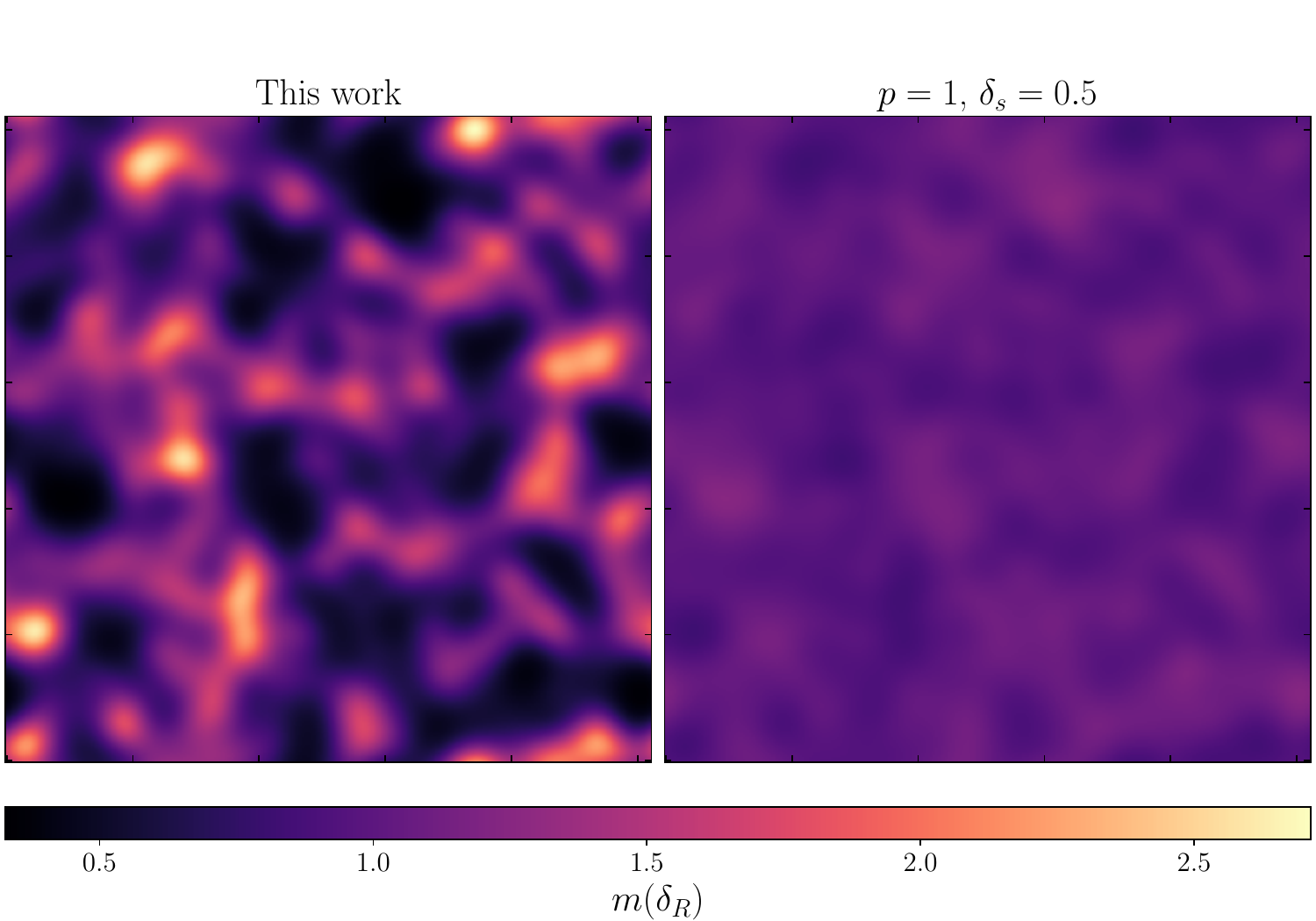}
    \caption{ \justifying   Mark functions $m(\delta_R)$ obtained with our optimization scheme at $z = 0$, for $R = 10 \, {\rm Mpc}/h$ (upper) and $R = 30 \, {\rm Mpc}/h$ (lower). We also plot the result for the mark function from Eq.~\eqref{eq:massara_mark}, with the values used in the analysis of~\cite{Massara:2024cvu}.}
    \label{fig:mdR_z0}
\end{figure}

To gain further insight into how the mark functions modify the density field, we show in Fig.~\ref{fig:deltas_z0} the density field together with its smoothed counterpart at redshift $z = 0$, while the corresponding mark functions are displayed in Fig.~\ref{fig:mdR_z0} for the two cases $R = 10$ and $ 30 \, \mathrm{Mpc}/h$. For the case $R = 10\,\mathrm{Mpc}/h$, upper row of Fig.~\ref{fig:mdR_z0}, our method does not appear to significantly enhance the contrast between the under- and overdense regions. This behavior becomes particularly evident when compared to the mark proposed~\cite{Massara:2024cvu}, with $p = 1$ and $\delta_s = 0.1$. A plausible explanation is that our approach substantially restricts the allowed region of the parameter space $\{c_0, c_1, c_2, c_3\}$ through the perturbativity prior, which induces a much smoother mark function. Indeed, as one can see from Fig.~\ref{fig:OptMarkR30}, the mark obtained by optimizing the FoM closely resembles the functional form of~\cite{Massara:2024cvu}, albeit with a more limited dynamic range. This represents a limitation of our approach: for small values of the smoothing radius $R$, the theory becomes less convergent. As a consequence of this, we find that using the mark adopted in~\cite{Massara:2024cvu} gives $\sim 2-5\%$ better constraints on the three parameters, confirming the intuition that a larger contrast between environmental density contrasts yields better cosmological constraints.

For a large $R = 30\,\mathrm{Mpc}/h$, such perturbative limitation is no longer present: our optimal mark generates now a stronger contrast than the standard mark function with $p = 1$ and $\delta_s = 0.5$ does as shown in the lower plot of Fig.~\ref{fig:mdR_z0}. These results indicate once again that, increasing the weight function for regions where there is more structure, such as voids and nodes or filaments, is optimal for parameter constraints. When compared to the usual mark functions, with parameters $p = 1$ and $\delta_s = 0.5$, we find that our procedure provides $40-60\%$ better constraints on the cosmological parameters, and a similar trend is observed for $R = 15 \, \mathrm{Mpc}/h$.

These results should be viewed in their proper context: we are working with the dark matter field in real space, a particularly clean setting with well-controlled theory and minimal systematics. On realistic tracers such as galaxies in redshift space, ref.~\cite{Marinucci:2024bdq} showed that the information gain from $P+C+M$ is comparable to that from $P+B$. Nevertheless, all the observables considered here are two-point functions that can be measured from galaxy catalogs at minimal extra computational cost, and whose covariance matrices are far easier to estimate and model than those of the bispectrum or higher-order statistics. The marked power spectrum therefore offers a valuable compromise: it captures part of the non-Gaussian information encoded in the bispectrum while retaining the simplicity of a two-point analysis.

We note that we also investigated whether our polynomial mark framework can reproduce the optimal mark shapes found in~\cite{Cowell:2024wyl} via Gaussian processes. First, we repeat our optimization varying only the two parameters $\Omega_m$ and $\sigma_8$ for $R = 30 \, \text{Mpc}/h$ at $z = 0$ in order to have a direct comparison with~\cite{Cowell:2024wyl}. We find the mark parameters to be $\{c_0, c_1, c_2, c_3\} = \{ 1.0, -1.36, -0.65, -0.65\}$ when the FoM is minimized and $\{c_0, c_1, c_2, c_3\} = \{ 1.0, -1.56, -0.75, 0.25\}$ and $\{c_0, c_1, c_2, c_3\} = \{ 1.0, -3.37, -1.46, 1.66\}$ when the error bar on $\Omega_m$ and $\sigma_8$ are minimized, respectively. When a Gaussian covariance is considered, the improvement on the parameters are in excellent agreement with the findings of~\cite{Cowell:2024wyl}, $r(\Omega_m) = 4.45$ and $r(\sigma_8) = 1.20$, both for FoM and single optimization. By including the full non-Gaussian covariance obtained from 1000 realizations, $r(\Omega_m) = 1.66/1.73$ and $r(\sigma_8) = 2.09/1.50$ for FoM and single optimization respectively. We observe that qualitatively these results agree with previous findings: in particular while the improvement for $\Omega_m$ is degraded by non-Gaussian contribution to the covariance, the improvement on $\sigma_8$ is enhanced. However, the slight quantitative disagreement with the improvement obtained by~\cite{Cowell:2024wyl}, see their Table 2, can be understood in terms of the estimate of the non-Gaussian covariance: while we use $N_{\rm sims} \gg N_{\rm bins}$ in order to have a correct estimate of the full covariance, they used $\sim 10$ times less simulations to estimate it, possibly underestimating the non-Gaussian contribution~\cite{Hartlap:2006kj}.

Moreover, we identified combinations of $\{c_0, c_1, c_2, c_3\}$ that closely match the GP-derived mark functions of~\cite{Cowell:2024wyl}; the corresponding coefficients are $\{c_0, c_1, c_2, c_3\} = \{ -0.37, 2.11, 11.43, -43.20\}$ for the case with $R = 30 \,\text{Mpc}/h$ at $z = 0$. This demonstrates the viability of our polynomial basis as an approximator for the GP result, and establishes a formal connection between the two fitting procedures. However, this particular coefficient set violates the perturbativity prior described in Sec.~\ref{sec:PTprior}, and we therefore do not use it for actual parameter forecasts in this work. The two approaches are in this sense complementary: the GP method has greater freedom to explore mark shapes beyond the perturbative regime, while ours offers full analytical control within it. It is nonetheless encouraging that, under comparable conditions (Gaussian covariance, as adopted in~\cite{Cowell:2024wyl}), both coefficient sets yield similar parameter improvements, further validating the physical picture of environmental contrast as the driver of information gain.

\begin{table}[h]
\centering
\begin{tabular}{c c}

\begin{tabular}{|c|c|c|c|}
\hline
\multicolumn{4}{|c|}{
\parbox[c][1.2cm][c]{0.3\linewidth}{\centering \large{Optimization of FoM}}
}\\
\hline\hline
\multicolumn{4}{|c|}{
\parbox[c][1.cm][c]{0.3\linewidth}{\centering $R = 10 \,\text{Mpc}\, h^{-1}$}
}\\
\hline
\cell{1.7cm}{0.5cm}{Redshift} &
\cell{1.7cm}{0.7cm}{$r(\Omega_m)$} &
\cell{1.7cm}{0.7cm}{$r(\sigma_8)$} &
\cell{1.7cm}{0.7cm}{$r(h)$} \\
\hline
\cell{1.7cm}{0.5cm}{0} &
\cell{1.7cm}{0.7cm}{2.34} &
\cell{1.7cm}{0.7cm}{8.32} &
\cell{1.7cm}{0.7cm}{2.47} \\
\hline
\cell{1.7cm}{0.5cm}{0.5} &
\cell{1.7cm}{0.7cm}{1.53} &
\cell{1.7cm}{0.7cm}{3.11} &
\cell{1.7cm}{0.7cm}{1.50} \\
\hline
\cell{1.7cm}{0.5cm}{1} &
\cell{1.7cm}{0.7cm}{1.43} &
\cell{1.7cm}{0.7cm}{2.06} &
\cell{1.7cm}{0.7cm}{1.38} \\
\hline\hline
\multicolumn{4}{|c|}{
\parbox[c][1.cm][c]{0.3\linewidth}{\centering $R = 15 \,\text{Mpc}\, h^{-1}$}
}\\
\hline
\cell{1.7cm}{0.5cm}{Redshift} &
\cell{1.7cm}{0.7cm}{$r(\Omega_m)$} &
\cell{1.7cm}{0.7cm}{$r(\sigma_8)$} &
\cell{1.7cm}{0.7cm}{$r(h)$} \\
\hline
\cell{1.7cm}{0.5cm}{0} &
\cell{1.7cm}{0.7cm}{1.48} &
\cell{1.7cm}{0.7cm}{5.55} &
\cell{1.7cm}{0.7cm}{1.57} \\
\hline
\cell{1.7cm}{0.5cm}{0.5} &
\cell{1.7cm}{0.7cm}{1.89} &
\cell{1.7cm}{0.7cm}{2.79} &
\cell{1.7cm}{0.7cm}{1.83} \\
\hline
\cell{1.7cm}{0.5cm}{1} &
\cell{1.7cm}{0.7cm}{1.46} &
\cell{1.7cm}{0.7cm}{1.97} &
\cell{1.7cm}{0.7cm}{1.32} \\
\hline\hline

\multicolumn{4}{|c|}{
\parbox[c][1.cm][c]{0.3\linewidth}{\centering $R = 30 \,\text{Mpc}\, h^{-1}$}
}\\
\hline
\cell{1.7cm}{0.5cm}{Redshift} &
\cell{1.7cm}{0.7cm}{$r(\Omega_m)$} &
\cell{1.7cm}{0.7cm}{$r(\sigma_8)$} &
\cell{1.7cm}{0.7cm}{$r(h)$} \\
\hline
\cell{1.7cm}{0.5cm}{0} &
\cell{1.7cm}{0.7cm}{1.28} &
\cell{1.7cm}{0.7cm}{1.83} &
\cell{1.7cm}{0.7cm}{1.25} \\
\hline
\cell{1.7cm}{0.5cm}{0.5} &
\cell{1.7cm}{0.7cm}{1.23} &
\cell{1.7cm}{0.7cm}{1.25} &
\cell{1.7cm}{0.7cm}{1.16} \\
\hline
\cell{1.7cm}{0.5cm}{1} &
\cell{1.7cm}{0.7cm}{1.48} &
\cell{1.7cm}{0.7cm}{1.72} &
\cell{1.7cm}{0.7cm}{1.32} \\
\hline
\end{tabular}

&
\begin{tabular}{|c|c|c|c|}
\hline
\multicolumn{4}{|c|}{
\parbox[c][1.2cm][c]{0.3\linewidth}{\centering \large{Optimization of single parameters}}
}\\
\hline\hline
\multicolumn{4}{|c|}{
\parbox[c][1.cm][c]{0.3\linewidth}{\centering $R = 10 \,\text{Mpc}\, h^{-1}$}
}\\
\hline
\cell{1.7cm}{0.5cm}{Redshift} &
\cell{1.7cm}{0.7cm}{$r(\Omega_m)$} &
\cell{1.7cm}{0.7cm}{$r(\sigma_8)$} &
\cell{1.7cm}{0.7cm}{$r(h)$} \\
\hline
\cell{1.7cm}{0.5cm}{0} &
\cell{1.7cm}{0.7cm}{2.34} &
\cell{1.7cm}{0.7cm}{8.32} &
\cell{1.7cm}{0.7cm}{2.47} \\
\hline
\cell{1.7cm}{0.5cm}{0.5} &
\cell{1.7cm}{0.7cm}{1.53} &
\cell{1.7cm}{0.7cm}{3.11} &
\cell{1.7cm}{0.7cm}{1.50} \\
\hline
\cell{1.7cm}{0.5cm}{1} &
\cell{1.7cm}{0.7cm}{1.38} &
\cell{1.7cm}{0.7cm}{2.06} &
\cell{1.7cm}{0.7cm}{1.27} \\
\hline\hline
\multicolumn{4}{|c|}{
\parbox[c][1.cm][c]{0.3\linewidth}{\centering $R = 15 \,\text{Mpc}\, h^{-1}$}
}\\
\hline
\cell{1.7cm}{0.5cm}{Redshift} &
\cell{1.7cm}{0.7cm}{$r(\Omega_m)$} &
\cell{1.7cm}{0.7cm}{$r(\sigma_8)$} &
\cell{1.7cm}{0.7cm}{$r(h)$} \\
\hline
\cell{1.7cm}{0.5cm}{0} &
\cell{1.7cm}{0.7cm}{1.48} &
\cell{1.7cm}{0.7cm}{5.55} &
\cell{1.7cm}{0.7cm}{1.57} \\
\hline
\cell{1.7cm}{0.5cm}{0.5} &
\cell{1.7cm}{0.7cm}{1.20} &
\cell{1.7cm}{0.7cm}{1.19} &
\cell{1.7cm}{0.7cm}{1.15} \\
\hline
\cell{1.7cm}{0.5cm}{1} &
\cell{1.7cm}{0.7cm}{1.09} &
\cell{1.7cm}{0.7cm}{2.21} &
\cell{1.7cm}{0.7cm}{1.23} \\
\hline\hline

\multicolumn{4}{|c|}{
\parbox[c][1.cm][c]{0.3\linewidth}{\centering $R = 30 \,\text{Mpc}\, h^{-1}$}
}\\
\hline
\cell{1.7cm}{0.5cm}{Redshift} &
\cell{1.7cm}{0.7cm}{$r(\Omega_m)$} &
\cell{1.7cm}{0.7cm}{$r(\sigma_8)$} &
\cell{1.7cm}{0.7cm}{$r(h)$} \\
\hline
\cell{1.7cm}{0.5cm}{0} &
\cell{1.7cm}{0.7cm}{1.72} &
\cell{1.7cm}{0.7cm}{2.50} &
\cell{1.7cm}{0.7cm}{1.69} \\
\hline
\cell{1.7cm}{0.5cm}{0.5} &
\cell{1.7cm}{0.7cm}{1.29} &
\cell{1.7cm}{0.7cm}{1.25} &
\cell{1.7cm}{0.7cm}{1.19} \\
\hline
\cell{1.7cm}{0.5cm}{1} &
\cell{1.7cm}{0.7cm}{1.36} &
\cell{1.7cm}{0.7cm}{1.67} &
\cell{1.7cm}{0.7cm}{1.26} \\
\hline
\end{tabular}
\end{tabular}
\caption{ \justifying Improvement in the cosmological parameter error bars obtained from the inclusion of the marked statistics. The improvement is quantified as the ratio $r(\theta) = \sigma_{P}(\theta)/\sigma_{P+C+M}(\theta)$, where $\sigma_{P}$ denotes the marginalized uncertainty from the power spectrum alone, and $\sigma_{P+C+M}$ corresponds to the joint analysis including cross- and marked-power contributions. Values larger than unity indicate a gain in constraining power relative to the standard analysis. The results shown here are based on the robust simulation-based analysis; a full comparison between Gaussian and simulation-based results is provided in the Appendix~\ref{app:further}.}
\label{tab:improv_all}
\end{table}

\section{Outlook and conclusions}
\label{sec:conclusions}

In this work, we have presented a fully analytical, perturbative framework to optimize the marked power spectrum, or more precisely, the joint statistics $\{P, C, M\}$ combining the standard power spectrum, the cross-spectrum, and the marked power spectrum at mildly nonlinear scales, working to one-loop order in the EFTofLSS. The mark function is written as a polynomial in the smoothed density field $\delta_R$, with free coefficients $\{c_1, c_2, c_3\}$ that are varied to maximize the Fisher information on any chosen set of cosmological parameters. This approach produces a polynomial fitting function for the optimal mark that is directly analogous to the Gaussian process ansatz of~\cite{Cowell:2024wyl}, but is obtained entirely analytically and without any $N$-body simulations.

A major practical advantage of the analytical approach is computational speed. Evaluating the full Fisher matrix grid over the mark parameter space takes minutes on a standard laptop, as opposed to the large sets of $N$-body simulations required to train a GP or a machine-learning model. The only step in our pipeline that requires simulations is the estimation of the non-Gaussian covariance matrix at the fiducial cosmology, needed to go beyond the Gaussian approximation; this requires a single set of realizations at a fixed cosmology, rather than the many simulation runs at varied cosmology that data-driven optimization methods typically demand. This makes our pipeline a natural candidate for incorporation into a future MCMC analysis of actual survey data, where the mark coefficients could be jointly varied alongside the cosmological parameters.

A second advantage of our approach is that marked statistics help break degeneracies between cosmological parameters and the EFTofLSS nuisance parameters. As we explicitly showed in Sec.~\ref{sec:improvs}, the strong correlation between $\sigma_8$ and the counterterm coefficient $c_s^2$ at the power spectrum level is largely broken by the inclusion of the cross-spectrum and the marked power spectrum, thanks to their distinct $k$-dependence. This is consistent with the findings of~\cite{Ebina:2024zkv, Marinucci:2024bdq}, where marked statistics were shown to alleviate degeneracies among cosmological and EFT parameters, a role traditionally played by the bispectrum.

Finally, a distinctive feature of our method is the necessity of imposing perturbativity priors on the mark coefficients. As discussed in Sec.~\ref{sec:PTprior}, mass conservation is not a symmetry of the marked field, so higher-loop contributions can enter at large scales already at one-loop rather than being suppressed by powers of $k$. Requiring that these contributions remain subdominant places non-trivial constraints on $\{c_1, c_2, c_3\}$, restricting the optimization to a theoretically controlled region of parameter space. While this boundary limits the accessible fitting space compared to a Gaussian process (GP) with no theoretical prior— (indeed, the GP-optimal coefficients of~\cite{Cowell:2024wyl} lie outside our perturbative region) we nonetheless recover information gains consistent with those of~\cite{Cowell:2024wyl} when using the same analysis specifications. The two approaches are therefore complementary: the GP method has more freedom to explore aggressive mark shapes, while ours provides full analytical control and a clear criterion for when the perturbative description is trustworthy.

We conclude with some remarks on caveats and future directions. A conceptually important result emerges from this optimization procedure. As first pointed out in~\cite{Cowell:2024wyl}, when the full data vector $\{P, C, M\}$ is used, the Fisher matrix is invariant under affine transformations of the mark, $m \to A\,m + B$, in the mildly nonlinear regime where the Gaussian likelihood approximation is valid. We apply this property here following~\cite{Cowell:2024wyl}; we note that at highly nonlinear scales, where the likelihood can develop significant non-Gaussianity, this invariance may no longer hold exactly. However, within our perturbative regime, it has a clear and important consequence: void-enhancing and overdensity-enhancing marks are equivalent in information content, and the optimizer has no reason to prefer one over the other. What it consistently selects for, across all parameters, smoothing scales, and redshifts considered, is a strong \emph{contrast} between under- and overdense regions, this being the only degree of freedom not removed by the invariance under affine symmetry of the Fisher matrix. The information gain from the marked power spectrum therefore does not reside in the special properties of voids per se, but in the environmental modulation of clustering captured by the mark. We stress that this conclusion is specific to the entire $\{P, C, M\}$ data vector; any analysis based on higher order functions such as the marked bispectrum (without including all the cross spectra) would not enjoy the same affine invariance, and in that case, the nature of the up-weighted regions would matter.

An important caveat is that the present analysis is restricted to the dark matter field in real space. Extending to biased tracers in redshift space is a necessary step before application to data and is currently a work in progress; preliminary results are encouraging.

Our framework also makes it straightforward to extend the analysis to cosmological parameters beyond vanilla $\Lambda$CDM, without requiring a dedicated simulation suite for each new parameter. Two extensions we are actively pursuing are the following. There is already a significant body of work showing that void-enhancing marked power spectra are particularly sensitive to neutrino masses~\cite{Massara:2020pli, Massara:2022zrf, Marinucci:2024bdq}, consistent with the physical expectation that neutrino free-streaming suppresses power preferentially in underdense regions. Within our framework, it will be interesting to derive the optimal $\{P, C, M\}$ combination for neutrino masses analytically, both to cross-check existing results and to ask whether the optimal mark for neutrinos differs qualitatively from the standard void-enhancing ansatz. We previously studied the performance of the marked power spectrum for primordial non-gaussianities $\fnl$ using the standard mark ansatz in~\cite{Marinucci:2024bdq}. That mark was originally designed with neutrinos and modified gravity in mind (see also~\cite{Karcher:2024twr}), and it is far from obvious that it is optimal for $\fnl$. The shapes of primordial non-Gaussianity couple very different configurations of the bispectrum (squeezed for local $\fnl$, equilateral or folded for other shapes), and the optimal environmental weighting may be qualitatively different. Applying our optimization pipeline to $\fnl$ for different primordial shapes is therefore a genuinely open question that we plan to address in forthcoming work.

More broadly, the marked power spectrum can be understood as a practical compression scheme for information from higher order functions: at one-loop, it provides a weighted integral over bispectrum triangle configurations, where the weight function is determined by the mark. On mildly nonlinear scales, where the bispectrum dominates the non-Gaussian signal, an optimized marked power spectrum can, in principle, perform comparably to a full $P + B$ analysis~\cite{Marinucci:2024bdq, Ebina:2026qzf}, while being considerably cheaper to compute and model. At more nonlinear scales, the mark also begins to compress information from higher-order correlators, while still retaining the simplicity of a two-point function. This is the deeper motivation for our interest in marked statistics as an efficient and practical summary for nonlinear LSS, and it underlies all the extensions described above. At those more nonlinear scales, however, a purely perturbative treatment is no longer sufficient, and one must rely on simulations and/or machine-learning emulators, as it has been done for $\fnl$ in~\cite{Jung:2024esv}.

\section*{Acknowledgements}

The authors thank Massimo Pietroni and Francesco Spezzati for useful discussions. ML and MM acknowledge support
by the MIUR Progetti di Ricerca di Rilevante Interesse Nazionale (PRIN) Bando 2022 - grant 20228RMX4A.
FS is partly supported by ICSC - Centro Nazionale di Ricerca in High Performance Computing, Big Data and Quantum Computing, funded by European Union - NextGenerationEU.

\appendix

\section{Validity of perturbation theory for marked fields}
\label{app:PTloop}
In this appendix we further investigate the accuracy of the perturbative approach adopted in this work for the description of marked fields. The mark function we use in this work is
\begin{equation}
    m(\delta_R) = c_0 + c_1 \delta_R + c_2 \delta_R^2 + c_3 \delta_R^3\,.
\end{equation}
One may wonder if the arbitrariness of the $c_i$'s coefficient may invalidate the assumption that the new weighted field is perturbative. Intuitively, the answer to this question depends on the underlying theoretical framework. In the perturbative context of the EFTofLSS, the coefficients are required to satisfy constraints dictated by perturbativity and by the convergence of the theory. In particular, we already know that some one-loop terms can give non-negligible contributions on very large scales, as already pointed out in~\cite{Philcox:2020fqx, Philcox:2020srd, Ebina:2024zkv}. This is due to the fact that mass conservation is not a symmetry of the marked field, even for dark matter, and the usual behavior $\sim k^2$ as $k\to 0$ is not guaranteed anymore, similarly to what happens for standard dark matter tracers~\cite{DAmico:2021rdb, Marinucci:2024add, Biselli:2026yte}. In addition, higher loop terms affecting large scales is a distinct feature of marked correlators: we are bringing information from nonlinear to linear scales, enhancing the effect of short modes on long ones. This is where the information gain resides, as already stated in previous works~\cite{Philcox:2020fqx, Philcox:2020srd, Ebina:2024zkv, Marinucci:2024bdq}. Even though this looks appealing, we need to require that the theory is still under perturbative control: since we stop at the one-loop order, we need to impose that the contributions from higher loops is negligible compared to the lower ones. This idea was previously introduced in the context of the EFTofLSS in~\cite{Braganca:2023pcp}, see also~\cite{Spezzati:2025zsb}. Its purpose is to restrict the parameter volume to regions of the bias-parameter space where higher-loop contributions remain under theoretical control. In the context of marked correlators, this kind of prior becomes even more important as we will see, since higher loops contribute non-negligibly already at large scales.

\subsection{Perturbativity prior from scaling arguments}

It is useful to evaluate the large scale limit of the one-loop contributions and compare them with the linear terms of the marked power spectrum. We know that the marked power spectrum at one loop at large scales is dominated by two pieces
\begin{equation}
    M_{22}(k) = 2 \int_{\bq}\big[H_2(\bk - \bq, \bq)\big]^2 P_{\rm lin}(q)P_{\rm lin}(|\bk - \bq|)\,,
\end{equation}
and
\begin{equation}
    M_{13}(k) = 6 C_{\delta_M}(k) P_{\rm lin}(k)\int_\bq H_3(\bk, \bq, -\bq)P_{\rm lin}(q)\,.
\end{equation}
Let us consider $M_{13}$ first: its ratio with the linear contribution $M_{11}$ is given by
\begin{equation}
    \begin{aligned}
        \frac{M_{13}(k)}{M_{11}(k)} = 6 \int_\bq F_3(\bk, \bq, -\bq)P_{\rm lin}(q) + \frac{8}{C_{\delta_M}(k)}\int_\bq C_{\delta_M^2}(q, |\bk - \bq|)F_2(\bk, -\bq)P_{\rm lin}(q) + \frac{6}{C_{\delta_M}(k)}\int_\bq C_{\delta_M^3}(k, q, q)P_{\rm lin}(q)\,.
    \end{aligned}
    \label{eq:M13ratio}
\end{equation}
Recall that the weighting functions of order $n$ contain, roughly, $n$ window functions $W_R$, explicitly $C_{\delta_m^n}(k) \sim W_R^n(k)$. These terms are under perturbative control: the first piece in Eq.~\eqref{eq:M13ratio} is the same as in the standard power spectrum, and it behaves as $\sim \sqrt{\Delta^2(k)}$, while the other two pieces behave as $\sim\sqrt{\Delta_M^2(k)}$. For the $M_{22}$ piece, the story is different as it is the one responsible for the shot-noise-like terms that can be dominant already at large scales, where perturbation theory is expected to be valid. It reads
\begin{equation}
\begin{aligned}
    \frac{M_{22}(k)}{M_{11}(k)} =& \frac{2 \int_\bq \big[F_2(\bk - \bq, \bq)\big]^2 P_{\rm lin}(q)P_{\rm lin}(|\bk - \bq|)}{P_{\rm lin}(k)} + \frac{4 \int_\bq C_{\delta_M^2}(q, |\bk - \bq|) F_2(\bk - \bq, \bq)P_{\rm lin}(|\bk - \bq|)P_{\rm lin}(q)}{C_{\delta_M}(k) P_{\rm lin}(k)} \\
    &+ \frac{2 \int_\bq C_{\delta_M^2}^2(q, |\bk - \bq|)P_{\rm lin}(q)P_{\rm lin}(|\bk - \bq|)}{C_{\delta_M}^2(k)P_{\rm lin}(k)}\,.
\end{aligned}
\label{eq:M22ratio}
\end{equation}
The first and the second term are safe, as they behave as $\sim \sqrt{\Delta^2(k)}$ and $\sim \sqrt{\Delta_M^2(k)}$ respectively. The last term in the second line of Eq.~\eqref{eq:M22ratio} is the more problematic: if we take the large scale limit of such term we obtain 
\begin{equation}
\lim_{k\to0} \frac{2 \int_\bq C_{\delta_M^2}^2(q, |\bk - \bq|)P_{\rm lin}(q)P_{\rm lin}(|\bk - \bq|)}{C_{\delta_M}^2(k)P_{\rm lin}(k)} = \frac{2 \int_\bq C_{\delta_M^2}^2(q,q) P_{\rm lin}^2(q)}{(c_0 + c_1)^2 P_{\rm lin}(k)}\,,
\end{equation}
which becomes problematic since the numerator is constant and the denominator goes to zero as $\lim_{k\to 0}P_{\rm lin}(k)\sim k^{n_s}$.

\noindent Crucially, if we open up the expression for this limit we can see that it depends on the value of $c_2$
\begin{equation}
\lim_{k\to0} \frac{2 \int_\bq C_{\delta_M^2}^2(q, |\bk - \bq|)P_{\rm lin}(q)P_{\rm lin}(|\bk - \bq|)}{C_{\delta_M}^2(k)P_{\rm lin}(k)} = \frac{2\Big(c_2^2 \int_\bq W_R^4(q) P_{\rm lin}^2(q) + c_1^2 \int_\bq W_R^2(q) P_{\rm lin}^2(q) + 2 c_2 c_1 \int_\bq W_R^3(q) P_{\rm lin}^2(q) \Big)}{(c_0 + c_1)^2 P_{\rm lin}}\,,
\end{equation}
where the integrals evaluate to
\begin{equation}
    \begin{aligned}
        \int_\bq W_R^2(q) P_{\rm lin}^2(q) \simeq 275\,,\qquad \int_\bq W_R^3(q) P_{\rm lin}^2(q) \simeq 181\,, \qquad \int_\bq W_R^4(q) P_{\rm lin}^2(q) \simeq 132\,
    \end{aligned}
    \label{eq:SNR_M22}
\end{equation}
for $R = 30 \, \text{Mpc}/h$ and $z = 0$, and
\begin{equation}
\begin{aligned}
        \int_\bq W_R^2(q) P_{\rm lin}^2(q) \simeq 863\,,\qquad \int_\bq W_R^3(q) P_{\rm lin}^2(q) \simeq 643\,, \qquad \int_\bq W_R^4(q) P_{\rm lin}^2(q) \simeq 512\,
    \end{aligned}
\end{equation}
for $R = 15 \, \text{Mpc}/h$ and $z = 0$. 

The results of these calculations are shown in the main text in Fig.~\ref{fig:PTprior}. 

\subsection{Higher-loop contributions for marked power spectra}

\label{app:highloop}
 
We start from the expression for the marked overdensity field
\begin{equation}
    \delta_M(\bx) = \frac{m (\delta_R(\bx))}{\bar{m}} \left[1 + \delta(\bx)\right] - 1\,.
\end{equation}
At this stage, the expression is kept fully non-perturbative, since it corresponds to the quantity provided to simulations (or observational data) for the computation of the two-point correlation function of the marked field. We first evaluate the marked power spectrum in its full non-perturbative form, as obtained from simulations, and then compare it to the perturbative prediction.
In Fourier space this becomes 
\begin{align}
\label{eq:app_dM_NPT}
    \delta_M(\bk) = \frac{1}{\bar{m}}\bigg\{&c_0 \delta(\bk)+ c_1 \delta_R(\bk)\\
    & +\mathcal{I}_{\bk;\bq_1, \bq_2 }\left[c_2 \delta_R(\bq_1)\delta_R(\bq_2) + \frac{c_1}{2}\left(\delta_R(\bq_1) \delta(\bq_2) + \delta(\bq_1) \delta_R(\bq_2)\right)\right] \nonumber\\
    & +\mathcal{I}_{\bk; \bq_1, \bq_2, \bq_3} \left[ c_3 \delta_R(\bq_1)\delta_R(\bq_2)\delta_R(\bq_3) + \frac{c_2}{3}\left(\delta_R(\bq_1)\delta_R(\bq_2)\delta(\bq_3) + \text{2 perms.}\right)\right] \nonumber\\
    &+\mathcal{I}_{\bk; \bq_1, \bq_2, \bq_3}  c_3\delta_R(\bq_1)\delta_R(\bq_2)\delta_R(\bq_3)\delta(\bq_4)\bigg\}\,, \nonumber
\end{align}
where we have used the notation
\begin{equation}
    \mathcal{I}_{\bk; \bq_1, \dots \bq_n} \equiv \int \frac{d^3\bq_1}{(2\pi)^3} \dots \int \frac{d^3\bq_n}{(2\pi)^3} (2\pi)^3\delta_D(\bk - \bq_{1\dots n})\,.
\end{equation}
We emphasize that this expression is precisely the one employed in the estimation of the power spectrum.
\begin{equation}
    \bar{m}^2\langle \delta_M(\bk)\delta_M(\bk')\rangle = (2\pi)^2 \delta_D(\bk + \bk')M(k)\,,
\end{equation}
and this expression is not perturbative, meaning that the density fields are the nonlinear ones. The expression for the full marked power spectrum presents many different pieces: we start with the one involving only one field (the first line of Eq.~\eqref{eq:app_dM_NPT})
\begin{equation}
    \bar{m}^2\langle\delta_M^{[1]}(\bk)\delta_M^{[1]}(\bk')\rangle' = (c_0 + c_1 W_R(\bk))^2P_{\rm NL} (k)
    \label{eq:app_PNL}
\end{equation}
where $\delta_M^{[1]}$ indicates the terms with only one field in Eq.~\eqref{eq:app_dM_NPT}, and $\langle\dots\rangle'$ is the correlator where the $(2\pi)^3\delta_D(\bk +\bk')$ has been removed. $P_{\rm NL}$ represents the complete nonlinear power spectrum for the dark matter density perturbations. In our perturbative perspective, $P_{\rm NL}$ would simply be the EFTofLSS prediction up to some order, 1-loop in this work. We could already argue that this piece, in a perturbative framework and at the scales considered in this work, should be the dominant one, as it represents the leading order. 
At large scales this piece behaves as $\sim (c_0 - c_1)^2 P_{\rm NL }(k)$, with $P_{\rm NL}(k) \simeq P_{\rm lin}(k) \to k^{n_s}$ for $k\to 0$. Imposing that this piece has to be the dominant one will turn out to be a constrain on the possible values for $c_2$ and $c_3$ compared to $c_1$ as we will see below. 
Moving on to contributions to the marked power spectrum that involve on and two fields we have
\begin{equation}
    \bar{m}^2\langle\delta_M^{[1]}(\bk)\delta_M^{[2]}(\bk')\rangle = C_{\deltam}(k) \int \frac{d^3\bq}{(2\pi)^3} C_{\deltam^2}(q, |\bk + \bq|) B_{\rm NL}(\bk, \bq, -\bk - \bq)\,,
\end{equation}
and 
\begin{equation}
    \bar{m}^2\langle\delta_M^{[1]}(\bk')\delta_M^{[2]}(\bk)\rangle = C_{\deltam}(k') \int \frac{d^3\bq}{(2\pi)^3} C_{\deltam^2}(q, |\bk - \bq|) B_{\rm NL}(-\bk, \bq, \bk - \bq)\,.
\end{equation}
This interesting exercise show explicitly that the two-point correlation function of the marked field contains pieces of higher order correlators, such as the (convolved) three-point correlation function, as shown in the above expressions. 
Considering the configurations with four fields ($[2]-[2]$ and $[1]-[3]$) we obtain
\begin{equation}
    \bar{m}^2\langle\deltam^{[2]}(\bk)\deltam^{[2]}(\bk')\rangle = \mathcal{I}_{\bk;\bq_1, \bq_2} \mathcal{I}_{\bk';\bp_1, \bp_2}C_{\deltam^2}(q_1, q_2)C_{\deltam^2}(p_1, p_2)\langle\delta(\bq_1)\delta(\bq_2)\delta(\bp_1)\delta(\bp_2)\rangle\,,
    \label{eq:app_d2d2}
\end{equation}
where the four-point function has a disconnected part
\begin{equation}
\langle\delta(\bq_1)\delta(\bq_2)\delta(\bq_3)\delta(\bq_4)\rangle_{\rm dc} = 2(2\pi)^6 \delta_D(\bq_{13}) \delta_D(\bq_{24}) P_{\rm NL}(q_1)P_{\rm NL}(q_2)\,,
\end{equation}
where we have already neglected the term that gives a disconnected contribution in Eq.~\eqref{eq:app_d2d2}, and a connected piece
\begin{equation}
    \langle\delta(\bq_1)\delta(\bq_2)\delta(\bq_3)\delta(\bq_4)\rangle_{\rm c} = (2\pi)^3 \delta_D(\bq_{1234}) T_{\rm NL}(\bq_1, \bq_2, \bq_3, \bq_4)\,.
\end{equation}
These two give
\begin{equation}
\begin{aligned}
    \bar{m}^2\langle\deltam^{[2]}(\bk)\deltam^{[2]}(\bk')\rangle = (2\pi)^3\delta_D(\bk + \bk') \Bigg[&2\int \frac{d^3 \bq}{(2\pi)^3}C_{\delta_M^2}^2(q, |\bk - \bq|) P_{\rm NL}(q)P_{\rm NL}(|\bk - \bq|) + \\
    &\int\frac{d^3 \bq_1}{(2\pi)^3}\frac{d^3 \bq_2}{(2\pi)^3} C_{\delta_M^2}(q_1, |\bk - \bq_1|) C_{\delta_M}(q_2, |\bk + \bq_2|) T_{\rm NL}(\bq_1, \bk - \bq_1, \bq_2, - \bk -     \bq_2)\Bigg]\,.
    \label{eq:app_tri}
\end{aligned}
\end{equation}
The other piece with four fields is given by
\begin{equation}
\begin{aligned}
\bar{m}^2(\langle\delta_M^{[1]}(\bk)\delta_M^{[3]}(\bk')\rangle + \langle\delta_M^{[3]}(\bk)\delta_M^{[1]}(\bk')\rangle) = &(2\pi)^3 \delta_D(\bk + \bk')\Bigg[6C_{\deltam}(k) P_{\rm NL}(k)\int\frac{d^3\bq}{(2\pi)^3} C_{\deltam^3}(k, q, q)P_{\rm NL}(q) \\
& + C_{\deltam}(k)\int\frac{d^3\bq_1}{(2\pi)^3}\frac{d^3\bq_2}{(2\pi)^3} \Big(C_{\deltam^3}(q_1, q_2, |\bk + \bq_{12}|)T_{\rm NL}(\bk, \bq_1, \bq_2, -\bk - \bq_{12}) \\
&+ C_{\deltam^3}(q_1, q_2, |\bk - \bq_{12}|)T_{\rm NL}(-\bk, \bq_1, \bq_2, \bk - \bq_{12})\Big)\Bigg].
\end{aligned}
\end{equation}
Note that the terms in the first line of Eq.~\eqref{eq:app_tri} are those entering our 1-loop expression for the marked power spectrum.

In what follows we focus on higher-loop contributions that can contaminate the large scale limit. The leading terms of this type arise at two-loop order and originate from the two-loop power spectrum, the one-loop bispectrum, and the tree-level trispectrum. These represent the lowest-order higher-point contributions that can propagate to large scales and therefore potentially spoil the expected large scale behavior. Higher-order correlators, or higher perturbative orders of the same correlators, involve an increasing number of factors of the linear power spectrum. In particular, such terms typically contain at least four powers of the linear spectrum. Their contribution is therefore parametrically suppressed once the conditions that we impose below are enforced. These conditions effectively ensure that the perturbative expansion remains well behaved and that higher-order terms do not generate sizeable contributions at large scales. For this reason, it is sufficient for our purposes to restrict the analysis to the terms listed above, which capture the dominant sources of higher-loop contamination within the perturbative regime. Contributions from higher-order correlators or higher-loop corrections to the same correlators are expected to be subleading under the assumptions adopted here.

The first term we compute is the one involving three fields:
\begin{equation}
    C_{\delta_M}(k) \int\frac{d^3\bq}{(2\pi)^3} C_{\delta_M^2}(q, |\bk - \bq|)B_{\rm NL}(-\bk, \bq, \bk - \bq)\,.
\end{equation}
In the EFTofLSS, the nonlinear bispectrum up to one-loop is given by
\begin{equation}
    B_{\rm NL }(\bk_1, \bk_2, \bk_3) = B_{\rm tree}(\bk_1, \bk_2, \bk_3) + B_{\rm 1loop}(\bk_1, \bk_2, \bk_3)\,,
\end{equation}
where
\begin{equation}
    B_{\rm tree}(\bk_1, \bk_2, \bk_3) = 2 F_2(\bk_1, \bk_2) P_L(k_1)P_L(k_2) + \text(2 perms. )\,,
\end{equation}
and the one-loop part is given by
\begin{equation}
    B_{\rm 1loop}(\bk_1, \bk_2, \bk_3) = B_{222}(\bk_1, \bk_2, \bk_3) + B_{321}^{(I)}(\bk_1, \bk_2, \bk_3) + B_{321}^{(II)}(\bk_1, \bk_2, \bq_3) + B_{411}(\bk_1, \bk_2, \bk_3)\,,
\end{equation}
with
\begin{equation}
\begin{aligned}
    B_{222}(\bk_1, \bk_2, \bk_3) &= 8 \int_\bq F_2(-\bq, \bk_1 + \bq) F_2(\bk_1 + \bq, \bk_2 - \bq)F_2(\bk_2 - \bq, \bq)P_L(q)P_L(|\bk_1 + \bq|)P_L(|\bk_2 - \bq|)\,,\\
    B_{321}^{(I)}(\bk_1, \bk_2, \bk_3) &= 6 P_L(k_1)\int_\bq F_3(-\bq, -\bk_2 + \bq, -\bk_1)F_2(\bq, \bk_2 - \bq) P_L(q) P_L(|\bk_2 - \bq|) + \text{5 perms. }\,,\\
    B_{321}^{(II)}(\bk_1, \bk_2, \bk_3) &= 6 P_L(k_1)P_L(k_2)F_2(\bk_1, \bk_2)\int_\bq F_3(\bk_1, \bq, -\bq) P_L(q) + \text{5 perms. }\,,\\
    B_{411}(\bk_1, \bk_2, \bk_3) &= 12 P_L(k_1)P_L(k_2) \int_\bq F_4(\bq, -\bq, -\bk_1, -\bk_2)P_L(q) + \text{2 perms.}\,. 
\end{aligned}
\end{equation}

The tree-level contribution is already contained in the one-loop expression for the marked power spectrum and therefore does not represent an additional independent contribution.

The large scale limit ($k\to0$) of the one-loop $321(I)$ term is
\begin{equation}
\begin{aligned}
    \lim_{k\to 0} C_{\delta_M}(k) &\int_\bq C_{\delta_M^2}(q, |\bk - \bq|)B_{321}^{(I)}(-\bk, \bq, \bk - \bq) = \\
    &= 6 (c_0 + c_1)\int_\bq \big(c_2 W_R(q)^2 + c_1 W_R(q)\big)P_L^2(q)\int_\bp F_3(- \bp , \bq + \bp, -\bq)F_2(\bp, - \bq - \bp) P_L(|\bq + \bp|)\nonumber\\
    &\equiv 6 (c_0 + c_1) \big(c_2 \,\mathcal{K}_1 + c_1 \,\mathcal{K}_2\big) = \text{const.}
\end{aligned}
\end{equation}

For the term with $B_{321}^{(II)}$, in the $k\to 0 $ limit, we obtain
\begin{equation}
\begin{aligned}
    \lim_{k\to 0} C_{\delta_M}(k) &\int_\bq C_{\delta_M^2}(q, |\bk - \bq|)B_{321}^{(II)}(-\bk, \bq, \bk - \bq) = 6 C_{\delta_M}(k) \int_\bq C_{\delta_M^2}(q,q)\times \\
    &\times\Bigg\{ P_L(k)P_L(q) F_2(\bq, 0)\int_\bp F_3(\bq, -\bp, \bp) P_L(p) + P_L(k) P_L(q) F_2(\bq, 0)\int_\bp F_3(0, \bp, -\bp) P_L(p)\\
    & + P_L(k) P_L(q) F_2(0, -\bq)\int_\bp F_3(-\bq, \bp, -\bp) P_L(p) + P_L(k) P_L(q) F_2(0, -\bq) \int_\bp F_3(\bp, \bp, -\bp)P_L(p)\Bigg\}\,.
\end{aligned}
\end{equation}
This result shows that this piece goes like $\sim C\times P_L(k)$ which safely goes to zero for $k\to 0$.
The $B_{222}$ part gives results similar to $B_{321}^{(I)}$
\begin{equation}
\begin{aligned}
        \lim_{k\to 0} C_{\delta_M}(k) \int_\bq C_{\delta_M^2}(q, |\bk - \bq|)B_{222}(-\bk, \bq, \bk - \bq) \simeq  8 (c_0 + c_1)\big(c_2 \,\mathcal{K}_3 + c_1 \,\mathcal{K}_4\big)\,.
\end{aligned}    
\end{equation}
These results show how higher order terms can give a non-negligible contribution already at large scales, depending on the mark parameters $\{c_0, c_1, c_2\}$.

Let us consider now 4 fields
\begin{align}
\label{eq:app_22}
    \bar{m}^2 \langle \delta_M^{[2]}(\bk)\delta_M^{[2]}(\bk')\rangle' =& 2 \int_\bq C_{\delta_M^2}^2(q, |\bk - \bq|) P_{\rm NL}(q) P_{\rm NL}(|\bk - \bq|)\\
    & + \int_{\bq_1, \bq_2}C_{\delta_M^2}(q_1, |\bk - \bq_1|)C_{\delta_M^2}(q_2, |\bk + \bq_2|) T_{\rm NL}(\bq_1, \bk - \bq_1, \bq_2, -\bk - \bq_2)\,.
\end{align} 
In the first line, the one involving the nonlinear power spectrum we expect, at the order we are interested in, contributions up to one-loop, meaning $~P_LP_L$ and $~P_LP_{\rm 1loop}$. The first contribution is simply the one already obtained with the one-loop calculation of the marked power spectrum
\begin{equation}
    2 \int_\bq C_{\delta_M^2}^2(q, |\bk - \bq|)P_L(q)P_L(|\bk - \bq|) \to 2 \int_\bq C_{\delta_M^2}^2(q, q)P_L^2(q) \sim \text{const.}\,.
\end{equation}
The $~P_LP_{\rm 1loop}$ terms give
\begin{align}
    2 \int_\bq C_{\delta_M^2}^2(q, |\bk - \bq|)\Bigg[&P_L(q) \bigg(2 \int_\bp F_2^2(\bk- \bq -\bp , \bp) P_L(p)P_L(|\bk - \bq - \bp|) + 6 \int_\bp F_3(\bk - \bq, \bp, -\bp) P_L(p) P_L(|\bk - \bq|)\bigg)\\
    & +P_L(|\bk - \bq|) \bigg(2 \int_\bp F_2^2(\bq -\bp , \bp) P_L(p)P_L(|\bq - \bp|) + 6 \int_\bp F_3(\bq, \bp, -\bp) P_L(p) P_L(q)\bigg) \Bigg]\,,
\end{align}
which becomes, in the limit $k\to 0$
\begin{align}
    &8 \int_\bq C_{\delta_M^2}^2(q,q) \Bigg[P_L(q) \bigg(\int_\bp F_2^2(\bp, -\bp - \bq) P_L(p) P_L(|\bq + \bp|) + 3 \int_\bp F_3(-\bq, -\bp, \bp)P_L(p) P_L(q)\bigg)\Bigg]= \\
    & = c_2^2 \mathcal{K}_{n_1}+ 2 c_1 c_2 \mathcal{K}_{n_2} + c_1^2 \mathcal{K}_{n_3}\,.
\end{align}
For the second piece in Eq.~\eqref{eq:app_22}, the one involving the nonlinear trispectrum, we just need to stop at the tree-level, as it already involves three power spectra
\begin{equation}
\begin{aligned}
    T_{\rm NL}(\bq_1, \bq_2, \bq_3, \bq_4) = T_{\rm tree}(\bq_1, \bq_2, \bq_3, \bq_4) =& \; 4 P_L(q_1) P_L(q_2) \big( F_2(\bq_1, -\bq_1 - \bq_3) F_2(\bq_2, \bq_1 + \bq_3) P_L (|\bq_! + \bq_3|) \\
&  + F_2(\bq_1, -\bq_1 - \bq_4) F_2(\bq_2, \bq_1 + \bq_4) P_L(|\bq_1 + \bq_4|) \big) + 5\, \text{perm.} \\
& + 6 \, F_3(\bq_1, \bq_2, \bq_3) P_L(q_1)P_L(q_2) P_L(q_3)+ 3 \, \text{perm.}\; ,
\end{aligned}
\end{equation}

The piece with $F_3$ gives
\begin{equation}
    \begin{aligned}
        6 \int_{\bq_1}\int_{\bq_2}C_{\delta_M^2}(q_1, |\bk - \bq_1|)C_{\delta_M^2}(q_2, |\bk + \bq_2|) \Big[&F_3 (\bq_1, \bk - \bq_1, \bq_2)P_L(q_1)P_L(q_2)P_L(|\bk - \bq_1|)\\
        &+F_3 (\bq_1, \bk - \bq_1, -\bk-\bq_2)P_L(q_1)P_L(|\bk - \bq_1|)P_L(|\bk + \bq_2|)\\
        &+F_3 (\bq_1, \bq_2, -\bk-\bq_2)P_L(q_1)P_L(q_2)P_L(|\bk + \bq_2|)\\
        &+F_3 (\bk-\bq_1, \bq_2, -\bk-\bq_2)P_L(q_2)P_L(|\bk - \bq_1|)P_L(|\bk + \bq_2|)\Big]\,,
    \end{aligned}
\end{equation}
which, in the limit $k \to 0$ becomes
\begin{equation}
    \begin{aligned}
        &12 \int_{\bq_1}\int_{\bq_2}C_{\delta_M^2}(q_1, q_1)C_{\delta_M^2}(q_2, q_2) \Big[F_3 (\bq_1, -\bq_1, \bq_2)P_L(q_1)^2P_L(q_2)\\
        &\hspace{5.6cm}+F_3 (\bq_1, \bq_2, -\bq_2)P_L(q_1)P_L(q_2)^2\Big] = \\
        &= c_2^2 \mathcal{K}_{x}+c_2 c_1\mathcal{K}_{y}+c_1^2 \mathcal{K}_{z}\,.
    \end{aligned}
\end{equation}
Finally, let us consider the first term involving the mark coefficient $c_3$, $[3]-[3]$, see Eq.~\eqref{eq:app_dM_NPT}
\begin{equation}
\begin{aligned}
    \bar{m}^2
    \langle\delta_M^{[3]}(\bk)\delta_M^{[3]}(\bk')\rangle = \mathcal{I}_{\bk;\bq_1, \bq_2, \bq_3}\mathcal{I}_{\bk';\bp_1, \bp_2, \bp_3} H_3(\bq_1, \bq_2, \bq_3)H_3(\bp_1, \bp_2, \bp_3) \langle\delta_{\bq_1}\delta_{\bq_2}\delta_{\bq_3}\delta_{\bp_1}\delta_{\bp_2}\delta_{\bp_3}\rangle\,.
\end{aligned}
\end{equation}
The correlator inside the integral gives a different kind of contribution, $\sim PPP, BB, PT, 6pt$: here we focus on the contribution $\sim PPP$ which is nothing but a term coming from the 2loop power spectrum. It reads
\begin{equation}
    \begin{aligned}
        \bar{m}^2
    \langle\delta_M^{[3]}(\bk)\delta_M^{[3]}(\bk')\rangle'_{PPP} =& 9P_L(k)\int_{\bq, \bp} H_3(\bk, \bq, -\bq)H_3(-\bk, \bp, -\bp) P_L(q)P_L(p) \\
    & +6 \int_{\bq, \bp}\Big[H_3(\bq, \bp, \bk - \bq, -\bp)\Big]^2 P_L(q)P_L(p)P_L(|\bk - \bq - \bp|)\,.
    \end{aligned}
\end{equation}
The first line behaves as $\sim P_L(k)$ at large scales, while the second line gives constant contributions $\propto (c_3^2, c_2 c_3, c_2^2)$. As a consequence of this, both $c_2$ and $c_3$ have to undergo the perturbativity criteria delineated in Section~\ref{sec:PTprior}.

\section{Affine invariance of the Fisher matrix for higher-order marked correlators}
\label{app:affine}

In the main text, we stated that the affine invariance of the Fisher matrix extends beyond the marked power spectrum to any complete set of $n$-point correlators of the density field $\delta$ and the marked field $\Delta$. Here we provide the explicit proof.

\subsection*{General argument}

\noindent Consider a data vector $\mathbf{d}$ comprising all auto- and cross-spectra of $\delta$ and $\Delta$ up to order $n$. The Fisher matrix is
\begin{equation}
    F_{\alpha\beta} = \left(\partial_\alpha \mathbf{d}\right)^T 
    C^{-1}\, \partial_\beta \mathbf{d}\,,
\end{equation}
where $C$ is the covariance matrix, which we leave completely general. Under the affine transformation of the mark function $m' = Am + B$, the marked field transforms linearly as
\begin{equation}
    \Delta' = A\,\Delta + B\,\delta\,,
\end{equation}
since $\Delta = m(\delta_R)(1+\delta) - \langle m(\delta_R)(1+\delta)\rangle$ is linear in $m$ by construction. Because every element of $\mathbf{d}$ is multilinear in the fields, each $n$-point correlator transforms as a polynomial in $A$ and $B$ of degree at most $n$. The data vector therefore transforms linearly as $\mathbf{d}' = \mathbf{T}\,\mathbf{d}$, where $\mathbf{T}$ is a square matrix whose entries are polynomials in $A$ and $B$. The covariance and derivatives transform accordingly as
\begin{equation}
    C' = \mathbf{T}\, C\, \mathbf{T}^T\,, \qquad 
    \partial_\alpha \mathbf{d}' = \mathbf{T}\,\partial_\alpha 
    \mathbf{d}\,.
\end{equation}
Substituting into the Fisher matrix,
\begin{equation}
    F'_{\alpha\beta} 
    = \left(\partial_\alpha \mathbf{d}\right)^T \mathbf{T}^T 
      \left(\mathbf{T}\, C\, \mathbf{T}^T\right)^{-1} 
      \mathbf{T}\,\partial_\beta \mathbf{d}
    = \left(\partial_\alpha \mathbf{d}\right)^T C^{-1} 
      \partial_\beta \mathbf{d}
    = F_{\alpha\beta}\,,
\end{equation}
where we used $\mathbf{T}^T(\mathbf{T}^T)^{-1} = \mathbf{I}$ and $\mathbf{T}^{-1}\mathbf{T} = \mathbf{I}$. 

The proof therefore reduces entirely to showing that $\mathbf{T}$ is invertible. This follows from the fact that $\mathbf{T}$ is lower-triangular, as can be seen by recalling that substituting $\Delta' = A\Delta + B\delta$ into any $n$-point correlator only generates correlators of equal or lower degree in $\Delta$, never higher. The determinant is therefore simply the product of the diagonal entries, each of which is a power of $A$, giving $\det \mathbf{T} = A^{n(n+1)/2} \neq 0$ for $A \neq 0$, as explicitly verified in the example below.

\subsection*{Explicit example: marked power spectrum and bispectrum}

We illustrate the argument with the data vector comprising the full set of marked power spectra and bispectra,
\begin{equation}
    \mathbf{d} = \left(P_{\delta\delta},\, P_{\delta\Delta},\, 
    P_{\Delta\Delta},\, B_{\delta\delta\delta},\, 
    B_{\delta\delta\Delta},\, B_{\delta\Delta\Delta},\, 
    B_{\Delta\Delta\Delta}\right)\,.
\end{equation}
Using $\Delta' = A\Delta + B\delta$, the transformation rules follow from the multilinearity of each correlator. For the power spectra:
\begin{align}
    P_{\delta\delta}' &= P_{\delta\delta}\,, \\
    P_{\delta\Delta}' &= A\,P_{\delta\Delta} + B\,P_{\delta\delta}\,, \\
    P_{\Delta\Delta}' &= A^2\,P_{\Delta\Delta} + 2AB\,P_{\delta\Delta} 
    + B^2\,P_{\delta\delta}\,.
\end{align}
For the bispectra, expanding $\Delta' = A\Delta + B\delta$ in each marked slot:
\begin{align}
    B_{\delta\delta\delta}' &= B_{\delta\delta\delta}\,, \\
    B_{\delta\delta\Delta}' &= A\,B_{\delta\delta\Delta} 
    + B\,B_{\delta\delta\delta}\,, \\
    B_{\delta\Delta\Delta}' &= A^2\,B_{\delta\Delta\Delta} 
    + 2AB\,B_{\delta\delta\Delta} + B^2\,B_{\delta\delta\delta}\,, \\
    B_{\Delta\Delta\Delta}' &= A^3\,B_{\Delta\Delta\Delta} 
    + 3A^2B\,B_{\delta\Delta\Delta} + 3AB^2\,B_{\delta\delta\Delta} 
    + B^3\,B_{\delta\delta\delta}\,.
\end{align}
The transformation matrix $\mathbf{T}$ is therefore lower-triangular,
\begin{equation}
    \mathbf{T} = \begin{pmatrix}
    1   & 0    & 0   & 0   & 0    & 0    & 0   \\
    B   & A    & 0   & 0   & 0    & 0    & 0   \\
    B^2 & 2AB  & A^2 & 0   & 0    & 0    & 0   \\
    0   & 0    & 0   & 1   & 0    & 0    & 0   \\
    0   & 0    & 0   & B   & A    & 0    & 0   \\
    0   & 0    & 0   & B^2 & 2AB  & A^2  & 0   \\
    0   & 0    & 0   & B^3 & 3AB^2& 3A^2B& A^3
    \end{pmatrix}\,,
\end{equation}
with determinant $\det \mathbf{T} = A^9\,$, which is non-vanishing for any $A \neq 0$. The matrix $\mathbf{T}$ is  therefore invertible, and the Fisher matrix is invariant under the affine transformation $F'_{\alpha\beta} = F_{\alpha\beta}$ for any  covariance structure $C$.

\section{Additional results}
\label{app:further}
We report here further results of our analysis. In table~\ref{tab:Cs_R30_R15} we report the mark coefficient obtained through the optimization procedure. The resulting mark functions are plotted in Fig.~\ref{fig:OptMarkR30}.

\begin{table}[h]
\centering
\begin{tabular}{c c c}

\begin{tabular}{|c|c|c|c|}
\hline

\multicolumn{4}{|c|}{
\parbox[c][1.0cm][c]{0.2\linewidth}{\centering $R = 10 \,\text{Mpc}\, h^{-1}$, $z=0$}
}\\
\hline

\cell{1.5cm}{0.6cm}{Parameter} & \cell{1.0cm}{0.6cm}{$c_1$} & \cell{1.0cm}{0.6cm}{$c_2$} & \cell{1.0cm}{0.6cm}{$c_3$} \\
\hline

\cell{1.5cm}{0.6cm}{FoM}  & \cell{1.0cm}{0.6cm}{-0.3} & \cell{1.0cm}{0.6cm}{0.02} & \cell{1.0cm}{0.6cm}{0.0036} \\
\hline

\cell{1.5cm}{0.6cm}{$\Omega_m$}  & \cell{1.0cm}{0.6cm}{-0.3} & \cell{1.0cm}{0.6cm}{0.02} & \cell{1.0cm}{0.6cm}{0.0036} \\
\hline

\cell{1.5cm}{0.6cm}{$\sigma_8$}  & \cell{1.0cm}{0.6cm}{-0.3} & \cell{1.0cm}{0.6cm}{0.02} & \cell{1.0cm}{0.6cm}{0.0036} \\
\hline

\cell{1.5cm}{0.6cm}{$h$}  & \cell{1.0cm}{0.6cm}{-0.3} & \cell{1.0cm}{0.6cm}{0.02} & \cell{1.0cm}{0.6cm}{0.0036} \\
\hline\hline

\multicolumn{4}{|c|}{
\parbox[c][1.0cm][c]{0.3\linewidth}{\centering $R = 10 \,\text{Mpc}\, h^{-1}$, $z=0.5$}
}\\
\hline

\cell{1.5cm}{0.6cm}{Parameter}  & \cell{1.0cm}{0.6cm}{$c_1$} & \cell{1.0cm}{0.6cm}{$c_2$} & \cell{1.0cm}{0.6cm}{$c_3$} \\
\hline

\cell{1.5cm}{0.6cm}{FoM}  & \cell{1.0cm}{0.6cm}{-0.4} & \cell{1.0cm}{0.6cm}{0.03} & \cell{1.0cm}{0.6cm}{0.0085} \\
\hline

\cell{1.5cm}{0.6cm}{$\Omega_m$}  & \cell{1.0cm}{0.6cm}{-0.4} & \cell{1.0cm}{0.6cm}{0.03} & \cell{1.0cm}{0.6cm}{0.0085} \\
\hline

\cell{1.5cm}{0.6cm}{$\sigma_8$}  & \cell{1.0cm}{0.6cm}{-0.4} & \cell{1.0cm}{0.6cm}{0.03} & \cell{1.0cm}{0.6cm}{0.0085} \\
\hline

\cell{1.5cm}{0.6cm}{$ h$}  & \cell{1.0cm}{0.6cm}{-0.4} & \cell{1.0cm}{0.6cm}{0.03} & \cell{1.0cm}{0.6cm}{0.0085} \\
\hline\hline

\multicolumn{4}{|c|}{
\parbox[c][1.0cm][c]{0.3\linewidth}{\centering $R = 10 \,\text{Mpc}\, h^{-1}$, $z=1$}
}\\
\hline

\cell{1.5cm}{0.6cm}{Parameter} & \cell{1.0cm}{0.6cm}{$c_1$} & \cell{1.0cm}{0.6cm}{$c_2$} & \cell{1.0cm}{0.6cm}{$c_3$} \\
\hline

\cell{1.5cm}{0.6cm}{FoM} & \cell{1.0cm}{0.6cm}{-0.6} & \cell{1.0cm}{0.6cm}{0.048} & \cell{1.0cm}{0.6cm}{0.021} \\
\hline

\cell{1.5cm}{0.6cm}{$\Omega_m$} & \cell{1.0cm}{0.6cm}{-0.4} & \cell{1.0cm}{0.6cm}{0.14} & \cell{1.0cm}{0.6cm}{0.062} \\
\hline

\cell{1.5cm}{0.6cm}{$\sigma_8$} & \cell{1.0cm}{0.6cm}{-0.6} & \cell{1.0cm}{0.6cm}{0.048} & \cell{1.0cm}{0.6cm}{0.021} \\
\hline

\cell{1.5cm}{0.6cm}{$h$} & \cell{1.0cm}{0.6cm}{-0.4} & \cell{1.0cm}{0.6cm}{0.14} & \cell{1.0cm}{0.6cm}{0.062} \\
\hline

\end{tabular}

&

\begin{tabular}{|c|c|c|c|}
\hline

\multicolumn{4}{|c|}{
\parbox[c][1.0cm][c]{0.3\linewidth}{\centering $R = 15 \,\text{Mpc}\, h^{-1}$, $z=0$}
}\\
\hline

\cell{1.5cm}{0.6cm}{Parameter}  & \cell{1.0cm}{0.6cm}{$c_1$} & \cell{1.0cm}{0.6cm}{$c_2$} & \cell{1.0cm}{0.6cm}{$c_3$} \\
\hline

\cell{1.5cm}{0.6cm}{FoM}  & \cell{1.0cm}{0.6cm}{-0.4} & \cell{1.0cm}{0.6cm}{-0.15} & \cell{1.0cm}{0.6cm}{0.068} \\
\hline

\cell{1.5cm}{0.6cm}{$\Omega_m$}  & \cell{1.0cm}{0.6cm}{-0.4} & \cell{1.0cm}{0.6cm}{-0.15} & \cell{1.0cm}{0.6cm}{0.068} \\
\hline

\cell{1.5cm}{0.6cm}{$\sigma_8$}  & \cell{1.0cm}{0.6cm}{-0.4} & \cell{1.0cm}{0.6cm}{-0.15} & \cell{1.0cm}{0.6cm}{0.068} \\
\hline

\cell{1.5cm}{0.6cm}{$h$}  & \cell{1.0cm}{0.6cm}{-0.4} & \cell{1.0cm}{0.6cm}{-0.15} & \cell{1.0cm}{0.6cm}{0.068} \\
\hline\hline

\multicolumn{4}{|c|}{
\parbox[c][1.0cm][c]{0.3\linewidth}{\centering $R = 15 \,\text{Mpc}\, h^{-1}$, $z=0.5$}
}\\
\hline

\cell{1.5cm}{0.6cm}{Parameter}  & \cell{1.0cm}{0.6cm}{$c_1$} & \cell{1.0cm}{0.6cm}{$c_2$} & \cell{1.0cm}{0.6cm}{$c_3$} \\
\hline

\cell{1.5cm}{0.6cm}{FoM}  & \cell{1.0cm}{0.6cm}{-3.3} & \cell{1.0cm}{0.6cm}{2.14} & \cell{1.0cm}{0.6cm}{1.39} \\
\hline

\cell{1.5cm}{0.6cm}{$\Omega_m$}  & \cell{1.0cm}{0.6cm}{2.1} & \cell{1.0cm}{0.6cm}{1.35} & \cell{1.0cm}{0.6cm}{0.23} \\
\hline

\cell{1.5cm}{0.6cm}{$\sigma_8$}  & \cell{1.0cm}{0.6cm}{2.5} & \cell{1.0cm}{0.6cm}{1.07} & \cell{1.0cm}{0.6cm}{-0.70} \\
\hline

\cell{1.5cm}{0.6cm}{$h$}  & \cell{1.0cm}{0.6cm}{0.8} & \cell{1.0cm}{0.6cm}{0.07} & \cell{1.0cm}{0.6cm}{-0.046} \\
\hline\hline

\multicolumn{4}{|c|}{
\parbox[c][1.0cm][c]{0.3\linewidth}{\centering $R = 15 \,\text{Mpc}\, h^{-1}$, $z=1$}
}\\
\hline

\cell{1.5cm}{0.6cm}{Parameter}  & \cell{1.0cm}{0.6cm}{$c_1$} & \cell{1.0cm}{0.6cm}{$c_2$} & \cell{1.0cm}{0.6cm}{$c_3$} \\
\hline

\cell{1.5cm}{0.6cm}{FoM}  & \cell{1.0cm}{0.6cm}{-2.5} & \cell{1.0cm}{0.6cm}{1.3} & \cell{1.0cm}{0.6cm}{0.64} \\
\hline

\cell{1.5cm}{0.6cm}{$\Omega_m$}  & \cell{1.0cm}{0.6cm}{4.1} & \cell{1.0cm}{0.6cm}{3.0} & \cell{1.0cm}{0.6cm}{2.55} \\
\hline

\cell{1.5cm}{0.6cm}{$\sigma_8$}  & \cell{1.0cm}{0.6cm}{-2.8} & \cell{1.0cm}{0.6cm}{2.5} & \cell{1.0cm}{0.6cm}{1.18} \\
\hline

\cell{1.5cm}{0.6cm}{$h$}  & \cell{1.0cm}{0.6cm}{1.7} & \cell{1.0cm}{0.6cm}{1.5} & \cell{1.0cm}{0.6cm}{-1.36} \\
\hline

\end{tabular}

&

\begin{tabular}{|c|c|c|c|}
\hline

\multicolumn{4}{|c|}{
\parbox[c][1.0cm][c]{0.3\linewidth}{\centering $R = 30 \,\text{Mpc}\, h^{-1}$, $z=0$}
}\\
\hline

\cell{1.5cm}{0.6cm}{Parameter}  & \cell{1.0cm}{0.6cm}{$c_1$} & \cell{1.0cm}{0.6cm}{$c_2$} & \cell{1.0cm}{0.6cm}{$c_3$} \\
\hline

\cell{1.5cm}{0.6cm}{FoM}  & \cell{1.0cm}{0.6cm}{3.6} & \cell{1.0cm}{0.6cm}{4.8} & \cell{1.0cm}{0.6cm}{0.4} \\
\hline

\cell{1.5cm}{0.6cm}{$\Omega_m$}  & \cell{1.0cm}{0.6cm}{-3.6} & \cell{1.0cm}{0.6cm}{4.0} & \cell{1.0cm}{0.6cm}{4.8} \\
\hline

\cell{1.5cm}{0.6cm}{$\sigma_8$}  & \cell{1.0cm}{0.6cm}{-2.4} & \cell{1.0cm}{0.6cm}{2.8} & \cell{1.0cm}{0.6cm}{2.8} \\
\hline

\cell{1.5cm}{0.6cm}{$h$}  & \cell{1.0cm}{0.6cm}{-3.6} & \cell{1.0cm}{0.6cm}{4.0} & \cell{1.0cm}{0.6cm}{4.8} \\
\hline\hline

\multicolumn{4}{|c|}{
\parbox[c][1.0cm][c]{0.3\linewidth}{\centering $R = 30 \,\text{Mpc}\, h^{-1}$, $z=0.5$}
}\\
\hline

\cell{1.5cm}{0.6cm}{Parameter}  & \cell{1.0cm}{0.6cm}{$c_1$} & \cell{1.0cm}{0.6cm}{$c_2$} & \cell{1.0cm}{0.6cm}{$c_3$} \\
\hline

\cell{1.5cm}{0.6cm}{FoM}  & \cell{1.0cm}{0.6cm}{4.4} & \cell{1.0cm}{0.6cm}{8.4} & \cell{1.0cm}{0.6cm}{3.6} \\
\hline

\cell{1.5cm}{0.6cm}{$\Omega_m$}  & \cell{1.0cm}{0.6cm}{0.4} & \cell{1.0cm}{0.6cm}{0.8} & \cell{1.0cm}{0.6cm}{1.2} \\
\hline

\cell{1.5cm}{0.6cm}{$\sigma_8$}  & \cell{1.0cm}{0.6cm}{4.4} & \cell{1.0cm}{0.6cm}{8.4} & \cell{1.0cm}{0.6cm}{3.6} \\
\hline

\cell{1.5cm}{0.6cm}{$h$}  & \cell{1.0cm}{0.6cm}{2.4} & \cell{1.0cm}{0.6cm}{4.8} & \cell{1.0cm}{0.6cm}{3.2} \\
\hline\hline

\multicolumn{4}{|c|}{
\parbox[c][1.0cm][c]{0.3\linewidth}{\centering $R = 30 \,\text{Mpc}\, h^{-1}$, $z=1$}
}\\
\hline

\cell{1.5cm}{0.6cm}{Parameter}  & \cell{1.0cm}{0.6cm}{$c_1$} & \cell{1.0cm}{0.6cm}{$c_2$} & \cell{1.0cm}{0.6cm}{$c_3$} \\
\hline

\cell{1.5cm}{0.6cm}{FoM}  & \cell{1.0cm}{0.6cm}{-2.8} & \cell{1.0cm}{0.6cm}{6.0} & \cell{1.0cm}{0.6cm}{8.8} \\
\hline

\cell{1.5cm}{0.6cm}{$\Omega_m$}  & \cell{1.0cm}{0.6cm}{1.6} & \cell{1.0cm}{0.6cm}{4.8} & \cell{1.0cm}{0.6cm}{4.0} \\
\hline

\cell{1.5cm}{0.6cm}{$\sigma_8$}  & \cell{1.0cm}{0.6cm}{2.4} & \cell{1.0cm}{0.6cm}{7.2} & \cell{1.0cm}{0.6cm}{-9.9} \\
\hline

\cell{1.5cm}{0.6cm}{$h$}  & \cell{1.0cm}{0.6cm}{3.6} & \cell{1.0cm}{0.6cm}{8.4} & \cell{1.0cm}{0.6cm}{-3.6} \\
\hline

\end{tabular}

\end{tabular}
\caption{ \justifying Values for the polynomial mark for smoothing radii $R = 15,\,30 \,\,\text{Mpc}/h$ at $z=0$ and $z=0.5$.}
\label{tab:Cs_R30_R15}
\end{table}

In table~\ref{tab:improv_all} we report the complete results of our analysis using both Gaussian and simulation-based covariances.
\begin{table}[h]
\centering
\begin{tabular}{c c}

\begin{tabular}{|c|c|c|c|}
\hline
\multicolumn{4}{|c|}{
\parbox[c][1.0cm][c]{0.3\linewidth}{\centering \large{Optimization of FoM}}
}\\
\hline
\hline
\multicolumn{4}{|c|}{\parbox[c][0.8cm][c]{0.3\linewidth}{\centering $R = 10 \,\text{Mpc}\, h^{-1}$ $z=0$}}\\
\hline
\cell{2.0cm}{0.5cm}{Covariance} & \cell{2.0cm}{0.5cm}{$r(\Omega_m)$} & \cell{2.0cm}{0.5cm}{$r(\sigma_8)$} & \cell{2.0cm}{0.5cm}{$r(h)$}  \\
\hline
\cell{2.0cm}{0.5cm}{Gaussian} & 1.54 & 14.69 & 1.60  \\
\hline
\cell{2.0cm}{0.5cm}{Simulations} & 2.34 & 8.32 & 2.47   \\
\hline
\multicolumn{4}{|c|}{\parbox[c][0.8cm][c]{0.3\linewidth}{\centering $R = 10 \,\text{Mpc}\, h^{-1}$ $z=0.5$}}\\
\hline
\cell{2.0cm}{0.5cm}{Covariance} & \cell{2.0cm}{0.5cm}{$r(\Omega_m)$} & \cell{2.0cm}{0.5cm}{$r(\sigma_8)$} & \cell{2.0cm}{0.5cm}{$r(h)$}  \\
\hline
\cell{2.0cm}{0.5cm}{Gaussian} & 1.20 & 2.75 & 1.23  \\
\hline
\cell{2.0cm}{0.5cm}{Simulations} & 1.53 & 3.11 & 1.50  \\
\hline
\multicolumn{4}{|c|}{\parbox[c][0.8cm][c]{0.3\linewidth}{\centering $R = 10 \,\text{Mpc}\, h^{-1}$ $z=1$}}\\
\hline
\cell{2.0cm}{0.5cm}{Covariance} & \cell{2.0cm}{0.5cm}{$r(\Omega_m)$} & \cell{2.0cm}{0.5cm}{$r(\sigma_8)$} & \cell{2.0cm}{0.5cm}{$r(h)$}  \\
\hline
\cell{2.0cm}{0.5cm}{Gaussian} & 1.12 & 1.86 & 1.13  \\
\hline
\cell{2.0cm}{0.5cm}{Simulations} & 1.43 & 2.06 & 1.38  \\
\hline
\hline
\multicolumn{4}{|c|}{\parbox[c][0.8cm][c]{0.3\linewidth}{\centering $R = 15 \,\text{Mpc}\, h^{-1}$ $z=0$}}\\
\hline
\cell{2.0cm}{0.5cm}{Covariance} & \cell{2.0cm}{0.5cm}{$r(\Omega_m)$} & \cell{2.0cm}{0.5cm}{$r(\sigma_8)$} & \cell{2.0cm}{0.5cm}{$r(h)$}  \\
\hline
\cell{2.0cm}{0.5cm}{Gaussian} & 2.17 & 12.05 & 2.30  \\
\hline
\cell{2.0cm}{0.5cm}{Simulations} & 1.48 & 5.55 & 1.57   \\
\hline
\multicolumn{4}{|c|}{\parbox[c][0.8cm][c]{0.3\linewidth}{\centering $R = 15 \,\text{Mpc}\, h^{-1}$ $z=0.5$}}\\
\hline
\cell{2.0cm}{0.5cm}{Covariance} & \cell{2.0cm}{0.5cm}{$r(\Omega_m)$} & \cell{2.0cm}{0.5cm}{$r(\sigma_8)$} & \cell{2.0cm}{0.5cm}{$r(h)$}  \\
\hline
\cell{2.0cm}{0.5cm}{Gaussian} & 1.75 & 3.97 & 1.81  \\
\hline
\cell{2.0cm}{0.5cm}{Simulations} & 1.89 & 2.79 & 1.83  \\
\hline
\multicolumn{4}{|c|}{\parbox[c][0.8cm][c]{0.3\linewidth}{\centering $R = 15 \,\text{Mpc}\, h^{-1}$ $z=1$}}\\
\hline
\cell{2.0cm}{0.5cm}{Covariance} & \cell{2.0cm}{0.5cm}{$r(\Omega_m)$} & \cell{2.0cm}{0.5cm}{$r(\sigma_8)$} & \cell{2.0cm}{0.5cm}{$r(h)$}  \\
\hline
\cell{2.0cm}{0.5cm}{Gaussian} & 1.42 & 2.50 & 1.44  \\
\hline
\cell{2.0cm}{0.5cm}{Simulations} & 1.43 & 2.06 & 1.38  \\
\hline
\hline
\multicolumn{4}{|c|}{\parbox[c][0.8cm][c]{0.3\linewidth}{\centering $R = 30 \,\text{Mpc}\, h^{-1}$ $z=0$}}\\
\hline
\cell{2.0cm}{0.5cm}{Covariance} & \cell{2.0cm}{0.5cm}{$r(\Omega_m)$} & \cell{2.0cm}{0.5cm}{$r(\sigma_8)$} & \cell{2.0cm}{0.5cm}{$r(h)$}  \\
\hline
\cell{2.0cm}{0.5cm}{Gaussian} & 2.01 & 11.17 & 2.04  \\
\hline
\cell{2.0cm}{0.5cm}{Simulations} & 1.28 & 1.83 & 1.25  \\
\hline
\multicolumn{4}{|c|}{\parbox[c][0.8cm][c]{0.3\linewidth}{\centering $R = 30 \,\text{Mpc}\, h^{-1}$ $z=0.5$}}\\
\hline
\cell{2.0cm}{0.5cm}{Covariance} & \cell{2.0cm}{0.5cm}{$r(\Omega_m)$} & \cell{2.0cm}{0.5cm}{$r(\sigma_8)$} & \cell{2.0cm}{0.5cm}{$r(h)$}  \\
\hline
\cell{2.0cm}{0.5cm}{Gaussian} & 1.97 & 4.12 & 2.00  \\
\hline
\cell{2.0cm}{0.5cm}{Simulations} & 1.23 & 1.25 & 1.16  \\
\hline
\multicolumn{4}{|c|}{\parbox[c][0.8cm][c]{0.3\linewidth}{\centering $R = 30 \,\text{Mpc}\, h^{-1}$ $z=1$}}\\
\hline
\cell{2.0cm}{0.5cm}{Covariance} & \cell{2.0cm}{0.5cm}{$r(\Omega_m)$} & \cell{2.0cm}{0.5cm}{$r(\sigma_8)$} & \cell{2.0cm}{0.5cm}{$r(h)$}  \\
\hline
\cell{2.0cm}{0.5cm}{Gaussian} & 1.35 & 2.21 & 1.42  \\
\hline
\cell{2.0cm}{0.5cm}{Simulations} & 1.48 & 1.72 & 1.32  \\
\hline

\end{tabular}
&
\begin{tabular}{|c|c|c|c|}
\hline
\multicolumn{4}{|c|}{
\parbox[c][1.0cm][c]{0.3\linewidth}{\centering \large{Optimization of single parameters}}
}\\
\hline
\hline
\multicolumn{4}{|c|}{\parbox[c][0.8cm][c]{0.3\linewidth}{\centering $R = 10 \,\text{Mpc}\, h^{-1}$ $z=0$}}\\
\hline
\cell{2.0cm}{0.5cm}{Covariance} & \cell{2.0cm}{0.5cm}{$r(\Omega_m)$} & \cell{2.0cm}{0.5cm}{$r(\sigma_8)$} & \cell{2.0cm}{0.5cm}{$r(h)$}  \\
\hline
\cell{2.0cm}{0.5cm}{Gaussian} & 1.54 & 14.169 & 1.60  \\
\hline
\cell{2.0cm}{0.5cm}{Simulations} & 2.34 & 8.32 & 2.47 \\
\hline
\multicolumn{4}{|c|}{\parbox[c][0.8cm][c]{0.3\linewidth}{\centering $R = 10 \,\text{Mpc}\, h^{-1}$ $z=0.5$}}\\
\hline
\cell{2.0cm}{0.5cm}{Covariance} & \cell{2.0cm}{0.5cm}{$r(\Omega_m)$} & \cell{2.0cm}{0.5cm}{$r(\sigma_8)$} & \cell{2.0cm}{0.5cm}{$r(h)$}  \\
\hline
\cell{2.0cm}{0.5cm}{Gaussian} & 1.21 & 2.75 & 1.23  \\
\hline
\cell{2.0cm}{0.5cm}{Simulations} & 1.53 & 3.11 & 1.50  \\
\hline
\multicolumn{4}{|c|}{\parbox[c][0.8cm][c]{0.3\linewidth}{\centering $R = 10 \,\text{Mpc}\, h^{-1}$ $z=1$}}\\
\hline
\cell{2.0cm}{0.5cm}{Covariance} & \cell{2.0cm}{0.5cm}{$r(\Omega_m)$} & \cell{2.0cm}{0.5cm}{$r(\sigma_8)$} & \cell{2.0cm}{0.5cm}{$r(h)$}  \\
\hline
\cell{2.0cm}{0.5cm}{Gaussian} & 1.12 & 1.86 & 1.13  \\
\hline
\cell{2.0cm}{0.5cm}{Simulations} & 1.38 & 2.06 & 1.27  \\
\hline
\hline
\multicolumn{4}{|c|}{\parbox[c][0.8cm][c]{0.3\linewidth}{\centering $R = 15 \,\text{Mpc}\, h^{-1}$ $z=0$}}\\
\hline
\cell{2.0cm}{0.5cm}{Covariance} & \cell{2.0cm}{0.5cm}{$r(\Omega_m)$} & \cell{2.0cm}{0.5cm}{$r(\sigma_8)$} & \cell{2.0cm}{0.5cm}{$r(h)$}  \\
\hline
\cell{2.0cm}{0.5cm}{Gaussian} & 2.17 & 12.05 & 2.30  \\
\hline
\cell{2.0cm}{0.5cm}{Simulations} & 1.48 & 5.55 & 1.57   \\
\hline
\multicolumn{4}{|c|}{\parbox[c][0.8cm][c]{0.3\linewidth}{\centering $R = 15 \,\text{Mpc}\, h^{-1}$ $z=0.5$}}\\
\hline
\cell{2.0cm}{0.5cm}{Covariance} & \cell{2.0cm}{0.5cm}{$r(\Omega_m)$} & \cell{2.0cm}{0.5cm}{$r(\sigma_8)$} & \cell{2.0cm}{0.5cm}{$r(h)$}  \\
\hline
\cell{2.0cm}{0.5cm}{Gaussian} & 1.81 & 5.10 & 1.65  \\
\hline
\cell{2.0cm}{0.5cm}{Simulations} & 1.20 & 1.19 & 1.15  \\
\hline
\multicolumn{4}{|c|}{\parbox[c][0.8cm][c]{0.3\linewidth}{\centering $R = 15 \,\text{Mpc}\, h^{-1}$ $z=1$}}\\
\hline
\cell{2.0cm}{0.5cm}{Covariance} & \cell{2.0cm}{0.5cm}{$r(\Omega_m)$} & \cell{2.0cm}{0.5cm}{$r(\sigma_8)$} & \cell{2.0cm}{0.5cm}{$r(h)$}  \\
\hline
\cell{2.0cm}{0.5cm}{Gaussian} & 1.58 & 2.61 & 1.98  \\
\hline
\cell{2.0cm}{0.5cm}{Simulations} & 1.09 & 2.21 & 1.23  \\
\hline
\hline
\multicolumn{4}{|c|}{\parbox[c][0.8cm][c]{0.3\linewidth}{\centering $R = 30 \,\text{Mpc}\, h^{-1}$ $z=0$}}\\
\hline
\cell{2.0cm}{0.5cm}{Covariance} & \cell{2.0cm}{0.5cm}{$r(\Omega_m)$} & \cell{2.0cm}{0.5cm}{$r(\sigma_8)$} & \cell{2.0cm}{0.5cm}{$r(h)$}  \\
\hline
\cell{2.0cm}{0.5cm}{Gaussian} & 2.19 & 12.88 & 2.62  \\
\hline
\cell{2.0cm}{0.5cm}{Simulations} & 1.72 & 2.50 & 1.69  \\
\hline
\multicolumn{4}{|c|}{\parbox[c][0.8cm][c]{0.3\linewidth}{\centering $R = 30 \,\text{Mpc}\, h^{-1}$ $z=0.5$}}\\
\hline
\cell{2.0cm}{0.5cm}{Covariance} & \cell{2.0cm}{0.5cm}{$r(\Omega_m)$} & \cell{2.0cm}{0.5cm}{$r(\sigma_8)$} & \cell{2.0cm}{0.5cm}{$r(h)$}  \\
\hline
\cell{2.0cm}{0.5cm}{Gaussian} & 2.24 & 4.12 & 2.17  \\
\hline
\cell{2.0cm}{0.5cm}{Simulations} & 1.29 & 1.25 & 1.19  \\
\hline
\multicolumn{4}{|c|}{\parbox[c][0.8cm][c]{0.3\linewidth}{\centering $R = 30 \,\text{Mpc}\, h^{-1}$ $z=1$}}\\
\hline
\cell{2.0cm}{0.5cm}{Covariance} & \cell{2.0cm}{0.5cm}{$r(\Omega_m)$} & \cell{2.0cm}{0.5cm}{$r(\sigma_8)$} & \cell{2.0cm}{0.5cm}{$r(h)$}  \\
\hline
\cell{2.0cm}{0.5cm}{Gaussian} & 2.07 & 2.77 & 2.13  \\
\hline
\cell{2.0cm}{0.5cm}{Simulations} & 1.36 & 1.67 & 1.26  \\
\hline

\end{tabular}\end{tabular}
\caption{ \justifying Improvement on the parameter error bars for Gaussian and simulation-based covariances.}
\label{tab:improv_all_app}
\end{table}

\begin{figure}
    \centering
    \includegraphics[width=\textwidth]{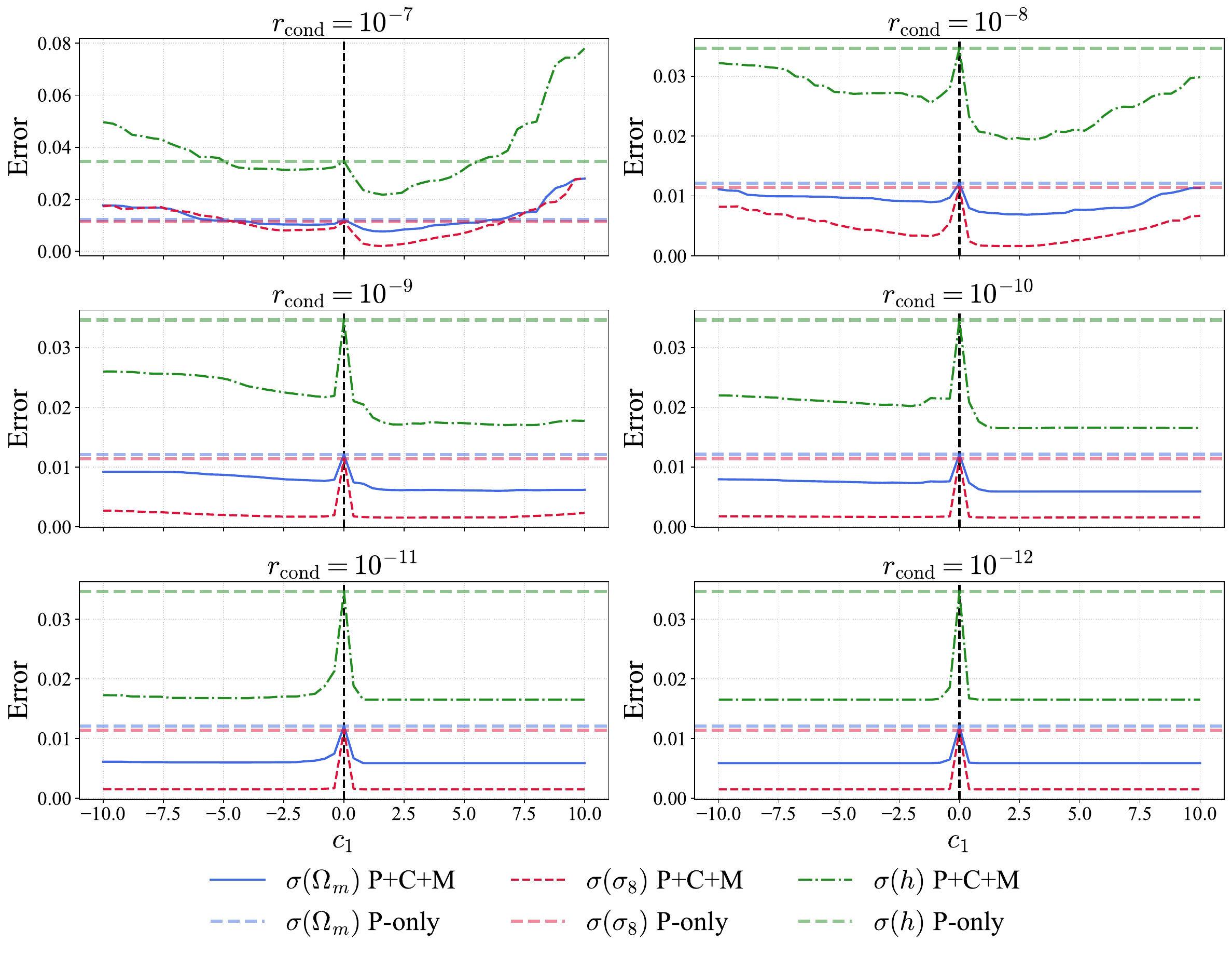}
    \caption{Dependence of the forecasted errors on the conditioning number.}
    \label{fig:fcond}
\end{figure}

We also show in Fig.~\ref{fig:fcond} a simple example used to estimate the value for $f_{\rm cond}$. We vary only $c_1$ and compare the error bars on the different cosmological parameters obtained using only $P$ and the combination $P+C+M$, with $c_s^2$ fixed to make it easier to visualize

\bibliography{ref}

\end{document}